\documentclass[twocolumn]{article}

\usepackage[subpreambles=true]{standalone}
\usepackage{import}

\usepackage[english]{babel}
\usepackage{authblk}

\usepackage[a4paper,top=2cm,bottom=2cm,left=1.5cm,right=1.5cm,marginparwidth=1.75cm]{geometry}

\usepackage[style=numeric-comp, sorting=none]{biblatex}

\renewbibmacro*{issue+date}{%
    \iffieldundef{issue}{}{%
        \printtext[parens]{\printfield{issue}}%
        \setunit*{\addspace}%
    }%
    \printtext[parens]{\printfield{year}} 
}

\usepackage{csquotes}
\usepackage{amsmath}
\usepackage{amssymb}
\usepackage{braket}
\usepackage{mathtools}
\usepackage{multirow}
\usepackage{booktabs}
\usepackage{makecell}
\usepackage{bm}
\usepackage{siunitx}
\usepackage{graphicx}
\usepackage{rotating}
\usepackage{quantikz}
\usepackage{adjustbox}
\usepackage{float}

\usepackage{url}
\usepackage[colorlinks=true, allcolors=blue]{hyperref}
\usepackage{orcidlink}

\usepackage{caption}
\usepackage{subcaption}

\usepackage{dblfloatfix}
\usepackage{calc}

\makeatletter
\def\Cline#1#2{\@Cline#1#2\@nil}
\def\@Cline#1-#2#3\@nil{%
  \omit
  \@multicnt#1%
  \advance\@multispan\m@ne
  \ifnum\@multicnt=\@ne\@firstofone{&\omit}\fi
  \@multicnt#2%
  \advance\@multicnt-#1%
  \advance\@multispan\@ne
  \leaders\hrule\@height#3\hfill
  \cr}
\makeatother

\newcommand{\oplusr}{\mathbin{\raisebox{0.2ex}{$\oplus$}}}

\title{Efficient Preparation of Resource States \\ for Hamiltonian Simulation and Universal Quantum Computation}
\author[1,2]{Thierry N. Kaldenbach~\orcidlink{0009-0008-5607-4427}*}
\author[2]{Isaac D. Smith~\orcidlink{0000-0002-2598-1656}}
\author[2]{\\Hendrik Poulsen Nautrup~\orcidlink{0000-0001-7815-7006}}
\author[3,4]{Matthias Heller~\orcidlink{0000-0002-4774-5072}}
\author[2]{Hans J. Briegel~\orcidlink{0000-0002-9065-1565}}

\affil[1]{Institute of Materials Research, German Aerospace Center (DLR), Cologne, Germany}
\affil[2]{University of Innsbruck, Department of Theoretical Physics, Innsbruck, Austria}
\affil[3]{Fraunhofer Institute for Computer Graphics Research IGD, Darmstadt, Germany}
\affil[4]{Technical University of Darmstadt, Interactive Graphics Systems Group, Darmstadt, Germany}

\date{\today}

\begin{document}

\twocolumn[
    \begin{@twocolumnfalse}
    \maketitle
    \vspace{-0.7cm}
    \begin{abstract}
    \begin{center}
    \begin{minipage}{0.8\textwidth}
        The direct compilation of algorithm-specific graph states in
        measurement-based quantum computation (MBQC) can lead to resource reductions in terms of circuit depth, entangling
        gates, and even the number of physical qubits. In this work, we
        extend previous studies on algorithm-tailored graph states to periodic
        sequences of generalized Pauli rotations, which commonly appear in, e.g., Trotterized Hamiltonian simulation. We first implement an enhanced simulated-annealing-based algorithm to find optimal periodic graph states within local-Clifford equivalent MBQC resources.
        In addition, we derive a novel scheme for the preparation of resource states based on a graph state and a ladder of CNOT gates, which we term anticommutation-based MBQC, since it uncovers a direct relationship between the graph state and the anticommutation matrix for the set of Hamiltonians generating the computation.
        We also deploy our two approaches to derive universal
        resource states from minimal universal sets of generating Hamiltonians.
        Finally, we demonstrate
        and compare both of our methods based on various examples from condensed matter physics and universal quantum computation. 
        \\ \\
        DOI: \href{https://doi.org/10.48550/arXiv.2509.05404}{10.48550/arXiv.2509.05404} \\ *Contact author: \url{thierry.kaldenbach@dlr.de}
    \end{minipage}
    \end{center}
    \end{abstract}
    \vspace{0.5cm}
    \end{@twocolumnfalse}
]

\tableofcontents

\clearpage

\section{Introduction}

    Quantum computation can be described through a number of different models, with the most frequently adopted one being the {\em quantum circuit model}, in which unitary operations are realized through quantum gates and measurements typically only occur at the end. 
    
    Alternatively, there is the model of {\em measurement-based quantum computation} (MBQC) \cite{Raussendorf20011WQC, briegel2009measurement}, which implements unitary operations through adaptive single-qubit measurements and corrections acting on an entangled resource state, the paradigmatic example of which is the 2D cluster state \cite{Briegel2001Persistent, Raussendorf2003Cluster}, a special type of graph state \cite{hein2006entanglement}. 
    
    Both of these models can straightforwardly be used for Hamiltonian simulation, i.e.,~to emulate the time evolution of quantum mechanical systems. This application led to the proposal of quantum computing in the first place \cite{Feynman1982Simulating}. Hamiltonian simulation is commonly achieved using Trotter-Suzuki decompositions \cite{trotter1959product, suzuki1976generalized} of the time evolution operator, commonly referred to as {\em Trotterization}. Such a decomposition typically leads to a periodic sequence of generalized Pauli rotations, which manifests itself as a repeating circuit or measurement pattern within the respective quantum computing models. 
    
    In the case of MBQC, a computation will typically include Pauli measurements on certain parts of the resource state (effectively, the Clifford parts of the corresponding circuit), and non-Pauli measurements on the rest. However, it is known that, for MBQC using graph states, (i) all Pauli measurements can be performed simultaneously and in the first round of measurement \cite{browne2007generalized,mhalla2022characterising}, and (ii) that these measurements induce a graphical transformation on the underlying graph states \cite{Hein2004Multiparty}. The classical preprocessing of such operations is efficient due to the Gottesman-Knill theorem \cite{Gottesman1998Heisenberg}.
    
    Accordingly, we can consider the resource state that results {\em after} all the Pauli measurements have been applied for a given computation. This resource state, which we refer to as {\em algorithm-specific}, is local Clifford equivalent to a graph state, and it is the direct compilation of these states that concerns us in this work. It is worth pointing out that the graph transformation rules employed to obtain the algorithm-specific graph state are not unique and may not inherit the periodic structure of the underlying computation.
    
    A compelling alternative to the classical preprocessing in terms of graphical rules lies in the {\em direct compilation} of the algorithm-specific graph state for LC-MBQC \cite{Vijayan2024Compilation, Kaldenbach2025Mapping}, which eliminates the need for treating Pauli measurements both on the classical- and the quantum computer.
    The computation can then be carried out using a sequence of non-Pauli measurements, thus employing some of the truly \enquote{quantum} aspects of the quantum computer \cite{veitch2014resource,PhysRevX.6.021043,li2024measurement}.
    Such techniques essentially rely on a conversion from the circuit model to MBQC using gate teleportation of non-Clifford gates and have already been demonstrated in the context of Clifford+$T$ circuits in Ref.~\cite{Vijayan2024Compilation}, and for generalized Pauli rotations in Refs.~\cite{Kaldenbach2025Mapping, PhysRevLett.132.220602}.
    While both classes of circuits are universal, for the purpose of this work we stick to Pauli rotations as they naturally arise in Trotter circuits. This gives a one--to--one correspondence between the number of Pauli rotations and number of auxiliary qubits employed in the measurement pattern.
    
    The work from \cite{Kaldenbach2025Mapping} makes use of the non-deterministic algorithm to simulate Clifford operations on graph states from Ref.~\cite{Anders2006Fast} in order to obtain a graph state representation of the algorithm-specific resource state. 
    This graph state is not unique and can be replaced by any {\em local-Clifford-equivalent} (LC-equivalent) graph state. 
    Given that any two LC-equivalent graph states can be transformed into each other using a sequence of local complementations \cite{Nest2004Graphical}, one can use this operation to span the entire graph state orbit \cite{Adcock2020Mapping}.
    This degree of freedom has been explored to minimize the number of edges using simulated annealing \cite{Kirkpatrick1983Optimization}; see,  e.g.,~Refs.~\cite{Kaldenbach2025Mapping} and \cite{Sharma2025Minimizing}.
    We refer to this computational model entailing purely non-Pauli measurements on a graph state which is optimized within its local complementation orbit as \textit{LC-MBQC}.
    
    Both graphical and direct compilation schemes for algorithm-specific graph states suffer from producing non-periodic graphs, which in addition contain long-ranging edges. Given that in MBQC, only a subset of qubits in the whole graph need to be active at each time step, these long-ranging edges unnecessarily grow the number of {\em active qubits} in LC-MBQC. The number of these active qubits depends on the entanglement properties of the graph \cite{browne2007generalized,mhalla2022characterising,PhysRevA.74.052310,Booth_2023}.
    
    Motivated by the drawbacks of such compilation schemes, our work adresses the following questions: 
    can we directly compile an algorithm-specific graph state that is both periodic {\em and } resource efficient? That is, a graph state which (i) has a periodically repeating structure, (ii) has a division into ``blocks'' with no edge spanning multiple blocks, and (iii) can implement the desired computation with only a few physical qubits ``active'' at a given time?
    Here, we provide a partial answer: we find many instances where our approach does indeed produce a periodic resource state requiring few active qubits, but there are other examples where no such state has been found. If, however, we allow for slightly more general processing, i.e., using some two-local Clifford operations, then the answer is \enquote{yes} and
    we demonstrate a scheme which works for any Hamiltonian. In more detail, our contributions to answer these questions are as follows:
    \begin{itemize}
        \item We provide a method for deterministically producing an algorithm-specific resource state in terms of its stabilizers, but also in terms of a graph state representation. The latter presents an explicit solution to the stabilizer-to-graph algorithm from Ref.~\cite{Nest2004Graphical}.
        \item Given a resource state produced in the manner above, we provide an enhanced optimization algorithm based on simulated annealing with local complementations, which permits one to find periodic graph states which minimize the number of edges and active qubits simultaneously. This optimization is non-deterministic and not guaranteed to converge. In fact, we provide explicit examples where such a periodic and efficient solution can not be found.
        At the same time, we provide workarounds to the above convergence issue and recover resource efficient graphs based on hybrid simulation \cite{Kaldenbach2025Mapping} and graph state embedding theory \cite{Hoyer2006Resources}. 
        \item We further provide a new deterministic preparation scheme for the resource state, which is obtained from a graph state by applying a ladder of CNOT gates. This scheme is guaranteed to work by construction for {\em any} Hamiltonian. We show how the underlying graph state boils down to the Pauli anticommutation graph of the Hamiltonian governing the time evolution, i.e.,~the graph obtained by computing the pairwise anticommutation of Hamiltonian terms in a Pauli-string decomposition.
        We refer to this new anticommutation-based approach as \textit{AC-MBQC}. 
    \end{itemize}
    
    In addition to the use-case of designing algorithm-specific resource states for Hamiltonian simulation, we also consider a further application of our methods, namely in producing universal resource states from {\em minimal generating sets} of Pauli-strings \cite{smith2025optimally}. By employing such sets as Hamiltonians, we can compile universal resource states using both the LC- and AC-MBQC approaches. 
    The utility of our approach stems from the {\em compilation rate}, which, in the context of MBQC, essentially corresponds to the resources required to implement each Pauli-string rotation. It turns out that some sets are less resource-intensive than others, which essentially boils down to the number of anticommuting pairs of elements from the set \cite{smith2025optimally}. As a consequence, we can deploy our algorithm to construct resource states which can natively implement rotations of the elements from such a set, and thus achieve universality with on average fewer measurements than, e.g., the cluster state model. 
    
    
    The remainder of this paper is structured as follows. In Sec.~\ref{sec:Preliminaries}, we introduce core concepts such as the stabilizer formalism and its binary representation (Sec.~\ref{subsec:Stabilizer}), the graph state representation of stabilizer states (Sec.~\ref{subsec:GraphState}), and a short introduction into MBQC (Sec.~\ref{subsec:MBQC}). Readers familiar with those subjects are encouraged to skip ahead to Sec.~\ref{sec:graph_state_algorithm} where we introduce the deterministic graph state construction. In Sec.~\ref{subsec:ResourceStabilizers}, we fully characterize the algorithm-specific resource state in terms of its stabilizers. Based on that, we explicitly convert these stabilizers to a graph state in Sec.~\ref{subsec:Resource_GraphState} and generalize our results to periodic operations. In Sec.~\ref{sec:ResourcePrep}, we introduce the two resource-state preparation schemes. The first scheme, LC-MBQC (Sec.~\ref{subsec:LC-MBQC}), is based on simulated annealing, and the second scheme, AC-MBQC (Sec.~\ref{subsec:AC-MBQC}), is based on the anticommutation graphs. In Sec.~\ref{sec:Hybrid}, we recap the hybrid simulation scheme from Ref.~\cite{Kaldenbach2025Mapping} and detail how to compute resources for specific observables based on our new findings. After establishing the theory, in Sec.~\ref{sec:Applications} we apply our procedures to various examples including time evolution in condensed matter physics and topological quantum error correction (Sec.~\ref{subsec:TimeEvolution}), as well as different realizations of universal quantum computation (Sec.~\ref{subsec:MinimalGeneratingSets}). Finally, we discuss the implications of our work and detail future directions of research in Sec.~\ref{sec:Discussion}. 

\newpage

\section{Preliminaries} \label{sec:Preliminaries}

In this work, we are predominantly concerned with unitary dynamics and projective measurements applied to pure states of qubit quantum systems. All state spaces, denoted $\mathcal{H}$, are taken to be $\mathbb{C}^{2^{N}}$ for some $N$ denoting the number of qubits. Pure states will be represented using ket-notation, i.e.,~$\ket{\psi} \in \mathcal{H}$ denotes an $N$-qubit state. When referring to unitary operators $U$, we will frequently make reference to the corresponding (time-independent) Hamiltonian, i.e.,~the Hermitian operator $H$ such that, for some $t \in \mathbb{R}$,
\begin{align}
U = e^{-iHt}.
\end{align}
In particular, the Hamiltonian will often take a particular form, namely that of a Pauli-string (up to a real scale factor). An $N$-qubit {\em Pauli-string} $\mathcal{P}$ is a Hermitian operator of the form
\begin{align}
    \mathcal{P} = P_{1} \otimes P_{2} \otimes \dots \otimes P_{N}
    \label{eq:PauliString}
\end{align}
where each $P_{j} \in \{I,X,Y,Z\}$, with the latter set denoting the single-qubit Pauli operators. 
Let us denote the set of all $N$-qubit Pauli-strings by $\mathbb{P}_{N}$.

As a large part of the focus of this work is on measurement-based quantum computation (MBQC - see below), we will make frequent reference to measurements. These measurements will typically be single-qubit projective measurements, which we will denote either with reference to an observable, such as a Pauli-$X$ measurement, or with reference to the corresponding projections, i.e., $\{\ket{+}\!\!\bra{+}, \ket{-}\!\!\bra{-} \}$.

\subsection{Stabilizer States} \label{subsec:Stabilizer}

A general $N$-qubit quantum state $\ket{\psi} \in \mathcal{H}$ requires exponentially many (in $N$) complex values to be fully specified. However, there are certain classes of quantum states, such as stabilizer states \cite{gottesman1997stabilizer,nielsen2010quantum} and tensor network states \cite{ORUS2014117}, that are efficiently representable.

A stabilizer state is defined with reference to the notion of the stabilizer group. Let $\mathcal{G}_{N}$ denote the the group of $N$-qubit Pauli-strings, that is,
\begin{align}
    \mathcal{G}_{N} := \{\pm 1, \pm i\} \times \{\mathcal{P} | \mathcal{P} \in \mathbb{P}_{N} \}
\end{align}
equipped with a group operation given by matrix multiplication. An $N$-qubit {\em stabilizer group} $\mathcal{S}_{0}$ is a subgroup of $\mathcal{G}_{N}$ that satisfies (i) $-I^{\otimes N} \notin \mathcal{S}_{0}$, (ii) all $A,B \in \mathcal{S}_{0}$ mutually commute (i.e., $[A,B] = 0$, where $[\cdot, \cdot]$ denotes the matrix commutator).  A consequence of condition (i) is that every element $A \in \mathcal{S}_{0}$ is of the form $\pm \mathcal{P}$ for some Pauli-string $\mathcal{P}$; that is, only real phases are possible. 

It is often convenient to specify $\mathcal{S}_{0}$ by a set of (independent) group generators\footnote{Note that the term ``generator'' will be used in a distinct fashion below, so when referring to the generators of the stabilizer group $\mathcal{S}_{0}$ we will typically use the term ``stabilizer''.}, i.e.
\begin{align}
    \mathcal{S}_{0} = \langle \mathcal{S}_{0}^{(1)}, \dots, \mathcal{S}_{0}^{(k)} \rangle,
\end{align}
where the operators on the right-hand side are such that any $A \in \mathcal{S}_{0}$ can be expressed as a (finite) product of the $\mathcal{S}_{0}^{(j)}$. In the case where $k = N$, we can define a stabilizer state: the {\em stabilizer state} for $\mathcal{S}_{0} = \langle \mathcal{S}_{0}^{(1)}, \dots, \mathcal{S}_{0}^{(N)}\rangle$ is the unique state $\ket{\psi} \in \mathcal{H}$ such that $A\ket{\psi} = \ket{\psi}$ for all $A \in \mathcal{S}_{0}$.

\paragraph{Binary Representation of Stabilizer States - The Stabilizer Tableau}
It is possible to represent any element $A \in \mathcal{S}_{0}$ using $2N+1$ binary values. Since $A = \pm \mathcal{P}$ for some Pauli-string $\mathcal{P}$, we can write
\begin{align}
A = (-1)^{r} P_{1} \otimes P_{2} \otimes \dots \otimes P_{N}
\end{align}
where $r \in \{0,1\}$ encodes whether the phase is $+1$ or $-1$. The remaining $2N$ binary values, denoted $x_{1},\dots,x_{N},z_{1},\dots,z_{N}$, specify the single-qubit Pauli-operators for each tensor factor. That is, for tensor factor $j \in \{1, \dots, N\}$, 
\begin{align}
x_{j},z_{j} = \begin{cases} 0,0 & \text{ if } P_{j} = I, \\
0,1 & \text{ if } P_{j} = Z, \\
1,0 & \text{ if } P_{j} = X, \\
1,1 & \text{ if } P_{j} = Y. 
\end{cases}
\end{align}
It is convenient to extend this representation also to the $N$ stabilizers $\mathcal{S}_{0}^{(1)},\dots,\mathcal{S}_{0}^{(N)}$ generating the stabilizer group $\mathcal{S}_{0}$, as follows. Collecting the binary vectors for each of the $N$ generators gives rise to an $N\times (2N+1)$ binary-valued matrix
\begin{equation}
    \left[
    \begin{array}{cccc|cccc|c}
        x_{11} & x_{12} & \cdots & x_{1N} & z_{11} & z_{12} & \cdots & z_{1N} & r_1 \\
        x_{21} & x_{22} & \cdots & x_{2N} & z_{21} & z_{22} & \cdots & z_{2N} & r_2 \\
        \vdots & \vdots & & \vdots & \vdots & \vdots & & \vdots & \vdots\\
        x_{N1} & x_{N2} & \cdots & x_{NN} & z_{N1} & z_{N2} & \cdots & z_{NN} & r_N
    \end{array}
    \right].
    \label{eq:StabilizerTableau}
\end{equation}
As the stabilizers $\mathcal{S}_{0}^{(1)},\dots,\mathcal{S}_{0}^{(N)}$ generate the group $\mathcal{S}_{0}$, which, in turn, uniquely specifies the corresponding stabilizer state $\ket{\psi}$, the above binary matrix is thus a compact representation of $\ket{\psi}$.

One can always obtain a new set of stabilizers using $\mathcal{S}_0^{n} \leftarrow \mathcal{S}_0^{(n)} \mathcal{S}_0^{(m)}$ for all $n$ with $n\neq m$. This changes the stabilizer string, but also potentially its real-valued phase.

Given the product of two single-qubit Pauli operators $P_1$ and $P_2$ with their respective binary encodings $[x_1|z_1]$ and $[x_2|z_2]$, one computes the resulting Pauli operator as $[x_1 \oplus x_2|z_1 \oplus z_2]$, and express the arising phase as $i^{g(x_1, z_1, x_2, z_2)}$, where $g$ follows the definition from Ref.~\cite{Aaronson2004Improved}:
\begin{equation}
    g(x_1, z_1, x_2, z_2) \coloneqq 
    \begin{cases}
        0           &\text{if } x_1=z_1=0,   \\
        z_2-x_2     &\text{if } x_1=z_1=1,   \\
        z_2(2x_2-1) &\text{if } x_1=1, z_1=0,\\
        x_2(1-2z_2) &\text{if } x_1=0, z_1=1.
    \end{cases}
    \label{eq:ExponentFunction}
\end{equation}
For the product of two $N$-qubit Pauli operators $(-1)^{r_j} \mathcal{P}^{(j)}$, $(-1)^{r_k} \mathcal{P}^{(k)}$, the two prefactors, as well as all qubit-wise phases arising from Eq.~\eqref{eq:ExponentFunction}, are taken into account, which yields the overall phase $i^{r'}$ with
\begin{equation}
    r' = 2r_j + 2 r_k + \sum_{l=1}^N g(x_{jl}, z_{jl}, x_{kl}, z_{kl}) 
    \label{eq:StringExponentFunction}
\end{equation}
Given two commuting Pauli-strings, as we always have for Pauli-strings corresponding to elements of the stabilizer group, the phase must be real and thus the exponent of $i$ is even, and we can express the resulting phase as $(-1)^r$ with
\begin{equation}
    r = r_j + r_k + \frac{1}{2}\sum_{l=1}^N g(x_{jl}, z_{jl}, x_{kl}, z_{kl}) 
    \label{eq:ExponentFunctionCommuting}
\end{equation}

\paragraph{Clifford Operations}

Alongside the set of $N$-qubit Pauli group $\mathcal{G}_{N}$, it is prudent to consider the set of $N$-qubit {\em Clifford operations}. Using the notation $C \mathcal{G}_{N} C^{\dagger} = \mathcal{G}_{N}$ to mean that, for every $i^{r}\mathcal{P} \in \mathcal{G}_{N}$, $i^{r}C\mathcal{P}C^{\dagger} \in \mathcal{G}_{N}$, we can define the set of $N$-qubit Clifford operations as the set of unitaries
\begin{align}
\textrm{Cl}_{N} := \{C | C \mathcal{G}_{N}C^{\dagger} = \mathcal{G}_{N}\}.
\end{align}
Formally, the Clifford group is the normalizer of the Pauli group. In effect, the Clifford operations map Pauli-strings to Pauli-strings. 

In this work, we mostly consider elements of $\mathcal{G}_{N}$ with a real phase, i.e., elements of the form $(-1)^{r}\mathcal{P}$ for some Pauli-string $r$. In terms of the binary representation of such elements, conjugation by a Clifford operator $C$ corresponds to a map $f_{C}:\{0,1\}^{2N+1} \rightarrow \{0,1\}^{2N+1}$: for $(-1)^{r}\mathcal{P}$ with corresponding binary string $x_{1},\dots,x_{N},z_{1},\dots,z_{N},r$, $(-1)^{r}C\mathcal{P}C^{\dagger}$ will have a corresponding binary string $f_{C}(x_{1},\dots,x_{N},z_{1},\dots,z_{N},r)$. A {\em local} Clifford operator $C \in \textrm{Cl}_{N}$ is such that, for all $\mathcal{P} \in \mathbb{P}_{N}$, $\mathcal{P}$ and $C \mathcal{P}C^{\dagger}$ differ in at most $1$ tensor factor.

Many of the Clifford operators considered in this work are local. This is, in part, due to the fact that each stabilizer state is equivalent to some graph state (see below) up to local Clifford operations on some qubits. However, we will also make frequent use of the two-qubit Clifford operation known as the CNOT or controlled Pauli-$X$ gate, which we write as $\Lambda_{c}(X_{t})$ for a CNOT between a control qubit $c$ and target qubit $t$. For later reference, it is worthwhile considering the effect of $\Lambda_{c}(X_{t})$ on  $(-1)^{r}\mathcal{P} \in \mathcal{G}_{N}$ in terms of the mapping of the corresponding binary string: if the binary string corresponding to $\mathcal{P}$ is $x_{1}, \dots, x_{N},z_{1}, \dots, z_{N}$, then the binary string corresponding to $(-1)^{r}\Lambda_{c}(X_{t})\mathcal{P}\Lambda_{c}^{\dagger}(X_{t})$ is given by 

\begin{align}
    x'_{j},z'_{j} \coloneqq \begin{cases} x_{j},z_{j} &\text{if } j \neq c,t, \\
    x_{c},z_{c} \oplus z_{t} &\text{if } j = c,\\ 
    x_{t} \oplus x_{c},z_{t} &\text{if } j = t. 
    \end{cases}
\end{align}

\subsection{Graph States} \label{subsec:GraphState}

Among all stabilizer states, there exists a canonical family known as graph states. These states, as the name indicates, are defined with respect to the mathematical notion of graphs. Let $G(V, E)$ denote a (connected, simple) graph with vertex set denoted by $V$ and edge set denoted by $E \subseteq V \times V$.\footnote{Despite denoting an edge as a tuple $(u,v) \in V \times V$, the ordering of the tuple is not taken into consideration, i.e. $(u,v)$ and $(v,u)$ correspond to the same element.} For each $v \in V$, the neighbourhood of $v$ in $G$ is given by
\begin{align}
    N_{G}(v) := \{u \in V| (u,v) \in E \}.
\end{align}

One way to go from a graph to a graph state is by defining a stabilizer group based on the former. For a given graph $G$, and for each $v \in V$, we can define the following $|V|$-qubit Pauli-string associated to $v$:
\begin{align}
    \mathcal{K}_{v} := X_{v} \otimes \bigotimes_{u \in N_G(v)} Z_{u},
    \label{eq:GraphStateStabilizerString}
\end{align}
where it is to be understood that all other tensor factors are the identity. Noting that $\langle \mathcal{K}_{v} | v\in V \rangle$ is a valid stabilizer group and that the $|V|$ operators $\mathcal{K}_{v}$ are independent generators of that group, we can define the {\em graph state} $\ket{G}$ corresponding to $G$ to be the stabilizer state corresponding to $\mathcal{S}_{0} := \langle \mathcal{K}_{v} | v\in V \rangle$. 

Recall from above that, given a set of stabilizers generating the stabilizer group $\mathcal{S}_{0}$, we can write down the corresponding $|V| \times (2|V|+1)$ binary matrix. For the stabilizers $\mathcal{K}_{v}, v \in V$ generating the stabilizer group for a graph state $\ket{\textrm{G}}$, this matrix takes an a particularly nice block form
\begin{equation}
    \left[
    \begin{array}{c|c|c}
        \textrm{I} & \Gamma & \bm{0} 
    \end{array}
    \right],
    \label{eq:GraphStateTableau}
\end{equation}
where $\textrm{I}$ denotes the $|V| \times |V|$ identity matrix\footnote{A comment regarding notation: throughout this work, the Pauli operators $X$ and $Z$, as well as the identity operator $I$, will be denoted using a slanted math font, while the binary block matrices related to the binary representation of Pauli strings will be denoted using an upright font, namely as $\textrm{X}$ and $\textrm{Z}$. If such a block is diagonal, it will be denoted using $\textrm{I}$.}, $\Gamma$ denotes the adjacency matrix of $G$, and $\bm{0}$ denotes the length-$|V|$ column vector of all $0$s.

\paragraph{From Stabilizer Tableau to Graph States}
It is with respect to the above binary matrix representation that the claim made above --- that graph states are canonical amongst stabilizer states --- is most clear. If we allow also for disconnected graphs, then {\em every} stabilizer state is equivalent to a graph state $\ket{G}$ for some $G$ under local Clifford operations \cite{Nest2004Graphical}. That is, every binary matrix of the form given in Eq.~\eqref{eq:StabilizerTableau} can be brought into the form of Eq.~\eqref{eq:GraphStateTableau} for some $\Gamma$, by following an algorithm closely related to Gaussian elimination. This algorithm was introduced in \cite{Nest2004Graphical}; see also Ref.~\cite{Vijayan2024Compilation} for a
pedagogical description of the algorithm. 
As we explicitly carry out this algorithm step-by-step in our work, we delay presenting further details to a later section (see Sec.~\ref{subsec:Resource_GraphState}).

\paragraph{Local Complementation}
Below, we will require the graph transformation rule known as local complementation \cite{bouchet1988transforming,bouchet1991efficient}. In the following, we mostly adopt the notation from Ref.~\cite{Adcock2020Mapping}. Local complementation at a vertex $v\in V$, denoted $\textrm{LC}_v$, modifies a graph $G(V, E)$ by complementing the subgraph formed by the neighborhood of $v$ -- that is, it removes existing edges in $N_G(v) \times N_G(v)$ and adds any that are missing. The action of a local complementation can be compactly denoted by first introducing the set of edges of the complete graph on the neighborhood $N_G(v)$, which we denote by $E_G(v)$. Then we have
\begin{align}
    \textrm{LC}_v(G(V, E)) \coloneqq G(V, E\Delta E_G(v)), 
    \label{eq:LocalComplementation}
\end{align}
where the new set of edges is computed via the symmetric difference $\Delta$, i.e., for sets $A$ and $B$, $A \Delta B := (A \cup B) \setminus (A \cap B)$. That is, the new edge set is the same as that for $G$ except for the local neighbourhood of $v$, which has been complemented. 

There is a straightforward relationship between the local complementation on a graph $G$, and local Clifford gates  acting on $\ket{G}$. By defining the unitary $U_G^{\textrm{LC}}(v)$ consisting of local Clifford gates via
\begin{align}
    U_G^{\textrm{LC}}(v) = \sqrt{iX_v} \bigotimes_{w\in N_G(v)} \sqrt{-i Z_w},
    \label{eq:LocalComplementationCliffords}
\end{align}
we get that $\ket{G}$ and $\ket{\textrm{LC}_v(G)}$ are related via
\begin{align}
    \ket{G} = U_G^\textrm{LC}(v) \ket{\textrm{LC}_v(G)}.
\end{align}
The notion of local complementation becomes even more powerful given the following statement: Any two graph states $\ket{G}$ and $\ket{G'}$ are LC-equivalent if and only if $G$ and $G'$ are related by a sequence of local complementations \cite{Nest2004Graphical}. In addition, without knowing the precise sequence of local complementations, it is possible to efficiently check whether two graph states $\ket{G}$ and $\ket{G'}$ are LC-equivalent, and if so, compute the local Clifford operator \cite{Nest2004Efficient, bouchet1991efficient}. 

Due to the importance of graph and stabilizer states for a range of quantum information processing tasks, there has been much effort spent on classifying the local Clifford equivalence classes of graph states (see, e.g., Refs.~ \cite{Hein2004Multiparty,hein2006entanglement,danielsen2008edge}) and various graphical properties that are invariant under local complementation are known (see, e.g., Refs.~\cite{PhysRevA.72.014307,PhysRevA.71.022310,claudet2024local,Burchardt2025foliagepartition}). 

\paragraph{Pauli Measurements}

An important subset of the allowed measurements for MBQC are those associated to the Pauli observables $X$, $Y$ and $Z$. There are two key results related to Pauli measurements on graph states that are relevant for our purposes. The first is that, for a given graph state $\ket{G}$ that supports deterministic computation, {\em all} the Pauli measurements can be performed simultaneously and as the first round of measurements in the partial order \cite{browne2007generalized,mhalla2022characterising}. The second is that performing a Pauli measurement on a graph state $\ket{G}$ produces, up to local Clifford operations, another graph state $\ket{G'}$ according to well-understood transformation rules associated to local complementation and vertex deletion \cite{hein2006entanglement}, which we review below. A consequence of these two facts is that, for a given measurement-based computation on $\ket{G}$ containing Pauli measurements, it is possible to perform a pre-processing step to obtain a different resource state consisting of fewer qubits upon which only non-trivial (i.e.,~non-Pauli) measurements are performed. These resultant algorithm-specific resource states are a primary object of interest in this work.

Let $\ket{G}$ be an $N$-qubit graph state, $v\in V$ a vertex of the corresponding graph $G(V, E)$ with a non-empty neighbourhood $N_G(v)$, and $w \in N_G(v)$ a distinguished (but arbitrary) neighbor. Let us temporarily use the notation $\ket{P, \pm}$ to denote the positive and negative eigenvectors for $P \in \{X,Y,Z\}$, i.e., $\ket{Z,+} = \ket{0}$, $\ket{Z,-} = \ket{1}$, $\ket{X,\pm} = \ket{\pm}$ and $\ket{Y,\pm} = \ket{\pm i} = (\ket{0} \pm i\ket{1})/\sqrt{2}$. A Pauli-$P$ measurement is then defined as a measurement in the $\{\ket{P,\pm}\!\!\bra{P, \pm}\}$ basis. 
We start with the simplest type of Pauli measurement on a graph state, namely the Pauli-$Z$ measurement. For that purpose, let us define the following unitaries:
\begin{align}
    U_{v}^{Z,+} \coloneqq I^{\otimes N-1}, \quad \textrm{and} \quad 
    U_{v}^{Z,-} \coloneqq \bigotimes_{w \in N_G(v)} Z_{w}.
    \label{eq:GraphStatePauliMeasurementZByproduct}
\end{align}
Let us also define the post-measurement graph $G'(v,P)$ via \cite{hein2006entanglement}
\begin{align}
    G'(v,Z) &= G \setminus \{v\},  \label{eq:GraphStatePauliMeasurementZ} \\
    G'(v,Y) &= \textrm{LC}_v(G) \setminus \{v\},  \label{eq:GraphStatePauliMeasurementY} \\ 
    G'(v,X) &= \textrm{LC}_v(\textrm{LC}_w(\textrm{LC}_v(G))) \setminus \{v\},  \label{eq:GraphStatePauliMeasurementX}
\end{align}
where $G \setminus \{v\}$ denotes the deletion of vertex $v$ and all edges incident to $v$ from the graph $G$. Note that for Pauli-$Y$ measurements, a local complementation of $v$ has to be performed prior to the deletion. This step introduces a local Clifford operator $\sqrt{-iZ_v}= S_v$, which effectively turns the measurement basis into $S_v^\dagger Y_v S_v=Z_v$. Similarly for Pauli-$X$, one performs an edge complementation \cite{Elliott2008Graphical, Elliott2010Graphical} along $(v, w)$, which effectively introduces a local Clifford operator $\sqrt{-iZ_v}\sqrt{iX_v}\sqrt{-iZ_v} = H_v$, and thus changes the measurement basis to $H_v X_v H_v=Z_v$. 

Following the (sequences of) local complementations in Eqs.~\eqref{eq:GraphStatePauliMeasurementY} and \eqref{eq:GraphStatePauliMeasurementX}, we define \cite{hein2006entanglement}
\begin{align}
     U_{v}^{Y,\pm} &\coloneqq  U_G^\textrm{LC}(v) U_{v}^{Z,\pm}, \label{eq:GraphStatePauliMeasurementYByproduct} \\
     U_{v}^{X,\pm} &\coloneqq  U_{\textrm{LC}_w(\textrm{LC}_v(G))}^\textrm{LC}(v) U_{\textrm{LC}_v(G)}^\textrm{LC}(w)U_G^\textrm{LC}(v) U_{v}^{Z,\pm} \label{eq:GraphStatePauliMeasurementXByproduct}
\end{align}
Then, as was demonstrated in Ref.~\cite{Hein2004Multiparty}, the measurement of some qubit $v$ in the Pauli-$Z$ basis, i.e.,~the application of the projectors $\ket{P,\pm}\!\!\bra{P, \pm}_{v}$ to the qubit $v$ of $\ket{G}$ followed by the partial trace over the same qubit affects the following mapping:
\begin{align}
    \bra{P,\pm}_{v}\ket{G} \mapsto U_{v}^{P,\pm}\ket{G'(v,P)}
    \label{eq:GraphStatePauliMeasurement}
\end{align}
for the $U_{v}^{P, \pm}$ and $\ket{G'}$ as above. For any of these Pauli-measurements performed on a non-isolated vertex, the measurement outcome is always $0$ or $1$ with equal probability. In case of isolated vertices, a Pauli-$X$ measurement always yields a 0, and only $Z$- and $Y$ measurements yield $0$ and $1$ with equal probability. 

\subsection{Measurement-Based Quantum Computation} \label{subsec:MBQC}

Above, we have introduced graph states and considered Pauli measurements upon them. If we allow for a specific, more general set of measurements, then it is possible, under certain circumstances, to implement a quantum computation. Computation implemented in this way is known as measurement-based quantum computation (MBQC) \cite{Raussendorf20011WQC,Raussendorf2003Cluster,briegel2009measurement}, which forms the basis for much of this work. In this subsection, we briefly introduce the required aspects of MBQC to be able to define a key notion for the rest of the manuscript, namely the notion of {\em active qubits}.

A measurement-based computation consists of three parts: (i) a graph state, typically with an assignment of certain qubits as relating to the computational input and other as relating to the computational output, (ii) an assignment of single-qubit projective measurements to each of the (non-output) qubits, and (iii) a correction method. Often, not all single-qubit projective measurements are allowed; they must come from one of three measurement planes (to be defined below), with the permissible measurement planes depending on the graph structure underlying the graph state. Furthermore, the desired computation is associated to only one of the outcomes of each measurement. As quantum mechanical measurements are indeterministic in general, this requires a method to {\em correct} for the undesired outcome that may occur for each measurement. The ability to correct all measurements, thereby allowing a deterministic computation, is a property of the underlying graph and choice of measurement planes, and has been characterized in, e.g., Refs.~\cite{browne2007generalized,mhalla2022characterising,PhysRevA.74.052310,Booth_2023}.

The measurement planes for MBQC are typically called the $XY$-, $XZ$- and $YZ$-planes of the Bloch sphere. For $A \neq B \in \{X,Y,Z\}$, the $AB$-plane refers to the states $\ket{\psi}$ lying on the great circle of the Bloch sphere containing the eigenstates of both $A$ and $B$. For example, $\ket{\psi}$ lies in the $XY$-plane of the Bloch sphere if it is of the form $\ket{\psi} = R_{Z}(\theta)\ket{+}$ for some $\theta \in \mathbb{R}$. Similarly, for $\ket{\psi}$ in the $XZ$-plane there is some $\theta$ such that $\ket{\psi} = R_{Y}(\theta)\ket{0}$ and for $\ket{\psi}$ in the $YZ$-plane, there is some $\theta$ such that $\ket{\psi} = R_{X}(\theta)\ket{0}$. The measurements corresponding to these planes are given by $\{\ket{\psi}\!\!\bra{\psi}, I - \ket{\psi}\!\!\bra{\psi} =: \ket{\psi^{\perp}}\!\!\bra{\psi^{\perp}} \}$ where $\ket{\psi}$ is a state lying in the chosen plane. Due to the parametrization of these states outlined above, once a measurement plane has been specified, it remains only to specify the angle $\theta$, referred to as the {\em measurement angle}.

There are two features of these measurements that are important for deterministic MBQC. First, for any $\ket{\psi}$ in any of the measurement planes, $\ket{\psi}$ and $\ket{\psi^{\perp}}$ are related by a single-qubit unitary appearing as a tensor factor in (a product of) the stabilizer generators $\mathcal{K}_{v}$ for graph states. For example, for $\ket{\psi}$ in the $YZ$-plane, which is the most relevant measurement plane for this work (see Fig.~\ref{fig:PauliGadget}), we have that $\ket{\psi^{\perp}} = X \ket{\psi}$. Accordingly, if a measurement of $\{\ket{\psi}\!\!\bra{\psi}, \ket{\psi^{\perp}}\!\!\bra{\psi^{\perp}} \}$ is performed on qubit $v$ of a graph state $\ket{G}$ and the undesired outcome (the one associated to $\ket{\psi^{\perp}}$) is obtained, we can use the stabilizer $\mathcal{K}_{v}$ of $\ket{G}$ to account for this as follows:
\begin{equation}
\begin{aligned}
\braket{\psi^{\perp}|G} &= \braket{\psi|X_{v}^{\dagger}|G} \\
&= \braket{\psi|X_{v}^{\dagger}\mathcal{K}_{v}|G}\\
&= \braket{\psi|\bigotimes_{w \in N_{G}(v)} Z_{w}|G}.
\end{aligned}
\end{equation}
That is, in effect, even if we obtain the undesired measurement outcome, it is equivalent to obtaining the correct outcome, but at the cost of the unitary $\bigotimes_{w \in N_{v}^{G}} Z_{w}$ acting on other qubits in $\ket{G}$. These are often known as {\em byproduct operators}.

The second feature of the measurements from these measurement planes relates to the ability to treat the byproduct operators appearing above. It turns out that, for any $\ket{\psi}$ in any of the measurement planes, the application of $X$, $Z$ or their product, which is to say, any tensor factor of the stabilizer generators $\mathcal{K}_{v}$ or their products, to $\ket{\psi}$ can be considered purely as an update to the measurement angle. For example, if $\ket{\psi}$ is in the $YZ$-plane, meaning that $\ket{\psi} = R_{X}(\theta)\ket{0}$, we have that $X\ket{\psi} = R_{X}(\theta + \pi)\ket{0}$, $Z\ket{\psi} = R_{X}(-\theta)\ket{0}$ and $XZ\ket{\psi} = R_{X}(-\theta + \pi)\ket{0}$. This fact, along with the previous feature of these measurements, indicates how measurement outcomes can be accounted for: an undesired measurement outcome for a measurement at $v$ in $\ket{G}$ can be corrected by updating the measurement angles for the measurements on other qubits in $\ket{G}$. 

The correction method outlined above requires that the qubits whose measurements may be updated conditional on the measurement outcome of a different qubit have not yet been measured. In particular, this leads to a measurement {\em sequence}. As stated earlier, the existence of a correction method for a given graph state $\ket{G}$ and assignment of measurement planes is a property of the underlying graph $G$. The full details of this characterization is beyond the scope of this work; the reader is referred to Refs.~\cite{browne2007generalized,mhalla2022characterising} for the details. The most important point for our present purposes is that the measurement sequence, corresponding to a partial order $<$ over the qubits $v$ of $\ket{G}$, is a requirement for deterministic MBQC, and moreover, one that is satisfied for all the resources considered in this work. 

The fact that measurements in MBQC are performed in sequence, opens up the possibility for reducing the number of physical qubits required for a given computation (in the measurement-based approach). That is, it is not necessary to prepare the full state $\ket{G}$ prior to commencing to measure. Suppose that the graph $G$ underlying $\ket{G}$ has vertex set $V$ and edge set $E$ and suppose that the partial order $<$ on $V$ is the coarsest possible while still permitting deterministic computation. The state $\ket{G}$ can be written as
\begin{align}
    \ket{G} = \prod_{(u,v) \in E} \Lambda_{u}(Z_{v})\ket{+}^{\otimes |V|}
\end{align}
where $\Lambda_{u}(Z_{v})$ denotes a controlled-$Z$ gate between the qubit $u$ and $v$. In particular, this means that, for any $v \in V$, the gates $\Lambda_{u}(Z_{w})$ for any $u,w \in V$ such that $u \neq v$ and $w \neq v$ commute with a measurement on $v$. That is, only at most the qubits in the neighbourhood of $v$ and the qubit $v$ itself must be actively entangled at the point where $v$ is measured. In general, if $v$ is not at the start of the measurement sequence, some of the neighbourhood of $v$ will have been measured prior to the measurement of $v$, so only $v$ and those qubits that are neighbours of $v$ and later than $v$ in the sequence are required to be entangled when $v$ is measured. This leads us to define the {\em active qubits} for the measurement of $v$ to be the set
\begin{align}
    \textrm{Act}_{v} := \{v\} \cup \{w \in N_{G}(v)| v < w\}.
\end{align}
Since the measurement sequence is a partial order, it is possibly to partition $V$ into subsets $V^{(0)}, \dots, V^{(t)}$ for some $t \in \mathbb{N}$ where $V^{(0)}$ is the set of all vertices that are lowest in the partial order, and $V^{(i+1)}$ is the set of vertices that covers the vertices in $V^{(i)}$ for $i = 0, \dots, t-1$\footnote{Recall, that an element $w$ covers an element $v$ in a partially ordered set $V$ if $v<w$ and there does not exist a $u \in V$ such that $v < u < w$.}. We can thus define the active qubits at a given step $i$ in the measurement sequence as 
\begin{align}
\textrm{Act}_{V^{(i)}} := \cup_{v \in V^{(i)}} \textrm{Act}_{v}.
\end{align}
The number $\max_{i} |\textrm{Act}_{V^{(i)}}|$ corresponds to the maximum number of physical qubits required to perform the desired measurement-based computation when following an alternating entangle-measure-entangle procedure. A resource state that permits a measurement sequence that (approximately) minimizes this quantity will be termed {\em resource efficient} or simply just {\em efficient}.

\section{Deterministic Graph State Algorithm}
\label{sec:graph_state_algorithm}
In the following, we are concerned with parameterized generalized Pauli rotations acting on an $N$-qubit system,
\begin{equation}
    R_\mathcal{P}(\theta)\coloneqq\exp\left(-i\frac{\theta}{2}\mathcal{P}\right),
    \label{eq:PauliRotation}
\end{equation}
where $\mathcal{P} \in \{I, X, Y, Z\}^{\otimes N}$ is an $N$-qubit Pauli-string. 
We consider an arbitrary sequence of $M$ (not necessarily distinct) Pauli rotations with generators $\mathcal{P}^{(1)}, \mathcal{P}^{(2)}, \dots, \mathcal{P}^{(M)}$,
\begin{equation}
    U(\bm \theta) = 
    \prod_{m=1}^{M} R_{\mathcal{P}^{(m)}}(\bm{\theta}_{m}), \label{eq:U_theta}
\end{equation}
where the rotation angles $\bm{\theta}_{m} \in [-\pi,\pi)$ can be arbitrary.
We apply $U(\bm \theta)$ to some initial state $N$-qubit system described by the state $\ket{\psi_0}$. Noting that any pure non-stabilizer state may be prepared by applying generalized Pauli rotations to some stabilizer state, we can, w.l.o.g., restrict ourselves to stabilizer-type input states $\ket{\psi_0}=\ket{\mathcal{S}_0}$. 

\begin{figure}[H]
    \centering
    \includegraphics[width=\linewidth]{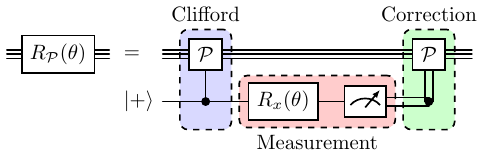}
    \caption{Circuit decomposition of the Pauli gadget $R_\mathcal{P}(\theta)$ in terms of MBQC. The auxiliary qubit is initialized in the $\ket{+}$ state before serving as a control of the generator $\mathcal{P}$ applied to the main system. The parameterized $R_x(\theta)$ rotation followed by the projective $Z$ measurement effectively implements the projection onto the $\mathcal{M}(\theta)$ basis. Last, the deterministic output is ensured through another (classically) controlled application of the generator $\mathcal{P}$ based on the measurement outcome.}
    \label{fig:PauliGadget}
\end{figure}

In the following, we derive a measurement pattern to implement $U(\bm \theta)\ket{\mathcal{S}_0}$. For that purpose, we consider a measurement-based implementation of the Pauli rotation $R_\mathcal{P}(\theta)$ within the quantum circuit model, in the literature often referred to as Pauli gadget \cite{Cowtan:2019loc,Chan:2023fwv}, as depicted in Fig.~\ref{fig:PauliGadget}. The gadget assumes a so-called standard form \cite{broadbent2009parallelizing} with minimal causal depth \cite{mhalla2008finding} consisting of three different layers; (i) a Clifford circuit implementing a controlled Pauli-string operator $\Lambda_a(\mathcal{P})$, (ii) a measurement in the $YZ$-plane of the Bloch sphere, i.e., in the basis
\begin{align}
    \mathcal{M}(\theta)= \{R_{X}(\theta)\ket{s}\!\bra{s}R_{X}^{\dagger}(\theta)~|~s =0,1\},
    \label{eq:MeasurementBasis}
\end{align}
with the classical measurement outcome $s\in \{0, 1\}$, and (iii) an adaptive Pauli correction $\mathcal{P}^s$ ensuring a deterministic output state 

To implement the unitary $U(\bm \theta)$, we can introduce one such gadget for each Pauli-string rotation in Eq.~\eqref{eq:U_theta}. The resulting circuit can be easily re-cast into a standard form. 
The general idea is best illustrated at hand for a sequence of two Pauli rotations with generators $\mathcal P$ and $\mathcal{P}'$. Given the following commutation rule for a Pauli-string $\mathcal{P}$ and a Pauli rotation $R_{\mathcal{P}'}(\theta)$ 
\begin{equation}
    \label{eq:correction_commutation}
    R_{\mathcal{P}'}(\theta) \mathcal{P} = \mathcal{P}
    \begin{cases}
        R_{\mathcal{P}'}(\theta), & \text{if $[\mathcal{P}, \mathcal{P'}]=0$,} \\
        R_{\mathcal{P}'}(-\theta), & \text{else,}
    \end{cases}
\end{equation}
we can commute the first correction $\mathcal{P}^s$ through the second rotation by flipping the measurement angle $\theta$ if the generators anticommute. The Clifford layer of the second gadget may then be (qubit-wise) commuted through the first measurement, thus recovering the standard form. 
By generalizing this procedure to implement $U(\bm \theta)$, we find that the Clifford part of the standard form is given by 
\begin{align}
    C \coloneqq \prod_{m=1}^M \Lambda_m(\mathcal{P}^{(m)}),
    \label{eq:FinalClifford}
\end{align}
where $\Lambda_m(\mathcal{P}^{(m)})$ denotes the $m$-th Pauli-string controlled by the $m$-th auxiliary qubit.
The system's state after applying the controlled Clifford gates, i.e.,
\begin{align}
    \ket{\mathcal{S}} \coloneqq C\ket{\mathcal{S}_0}\ket{+}^{\otimes M},
    \label{eq:ResourceState}
\end{align}
is what we refer to as the {\em resource state}. Since $C$ is Clifford, the resource state $\ket{\mathcal{S}}$ is again a stabilizer state and hence is LC-equivalent to a graph state. One contribution of this work consists in deriving a method for compiling these resultant graph states directly.
Concerning the measurement layer, the measurement angles are adjusted as follows:
\begin{align}
    \theta_m \leftarrow (-1)^{h_m} \theta_m, \text{ where } h_m\coloneqq \sum_{\substack{l<m\\ \{\mathcal{P}^{m}, \mathcal{P}^{l}\}=0}} s_{l}.
    \label{eq:AdaptivePattern}
\end{align}
The exponents $h_m$ take into account the angle flips due to all previous measurement results on auxiliary qubits $m'<m$ corresponding to generators $\mathcal{P}^{(m')}$ which anticommute with $\mathcal{P}^{(m)}$. As a direct consequence, one can measure the auxiliary qubits corresponding to subsequent commuting operations in parallel, since the measurement bases are independent of each other. One may exploit this property to minimize the depth of the pattern by partitioning the sequence of Pauli rotations into mutually commuting cliques \cite{Kaldenbach2025Mapping}.  
The final correction is given by
\begin{align}
    \prod_{m=1}^M (\mathcal{P}^{(m)})^{s_m}.
    \label{eq:FinalCorrection}
\end{align}
This is essentially the same controlled operation as the Clifford operator in Eq.~\eqref{eq:FinalClifford}, but conditioned on the classical measurement outcomes $s_m$ instead of the auxiliary qubits themselves. 

\subsection{Stabilizers of the Resource State} \label{subsec:ResourceStabilizers}

\begin{figure}[H]
\begin{flushleft}
    \begin{subfigure}{\linewidth}
        \includegraphics[width=\linewidth]{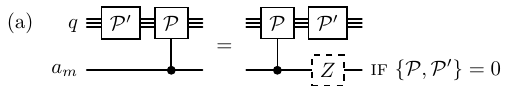}
    \end{subfigure}
    \par\vspace{1em} 
    \begin{subfigure}{\linewidth}
        \includegraphics[width=0.7\linewidth]{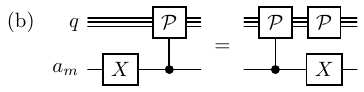}
    \end{subfigure}
\caption{Circuit representations of the two propagation rules. (a) The first propagation rule from Eq.~\eqref{eq:PropagationRule1}. The propagation of $\mathcal{P}'$ through $\Lambda_m(\mathcal{P})$ tags the auxiliary qubit $a_m$ with a Pauli $Z_m$ (dashed) if the generators $\mathcal{P}$ and $\mathcal{P}$' anticommute. (b) The second propagation rule from Eq.~\eqref{eq:PropagationRule2}. The propagation of $X_m$ through $\Lambda_m(\mathcal{P})$ tags the main qubits with the generator $\mathcal{P}$.}
\label{fig:PropagationRules}
\end{flushleft}
\end{figure}

To derive the structure of that stabilizer state, it is sufficient to consider two different Pauli propagation rules. First, when propagating some Pauli-string $\mathcal{P}'$ on the $N$ main qubits through some Pauli-string operator $\Lambda_m(\mathcal{P})$ controlled by the $m$-th auxiliary qubit, it evolves according to 
\begin{equation}
    \Lambda_m(\mathcal{P})\mathcal{P}'\Lambda_m^\dagger(\mathcal{P}) = \mathcal{P}' \otimes
    \begin{cases}
    I_{m}, & \text{if } [\mathcal{P}, \mathcal{P}'] =0, \\
    Z_{m}, & \text{if } \{\mathcal{P}, \mathcal{P}'\}=0.
    \end{cases} 
    \label{eq:PropagationRule1}
\end{equation}
In other words, when propagating $\mathcal{P}'$ through $C$, all the auxiliary qubits $a_m$ corresponding to rotations with generators $\mathcal{P}_m$ anticommuting with $\mathcal{P}'$ are tagged with a Pauli $Z_m$.  A visualization within the circuit model is provided in Fig.~\ref{fig:PropagationRules}~(a). With this rule, we can describe the evolution of the $N$ elements of the stabilizer group $\mathcal{S}_0$ corresponding to the initial state $\ket{\mathcal{S}_0}$.

To fully describe our system, we also need to evolve the stabilizers corresponding to the auxiliary qubits. The auxiliary qubits are all initialized in the $\ket{+}$ and thus initially stabilized by Pauli-$X$ operators. One can easily show that the stabilizer of the $m$-th auxiliary qubit $X_m$ is trivially propagated through all the previous $\text{C}_l\mathcal{P}_l$ with $l<m$ due to qubit-wise commutation. Therefore, we first consider the propagation rule
\begin{equation}
    \Lambda_m(\mathcal{P})X_m\Lambda_m(\mathcal{P}) = \mathcal{P} \otimes X_m,
    \label{eq:PropagationRule2}
\end{equation}
which is visualized in Fig.~\ref{fig:PropagationRules}~(b).
Note that the $X_m$ commutes with all the subsequent $\text{C}_l\mathcal{P}_l$, meaning that $\mathcal{P} \otimes X_m$ can be further propagated using Eq.~\eqref{eq:PropagationRule1}. This will once again tag all subsequent rotations $l>m$ with anticommuting generators with a Pauli $Z_l$.

We now introduce the anticommutation-string function $\mathcal{A}(\mathcal{P}, m_0)$, which encapsulates all the Pauli-$Z$ operations arising due to the first propagation rule from Eq.~\eqref{eq:PropagationRule1} as follows:
\begin{equation}
    \mathcal{A}(\mathcal{P}, m_0) \coloneqq \bigotimes_{m=m_0}^M 
    \begin{cases}
    I_{m}, & \text{if } [\mathcal{P}, \mathcal{P}^{(m)}] =0, \\
    Z_{m}, & \text{if } \{\mathcal{P}, \mathcal{P}^{(m)}\}=0.
    \end{cases}
    \label{eq:anticommutation_string_func}
\end{equation}
The argument $m_0$ is supplied to check for anticommutation starting from the $m_0$-th Pauli.
The state $\ket{\mathcal{S}_{0}}\ket{+}^{\otimes M}$ has a stabilizer $\mathcal{S}'$ generated by the elements of the set $\{S_{0}^{(n)} \otimes I^{\otimes M} | n = 1, \dots, N \} \cup \{ I^{\otimes N} \otimes X_{N+m} | m = 1, \dots, M \}$ where $X_{N+m}$ denotes the Pauli string comprised of a Pauli-$X$ operation on the $m$-th copy of the state $\ket{+}$ and identities on the other auxiliary qubits.
By applying both propagation rules, we compute the resultant stabilizer $\mathcal{S} := C \mathcal{S}'C^{\dagger}$. We find that the first $N$ stabilizer strings of the system $\mathcal{S}^{(n)}=C (\mathcal{S}_0^{(n)} \otimes I^{\otimes M}) C^\dagger$ are given by
\begin{equation}
    \mathcal{S}^{(n)} = \mathcal{S}_0^{(n)} \otimes 
    \mathcal{A}\left(\mathcal{S}_0^{(n)},1 \right), \label{eq:stab_main_qubits}
\end{equation}
where $n=1,\dots, N$, whereas the additional $M$ stabilizers $S^{(N+m)}=C (I^{\otimes N} \otimes  X_{m}) C^\dagger$ assume the form
\begin{equation}
    \mathcal{S}^{(N+m)}= \mathcal{P}^{(m)} \otimes X_{m} \otimes  \mathcal{A}\left(\mathcal{P}^{(m)}, m+1\right), 
    \label{eq:stab_aux_qubits}
\end{equation}
where $m=1,\dots, M$.
This gives us the complete description of the resource state $\ket{\mathcal{S}}$ 
in terms of $N+M$ independent stabilizer strings $\mathcal{S}=\langle S^{(1)},\dots ,S^{(N)}, S^{(N+1)}, \dots, S^{(N+M)}\rangle$. 

\subsubsection{Tableau Representation of the Stabilizers}

We now represent the resource state $\ket{\mathcal{S}}$ in terms of the stabilizer tableau (cf.~Eq.~\eqref{eq:StabilizerTableau}). 
First, we describe the initial state $\ket{\mathcal{S}_0}$ by the two $N\times N$ binary matrices $\textrm{X}_0, \textrm{Z}_0$ and the $N$-component binary phase exponent vector $\textrm{R}_0$. 
Based on Eq.~\eqref{eq:stab_main_qubits}, the first $N$ stabilizer generators $\mathcal{S}^{(n)}$ have the same phases $\textrm{R}_0$ as their initial counterparts $\mathcal{S}_0^{(n)}$.
Then, we describe the $M$ Pauli generators $\mathcal{P}^{(1)},\dots,\mathcal{P}^{(m)}$ using the two $M\times N$ matrices $\textrm{X}, \textrm{Z}$.
The additional $M$ stabilizer generators $\mathcal{S}^{(N+1)},\dots, \mathcal{S}^{(N+M)}$ from Eq.~\eqref{eq:stab_aux_qubits} have the same phases
as their corresponding Pauli generators $\mathcal{P}^{(m)}$, which by our definition of Pauli-strings (cf.~Eq.~\eqref{eq:PauliString}) are all zero.

Through the remainder of this work, we use the following definition for binary matrix multiplication with the symplectic inner product: for binary matrices $A, B$, this is given by
\begin{align}
    (AB)_{ij} &\coloneqq \bigoplus_k A_{ik}  B_{kj}, \label{eq:matmul_bin}
\end{align}
where $\oplus$ denotes the logical \texttt{XOR} operation. 
To capture the anticommutation strings in Eqs.~\eqref{eq:stab_main_qubits} and \eqref{eq:stab_aux_qubits}, we introduce the anticommutation matrices
\begin{align}
    \textrm{A}_0 \coloneqq \textrm{X}_0 \textrm{Z}^T \oplus \textrm{Z}_0 \textrm{X}^T, \quad \text{and} \quad \textrm{A} \coloneqq \textrm{X}\textrm{Z}^T \oplus \textrm{Z}\textrm{X}^T,
    \label{eq:AnticommutationMatrices}
\end{align}
where the superscript `$T$' denotes matrix transposition. The matrix $\textrm{A}_{0}$ is such that $(\textrm{A}_{0})_{nm}$ denotes if $\mathcal{S}_0^{(n)}$ and $\mathcal{P}^{(m)}$ anticommute, and the matrix $\textrm{A}$ is such that $(\textrm{A})_{mm'}$ denotes if $\mathcal{P}^{(m)}$ and $\mathcal{P}^{(m')}$ anticommute. Finally, we combine these definitions to specify all entries in the stabilizer tableau corresponding to $\ket{\mathcal{S}}$:
\begin{equation}
    \left[
    \begin{array}{cc|cc|c}
    \textrm{X}_0 & 0 & \textrm{Z}_0 & \textrm{A}_0 & \textrm{R}_0 \\
    \textrm{X} & \textrm{I} & \textrm{Z} & \text{UT}(\textrm{A}) & \bm 0
    \end{array}
    \right]
    \label{eq:StabilizerTableauResource}
\end{equation}
where $\text{UT}(\cdot)$ denotes the strict upper triangle of $\textrm{A}$. Since in Eq.~\eqref{eq:stab_aux_qubits}, anticommutation is only tagged with respect to all following generators, considering the strict upper triangle of $\textrm{A}$ is sufficient. Similarly, the adaptive measurement pattern from Eq.~\eqref{eq:AdaptivePattern} only entails previous anticommuting strings, and we can compactly rewrite it using the anticommutation matrix $\textrm{A}$
\begin{align}
    \bm \theta \leftarrow (-1)^{\textrm{LT}(\textrm{A}) \bm s} \cdot \bm \theta,
\end{align}
where $\text{LT}(\cdot)$ denotes the strict lower triangular matrix, and $\bm s$ is the $M$-component vector of measurement outcomes $s_1,\dots, s_M$ on the auxiliary qubits.

\subsection{Graph State Representation of the Resource State} \label{subsec:Resource_GraphState}

In the following, using the LC-equivalence between stabilizer states and graph states, we express our input state $\ket{\mathcal{S}_0}= \mathcal{C}^{(0)}\ket{G_0}$ as some graph state $\ket{G_0}$ up to local Clifford operations $\mathcal{C}^{(0)}$, which can be achieved using the algorithm from \cite{Nest2004Graphical}. Interchanging these LC operators with the unitary $U(\bm\theta)$ gives rise to some other sequence of Pauli rotations with generators $\mathcal{P}'^{(m)}$ according to
\begin{align}
    \mathcal{C}^{(0)\dagger} \mathcal{P}^{(m)} \mathcal{C}^{(0)} = (-1)^{\textrm{R}_m} \mathcal{P}'^{(m)}.
\end{align}
Given that we want the measurement angles $\bm \theta_m$ from Eq.~\eqref{eq:MeasurementBasis} to be identical to the rotation angles defined by $U(\bm \theta)$ in Eq.~\eqref{eq:U_theta}, we need to absorb the phases $(-1)^{\textrm{R}}$ into the resource state, which yields $\mathcal{S}^{(N+m)} \leftarrow (-1)^{\textrm{R}_m} \mathcal{S}^{(N+m)}$.  
Below, we redefine the matrices $\textrm{X}, \textrm{Z}$, and introduce $\textrm{R}$, which together now describe the $M$ generators $\mathcal{P}'^{(1)},\dots, \mathcal{P}'^{(M)}$ of the $N$-qubit Pauli rotations {\em after} the LC transformation. Due to the stabilizer structure of graph states (cf.~Eq.~\eqref{eq:GraphStateStabilizerString}), we have $\textrm{X}_0 = \textrm{I}$ and $\textrm{Z}_0 = \Gamma_0$, where $\Gamma_0$ is the adjacency matrix of the initial graph state $\ket{G_0}$. This simplifies the first anticommutation matrix to $\textrm{A}_0 = \textrm{Z}^T \oplusr \Gamma_0 \textrm{X}^T$. In addition, we have $\textrm{R}_0=\bm 0$.
The tableau representation $[{\bf X}|{\bf Z}|{\bf R}]$ then takes the form
\begin{equation}
    \left[\begin{array}{cc|cc|c}
    \textrm{I} & 
    0 & 
    \Gamma_0 & 
    \textrm{A}_0 & 
    \bm 0
    \\
    \textrm{X} & 
    \textrm{I} & 
    \textrm{Z} & 
    \text{UT}(\textrm{A}) & 
    \textrm{R}
    \end{array}\right].
    \label{eq:StabilizerTableauGraph_STEP0}
\end{equation}
It is worth mentioning that the anticommutation matrix $\textrm{A}$ is invariant under local Clifford transformations (unlike $\textrm{X}$ and $\textrm{Z}$). 

In order to derive the graph state of the full resource, we again use the algorithm from \cite{Nest2004Graphical}. The novelty here lies in the fact that we provide the explicit solution to this algorithm for arbitrary initial graphs and rotations. We first perform a row reduction on the second block-row, which is always possible since the rank of ${\bf X}$ is full by construction. The intermediate steps are discussed in Appendix \ref{app:BlockRow}. After the reduction, we have the tableau $[{\bf I}|{\bf Z}'|{\bf R}']$, with 
\begin{equation}
    {\bf Z}' =
    \begin{bmatrix}
    \Gamma_0 & \textrm{A}_0 \\
    \textrm{A}_0^T & \Gamma_{\textrm{X}} \oplusr \text{D}(\textrm{X}\textrm{Z}^T) \oplusr \text{UT}(\textrm{Z}\textrm{X}^T) \oplusr \text{LT}(\textrm{X}\textrm{Z}^T)
    \end{bmatrix},
    \label{eq:GraphStateResourceLoops}
\end{equation}
where $\Gamma_{\textrm{X}} \coloneqq \textrm{X}\Gamma_0 \textrm{X}^T$, and $\text{D}(\cdot)$ denotes the diagonal matrix.
Note that ${\bf Z}'$ is now symmetric, with the only diagonal entries given by $\text{D}(\textrm{X}\textrm{Z}^T)$. 
To obtain the proper adjacency matrix of the graph (without self-loops), we have to shift these entries into the local Clifford layer. 
This is achieved by applying an $S$ gate to every auxiliary qubit $m$ whose generator $\mathcal{P}'^{(m)}$ contains an odd number of $Y$'s. 

The reduction further gives rise to the new phases with ${\bf R}'=[0, \textrm{R}']^T$, where the phases of the auxiliary stabilizer strings are given by
\begin{equation}
    \textrm{R}' = \textrm{R} \oplusr \left\lceil \frac{\textrm{N}_Y}{2} \right\rceil \oplusr \frac{\Gamma_{\textrm{X}}^\circ}{2},
    \label{eq:GraphStateResourcePhases}
\end{equation}
where $\textrm{N}_Y$ is an $M$-component vector counting the number of $Y$'s in each generator, and $\Gamma_{\textrm{X}}^{\circ}$ is the vector given by the diagonal entries of $\textrm{X} \circ \Gamma_0 \circ \textrm{X}^T$, i.e., $(\Gamma_{\textrm{X}}^\circ)_{m} = (\textrm{X} \circ \Gamma_0 \circ \textrm{X}^T)_{mm} $. 
Here, we perform regular matrix multiplication, which we denote by $(A\circ B)_{ij} \coloneqq \sum_k A_{ik}  B_{kj}$.
The intermediate steps to keep track of the phases during the block-row reduction are  discussed in Appendix \ref{app:BlockRow}. The phases corresponding to the entries of $\textrm{R}$ containing a $1$, can be effectively removed by applying Pauli-$Z$ gates to the corresponding nodes with $\textrm{R}_m=1$.

Conveniently, one can simultaneously account for the self-loops and phases by using the following local Clifford operation which solely entails phase gates acting on the auxiliary qubits:
\begin{align}
    \mathcal{C}_\textrm{aux.} = \bigotimes_{m=1}^M S_m^{(2\textrm{R} + \Gamma_{\textrm{X}}^\circ - \textrm{N}_Y)_m},
    \label{eq:GraphStateResourceVOPs}
\end{align}
where $(2\textrm{R}+\Gamma_{\textrm{X}}^\circ -\textrm{N}_Y)_m$ is the $m$-th entry of the non-binary vector $2\textrm{R}+\Gamma_{\textrm{X}}^\circ -\textrm{N}_Y$.
For technical details, we refer the reader to Appendix~\ref{app:VOPs}.
The overall LC operation is thus given by $\mathcal{C} \coloneqq  \mathcal{C}^{(0)} \otimes \mathcal{C}_\textrm{aux.}$.
From now on, we are only concerned with the adjacency matrix $\Gamma$ of the graph state $\ket{\textrm{G}}$. As discussed, we obtain it by removing the diagonal entries of ${\bf Z}'$, which gives rise to
\begin{equation}
    \Gamma = 
    \begin{bmatrix}
         \Gamma_0 & \textrm{A}_0  \\
         \textrm{A}_0^T & \Gamma_{\textrm{X}} \oplusr \text{UT}(\textrm{Z}\textrm{X}^T) \oplusr \text{LT}(\textrm{X}\textrm{Z}^T)
    \end{bmatrix}.
    \label{eq:GraphStateResource}
\end{equation}
Finally, we can prepare the resource state by using $\mathcal{C}\ket{G} = \ket{\mathcal{S}}$.

\paragraph{Graph States for Disentangled Initial States:}
It is worth highlighting the graphs for fully disentangled input states with $\Gamma_0 = 0$:  
\begin{equation}
    \Gamma = 
    \begin{bmatrix}
         0 & \textrm{Z}^T \\
         \textrm{Z}  & \text{UT}(\textrm{Z}\textrm{X}^T) \oplusr \text{LT}(\textrm{X}\textrm{Z}^T) 
    \end{bmatrix}.
    \label{eq:GraphState_G0=0}
\end{equation}
Note that Eq.~\eqref{eq:GraphState_G0=0} is readily usable if the initial state is $\ket{+}^{\otimes N}$, which is itself a graph state of $N$ qubits, corresponding to the empty graph. For the initial state $\ket{0}^{\otimes N}$, we have to describe it as the $\ket{+}^{\otimes N}$ state up to local Cliffords $H^{\otimes N}$. This effectively interchanges the roles of $\textrm{X}$ and $\textrm{Z}$ in Eq.~\eqref{eq:GraphState_G0=0}. Similar principles hold for arbitrary stabilizer product states. For such states, the dependency of $\mathcal{C}$ on $\Gamma_{\textrm{X}}^\circ$ vanishes.

\paragraph{Graph State for Periodic Operations:}
To simulate the quantum dynamics governed by some time-dependent Hamiltonian $H(t)=\sum_{l=1}^L \alpha_l(t) \mathcal{P}^{(l)}$ in first-order Trotterization, we approximate the time evolution operator $U(t) =\exp_\mathcal{T}(-\frac{i}{2}\int_{0}^t d\tau H(\tau))$, where $\exp_\mathcal{T}$ denotes the time ordered operator exponential, via
\begin{align}
    U(t) &=\prod_{k=1}^K \prod_{l=1}^L R_{\mathcal{P}^{(l)}}\left(\alpha_l\left(\frac{k-\frac{1}{2}}{K} t\right)\right) + \mathcal{O}\left(\frac{t^2}{K}\right),
    \label{eq:HamiltonianSimulationTrotter}
\end{align}
where $K$ is the number of Trotter steps. By increasing $K$, the approximation error can be made arbitrarily small. The time evolution operator is thus expressed by $K$ repetitions of a sequence of $L$ generalized Pauli rotations, and therefore assumes the form of Eq.~\eqref{eq:U_theta} for $M=KL$.
We can reuse Eq.~\eqref{eq:GraphStateResource} to obtain
\begin{equation}
\begin{adjustbox}{width=\linewidth}$
    \Gamma= 
    \begin{array}{|c!{\vrule width 2pt}c|c|c|c|} 
    \hline 
    \Gamma_0 & \textrm{A}_0 & \textrm{A}_0 & \cdots & \textrm{A}_0 \\ \Xcline{1-5}{2pt} \rule{0pt}{10pt}
    & \Gamma_{\textrm{X}} & \Gamma_{\textrm{X}}  &  & \Gamma_{\textrm{X}} \\ 
    \textrm{A}_0^T & \oplusr \text{UT}(\textrm{Z}\textrm{X}^T)& \oplusr \textrm{Z}\textrm{X}^T & \cdots & \oplusr \textrm{Z}\textrm{X}^T \\ 
    & \oplusr \text{LT}(\textrm{X}\textrm{Z}^T)& & &\\  \hline 
    & \Gamma_{\textrm{X}} & \Gamma_{\textrm{X}} &  & \Gamma_{\textrm{X}} \\ 
    \textrm{A}_0^T & & \oplusr \text{UT}(\textrm{Z}\textrm{X}^T)& \cdots& \oplusr \textrm{Z}\textrm{X}^T \\ 
    &  \oplusr \textrm{X}\textrm{Z}^T & \oplus \text{LT}(\textrm{X}\textrm{Z}^T) & &\\ \hline 
    \vdots & \vdots & \vdots & \ddots & \vdots \\ \hline 
    & \Gamma_{\textrm{X}} & \Gamma_{\textrm{X}} & & \Gamma_{\textrm{X}}  \\
    \textrm{A}_0^T & & & \cdots & \oplusr \text{UT}(\textrm{Z}\textrm{X}^T) \\
    & \oplusr \textrm{X}\textrm{Z}^T & \oplusr \textrm{X}\textrm{Z}^T & & \oplusr \text{LT}(\textrm{X}\textrm{Z}^T) \\ 
    \hline 
    \end{array}
$\end{adjustbox}
\label{eq:GraphStateResourcePeriodic}
\end{equation}
This provides us with a graph state representation of the Hamiltonian-specific resource state. Note that in this representation of the resource state, the number of edges in the graph grows as $\mathcal{O}(K^2)$ in the number of Trotter steps $K$, which is different from the $\mathcal{O}(K)$ number of entangling gates one would obtain in, e.g.,~the quantum circuit model. At the same time, due to the \enquote{all-to-all} connectivity between the main- and auxiliary registers, the number of active qubits is $\mathcal{O}(K)$, which is again more costly than constant number of active qubits in the circuit model. 

\subsection{Comparison to Previous Methods} \label{subsec:Advantages}
With Eqs.~\eqref{eq:GraphStateResourceVOPs}, \eqref{eq:GraphStateResource}, and \eqref{eq:GraphStateResourcePeriodic}, our work provides a structured description of the resource state as a graph state up to the local Cliffords based on the initial state and the applied rotations. In this section, we briefly compare the complexity to compute this solution to the other approaches for the direct compilation of algorithm-specific graph states taken in Refs.~\cite{Vijayan2024Compilation, Kaldenbach2025Mapping}. 

Note that the algorithm we have employed to find the graph state \cite{Nest2004Graphical} is akin to Gaussian elimination which has a complexity of $\mathcal{O}(n^3)$, where $n=N+K L$ is the total number of qubits. This approach has for instance between taken in Ref.~\cite{Vijayan2024Compilation}, although the comparability suffers given that the authors did not explicitly consider periodic circuits. 
Due to the periodicity of our solution, it can however be computed in $\mathcal{O}(NL^2 + N^2L)$, which is due to $\Gamma_{\textrm{X}}$ involving two matrix multiplications. In case of initial product states ($\Gamma_0=0$), the complexity reduces to $\mathcal{O}(NL^2)$, which is then due to the computation of $\textrm{Z}\textrm{X}^T$. Most importantly, our solution can be computed independently of $K$. 

In the previous work from Ref.~\cite{Kaldenbach2025Mapping}, one would instead obtain some graph state which is LC-equivalent, without any obvious structure due to the fact that the graph state algorithm from \cite{Anders2006Fast} is based on a non-deterministic simulation scheme. 
Importantly, using the algorithm of \cite{Anders2006Fast}, the simulation time of two-qubit Clifford gates acting on an $n$-qubit graph state typically scales as $\mathcal{O}(n\log n)$ \cite{Anders2006Fast}. By sequentially applying the Clifford part of the Pauli gadgets (cf.~Fig.~\ref{fig:PauliGadget}), one can grow the graph state such that for the $m$-th rotation, it entails $N+m$ qubits. 
The time complexity to obtain the full graph state is thus given by $\mathcal{O}(K L(N+K L)\log(N+K L))$. 
While this algorithm offers a run-time advantage for arbitrary sequences of rotations compared to the tableau-to-graph algorithm from Ref.~\cite{Nest2004Graphical}, for multiple time steps it still scales with $K$, and further does not produce periodic solutions.

\section{Resource State Preparation Schemes} \label{sec:ResourcePrep}

In this section, we first improve existing non-deterministic optimization schemes to obtain periodic graph states with minimal number of active qubits. Then, we introduce a deterministic strategy based on two layers of entangling gates. Based on this, we prove that the graph state $\ket{G}$, despite having $\mathcal{O}(K^2)$ edges, can always be prepared with $\mathcal{O}((||\textrm{A}||_1+L) K)$ entangling gates and $N+L+1$ active qubits. This way, we join the benefits of the circuit model with the optimal parallelism and potential for hybrid simulation (cf.~Sec.~\ref{sec:Hybrid}) within MBQC. 

\subsection{LC-MBQC} \label{subsec:LC-MBQC}

Given some graph state $\ket{G}$, one can reach any other graph state in the LC orbit of $\ket{G}$ via some finite set of local complementations \cite{Nest2004Graphical}. Previous works \cite{Cabello2011Optimal, Vijayan2024Compilation, Kaldenbach2025Mapping, Sharma2025Minimizing} have explored this degree of freedom to find graph states which minimize, e.g., the total number of edges or the maximum degree. In particular, Ref.~\cite{Kaldenbach2025Mapping} introduces a simulated annealing scheme to minimize the edges in algorithm-specific graph states starting from some random graph state which is LC-equivalent to the graph state from Eq.~\eqref{eq:GraphStateResource}. Ref.~\cite{Sharma2025Minimizing} additionally proposes weighted-edge minimization. 

Given that our resource states are subject to some temporal order in the measurements, we now define a {\em distance matrix} $\textrm{D}$, which serves as a penalty for long-range edges in the graph state, i.e., edges between qubits which are separated in time by multiple measurement rounds in the pattern. One can then define the total edge weight of the graph
\begin{align}
    W(\Gamma, \textrm{D}) = \sum_{i=1}^{N+M} \sum_{j > i}^{N+M} \Gamma_{ij} \textrm{D}_{ij}.
\end{align}
The minimization of $W$ then reduces the number of active qubits in the measurement pattern. In practice, we choose $\textrm{D}$ such that edges between auxiliary qubits corresponding to the same or consecutive measurement rounds have distance 1. 
In case that the subsequent measurement round does not hold enough qubits to store the full intermediate quantum state, we include more subsequent measurement rounds with distance 1. 
Larger distances, corresponding to unnecessarily high number of active qubits, are weighted exponentially. 
Typically, we assume that the full quantum state can be stored using $N$ qubits. In case that the Hamiltonian possesses generic $\mathbb{Z}_2$ Pauli symmetries, one can use qubit tapering \cite{bravyi2017tapering, setia2020reducing} to remove one qubit for every symmetry. We use the same technique to determine if fewer qubits are sufficient, and thereby establish a distance matrix with tighter constraints. 
We provide the full algorithm for the construction of $\textrm{D}$, together with an explicit example, in Appendix \ref{app:DistanceMatrix}.

In practice, we observe that the minimization of $W$ does a good job at enforcing graph states which comply with the temporal order of the pattern. In many cases, these graphs are periodic, and therefore permit the extrapolation to larger values of $K$. However, if multiple configurations have identical or very similar total edge weights, the minimization of $W$ potentially leads to efficient, but non-periodic solutions. We can circumvent such cases by introducing a periodicity constraint to the objective function. We now introduce the {\em aperiodicity} $\Pi$ (only for multiple time steps $K>1$)
\begin{align}
    \Pi(\Gamma) = \sum_{i=1}^{M-L} \sum_{j>i}^{M-L} \Gamma_{i,j} \oplus \Gamma_{i+L, j+L}.
\end{align}
For a perfectly periodic graph state, this quantity is 0. In particular, we have $\Pi(\Gamma)=0$ for the graph state from Eq.~\eqref{eq:GraphStateResourcePeriodic}, the point of departure for our optimization. Based on Eq.~\eqref{eq:GraphStateResourcePeriodic}, one can only perform local complementations of the main qubits without violating the constraint. However, this domain is not sufficient to minimize $W$. Therefore, we implemented $\Pi$ as a weak constraint, which can be violated during the optimization, but is reinforced in the end. To this end, we use the total cost function 
\begin{align}
    f(\Gamma, D) = W(\Gamma, D) + \Pi^2(\Gamma)
    \label{eq:CostFunc}
\end{align}
for the optimization. From now on, we will simply denote it as $f(\Gamma)$, given that $D$ is fixed for a given optimization. The choice to square the constraint is not necessary as $\Pi$ is positive, but works well in practice.

To perform the actual optimization, we employ simulated annealing analogously to Ref.~\cite{Kaldenbach2025Mapping}. In particular, we compute the initial temperature dynamically using the standard deviation $\sigma$ of the cost function change $\Delta_v f(\Gamma) \coloneqq f(\textrm{LC}_v(\Gamma)) - f(\Gamma)$ with respect to to all possible local complementations acting on the initial graph
\begin{align}
    T_0 = \sigma \left( \left\{ \Delta_v f(\Gamma_{\textrm{init.}}) | 1 \leq v \leq N+M \right\} \right),
\end{align}
and use a geometric cooling scheme $T_n = T_0  \lambda^n$
with the cooling rate $\lambda \in (0, 1)$, which stops when $T_n < 1 - \lambda$.

We want to finish this subsection with some remarks on efficiently implementing this algorithm.
Computing the cost function from Eq.~\eqref{eq:CostFunc} is prohibitively expensive, as clearly the cost of evaluating $W$ is $\mathcal{O}\left((N+M)^2\right)$, and for $\Pi$ it is $\mathcal{O}(M^2)$. The change in the cost function caused by a local complementation can however be accessed much more efficiently. One can calculate $\Delta_v f(\Gamma)$ by looking up the neighborhood $N_G(v)$, and then simply evaluate $W$ and $\Pi$ on the set of edges in the neighborhood and its complement. Conveniently, the more the optimization progresses, i.e., the sparser $\Gamma$ gets, the more efficient the computation of $\Delta_v f(\Gamma)$ becomes.  

There are several options to account for the LC operators which accumulate due to the local complementations. For every accepted local complementation, one can immediately update the vertex operators (VOPs) (cf.~Ref.~\cite{Anders2006Fast}). To avoid the overhead of storing the VOPs during the search itself, one can simply keep track of the sequence of local complementations during the search, and then repeat the same sequence in the end to recover the VOPs. The third option, which we employed, lies in using the polynomial-time algorithm to check for LC-equivalence of graph states from Ref.~\cite{Nest2004Efficient}. Conveniently, besides the ability to check for LC-equivalence, the algorithm returns an explicit solution for the VOPs. 
Given that our graph states are always LC-equivalent by construction, we can robustly employ this algorithm to compute the VOPs and thereby avoid keeping track of anything during the search.

Having found a periodic graph state and a periodic sequence of VOPs, we can then extrapolate the pattern towards larger $K$. In practice, we observe that sometimes the first time step $k=0$, and the last step $k=K$ have graphs, which differ from the periodic pattern obtained on $k=1,\dots, K-1$. For such cases, we advise to run the annealing algorithm for $K=4$, which permits to safely extrapolate the pattern between $k=2$ and $k=3$. In other instances, it is sufficient to consider $K=2$ or $K=3$ to extrapolate. The key insight is that by pairing the annealing approach with the extrapolation scheme, one can avoid optimizing larger graphs for $K>4$. For structured Hamiltonians, such as in many lattice models, one can even perform spatial extrapolation. That is, we compute the graph states for a couple of small $N$, and from that derive the graph state and VOPs for arbitrary $N$.

\subsection{AC-MBQC} \label{subsec:AC-MBQC}

\begin{figure}[H]
    \centering
    \includegraphics[width=0.85\linewidth]{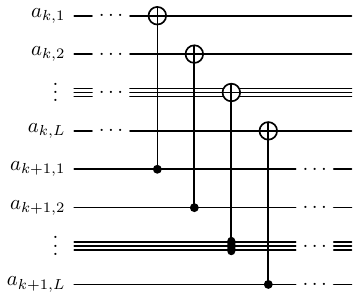}
    \caption{Quantum circuit diagram of the forward CNOT ladder from Eq.~\eqref{eq:CNOTLadderForward}. Note that all CNOT gates between two auxiliary registers $k$ and $k+1$ can be applied in parallel. The total circuit depth is thus $K-1$.}
    \label{fig:CNOTForwardLadder}
\end{figure}

In the following, we explore a two-fold preparation scheme for the resource state relying on i) the preparation of a graph state and ii) a ladder of CNOT gates acting on the auxiliary qubits. 
The approach works as follows: we insert an artificial identity operation on the auxiliary qubits, which we divide into a forward- and backward (inverse) CNOT ladder. 
We define the {\em forward ladder} as
\begin{equation}
    \Lambda (X)_{\text{FW}} \coloneqq \prod_{l=1}^L \prod_{k=1}^{K-1} \Lambda_{k+1, l}(X_{k, l}),
    \label{eq:CNOTLadderForward}
\end{equation}
where the product order is taken from right to left. For the sake of clarity, we show the $\Lambda (X)_{\textrm{FW}}$ gate in Fig.~\ref{fig:CNOTForwardLadder}. 
We absorb the backward ladder $\Lambda (X)_{\textrm{FW}}^\dagger$ into the resource state $\ket{\mathcal{S}}$ from which we obtain a new graph state $\ket{G}$, which is LC-equivalent to $\Lambda (X)_{\textrm{FW}}^\dagger \ket{\mathcal{S}}$. After some intermediate steps illustrated in Appendix \ref{subapp:Backward}, we obtain the new graph state $\ket{G}$ with the adjacency matrix
\begin{equation}
\begin{adjustbox}{width=\linewidth}$
    \Gamma = 
    \begin{array}{|c!{\vrule width 2pt}c|c|c|c|c|}
        \hline 
        \Gamma_0 & \textrm{A}_0 & 0 & 0 & \cdots & 0 \\ \Xcline{1-6}{2pt} \rule{0pt}{10pt}
        & \Gamma_{\textrm{X}} & & & & \\ 
        \textrm{A}_0^T & \oplusr \text{UT}(\textrm{Z}\textrm{X}^T) & \text{LT}(\textrm{A}) & 0 & \cdots & 0\\ 
        & \oplusr \text{LT}(\textrm{X}\textrm{Z}^T) & & & & \\ \hline 
        0 & \text{UT}(\textrm{A}) & \textrm{A} & \text{LT}(\textrm{A}) & \cdots & 0  \\ \hline 
        0 & 0 & \text{UT}(\textrm{A}) & \textrm{A} & \cdots & 0  \\ \hline 
        \vdots & \vdots & \vdots & \vdots & \ddots & \vdots \\ \hline 
        0 & 0 & 0 & 0 & \cdots & \textrm{A}  \\
        \hline
    \end{array},
$\end{adjustbox}
\label{eq:GraphState_BW}
\end{equation}
and the local Clifford operators
\begin{align}
    \mathcal{C}_\textrm{aux.} = \bigotimes_{l=1}^L S_{1,l}^{(2\textrm{R} + \Gamma_{\textrm{X}}^\circ - \textrm{N}_Y)_l}.
    \label{eq:GraphStateLadderVOPs}
\end{align}
Note that these are the same ones as for the LC-MBQC, but now acting solely on the first time step $k=1$, rather than the entire auxiliary system. Then, we undo the operation by performing the forward ladder explicitly on top. The full preparation of the resource state is thus decomposed as
\begin{align}
    \Lambda (X)_{\text{FW}}~\mathcal{C} \ket{G} = \ket{\mathcal{S}}.
    \label{eq:ResourceStatePrepLadder}
\end{align}

This approach leads to a graph state which is almost entirely determined by the anticommutation matrix $\textrm{A}$, reflecting the commutation structure of the Pauli rotations. For $K=1$, we recover the same result as for LC-MBQC, since no ladder is actually applied. For larger $K$, we find that the graph is simply a decoration of the solution for $K=1$ in terms of the anticommutation graphs and the ladder. 
In this scheme, one needs at least $N+L+1$ active qubits to implement the pattern, which is independent from $K$ as desired. This number is due to the necessity to apply $\Lambda_{k+1, l}(X_{k, l})$ prior to the measurement of some auxiliary qubit $a_{k, l}$. The cost of extending the resource state preparation from Eq.~\eqref{eq:ResourceStatePrepLadder} by one time-step is precisely given by $||\textrm{A}||_1 + L$ two-qubit entangling gates.

It is important to highlight that the order of the forward and backward CNOT ladders are not interchangeable. 
Our choice of absorbing the backward ladder and explicitly applying the forward ladder ensures that the resulting graph state {\em and} the CNOT ladder on top both align with the temporal order of the pattern. If we instead absorb a forward ladder (where the controls and targets are interchanged, cf.~Appendix~\ref{subapp:Backward}), we also obtain an efficient graph state, which is rendered inefficient by the backward ladder. This is because prior to the measurement of some auxiliary qubit $a_{k, l}$, one needs to perform all the CNOT gates $\Lambda_{k, l}(X_{k+1, l})\dots \Lambda_{K-1, l}(X_{K, l})$, thus requiring all auxiliary qubits to be active at once. 

\paragraph{Relation to Clifford Circuit Synthesis}

The graph state construction with a forward CNOT ladder is closely related to known compilation rules of Clifford circuits.
The synthesis of arbitrary Clifford circuits has been studied extensively in the literature~\cite{Bravyi2021cliffordcircuit,10.5555/2011763.2011767,8335339,9435351}.
In Ref.~\cite{8335339}, it has been shown that such circuits can be decomposed into $7$ stages, given by $\Lambda(X)$-$\Lambda(Z)$-$P$-$H$-$P$-$\Lambda(Z)$-$\Lambda(X)$, where, here, $P$ is used to denote any gate of the set consisting of $S$, $Z$, and $S^\dagger$.  
When dealing with the preparation of graph states starting from the product state $\ket{0}^{\otimes n}$, we immediately see that the first three layers are not needed, since they leave the $\ket{0}^{\otimes n}$ state invariant, and the synthesis reduces to the $4$ stages $H$-$P$-$\Lambda(Z)$-$\Lambda(X)$.
These stages are then equivalent to the compilation result from Eq.~\eqref{eq:ResourceStatePrepLadder}, given that the $P$ and $\Lambda(Z)$ stages are interchangeable. 

\begin{figure*}[hbt!]
    \centering
    \includegraphics[width=0.95\textwidth]{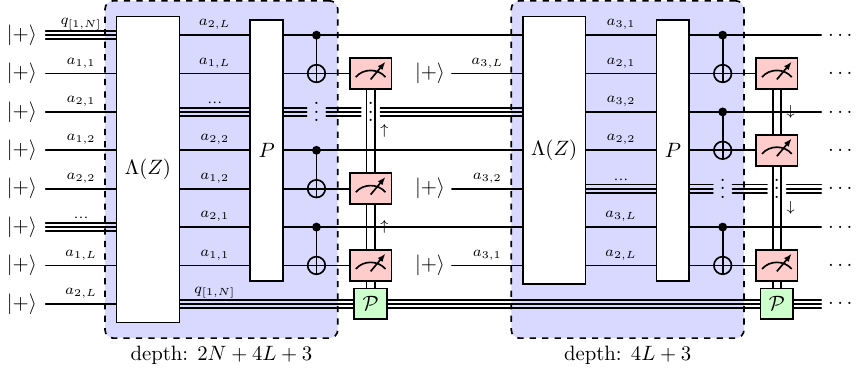}
    \caption{Implementation of the measurement pattern on a linear topology for $K>1$. Auxiliary qubits of subsequent time steps are arranged alternately, such that the respective part of the CNOT ladder can be implemented in depth one. The $\Lambda (Z)$-stages are decomposed onto the linear topology using the algorithm from Ref.~\cite{8335339}, which provides the depth bound. Note that for $K=1$, the $\Lambda Z$-stage of the first and only time step would only act on $N+L$ qubits and thus have a depth of $2N+2L+2$. Also note that the measurements (red) occur in the adaptive measurement bases from Eq.~\eqref{eq:AdaptivePattern}.}  
    \label{fig:LinearTopology}
\end{figure*} 

\paragraph{AC-MBQC on a Linear Topology}
From the results of Ref.~\cite{8335339}, we can also give a resource estimation for the compilation of the graph state on a quantum computer using $\Lambda(X)$ gates on a linear topology.
For an arbitrary $\Lambda(Z)$ stage on $n$ qubits, one needs a circuit of depth $(2n+2)$, however, only up to the complete reversal of all qubits.
If we arrange the auxiliary qubits such that two qubits implementing the same Pauli rotations at two subsequent time steps are next to each other, the respective part of the CNOT-ladder can be implemented in depth $1$. We illustrate this layout in Fig.~\ref{fig:LinearTopology}.
Thus, in the case where we apply $L$ Pauli rotations on $N$ qubits $K$ times (in which we need $N+2L$ active qubits on a linear topology) we have a preparation circuit of depth $2(N+2L)+3$ for the first step (in which the initial $N$ qubits are connected to $2L$ auxiliary qubits) and a $(K-1)$-times repeating circuit of depth $4L+3$ for the connection of $L$ auxiliary qubits in iteration $k$ to the ones in iteration $k+1$ on a linear topology. Overall, we can bound the depth in terms of two-qubit gates by
\begin{align}
    d = \begin{cases}
    2N + 2L + 2 & \textrm{for } K=1, \\
    2N + (K-1)(4L+3) & \textrm{for } K>1. \\\end{cases}
\end{align}

\section{Hybrid Simulation Scheme} \label{sec:Hybrid}

So far, we have presented resource states, which permit us to output states of the form $U(\bm \theta)\ket{\mathcal{S}_0}$ using MBQC. Depending on the goal of the computation, this evolved state might serve as an input for subsequent quantum operations. In other instances, one wants to directly estimate some observable $O$ on the evolved state, i.e.,~compute $\braket{O}=\braket{\mathcal{S}_0|U^\dagger(\bm \theta)  O U(\bm \theta) |\mathcal{S}_0}$. A straightforward approach then lies in decomposing the observable in the Pauli basis $O=\sum_i \alpha_i O_i$ such that each $O_i$ is a Pauli-string and all $\alpha_i$ are real. What follows is a recap of the hybrid simulation scheme introduced in Ref.~\cite{Kaldenbach2025Mapping}.

In standard form, our measurement patterns consist, as mentioned in Sec.~\ref{sec:graph_state_algorithm}, of a Clifford layer, a measurement pattern on the auxiliary qubits, and a final Pauli correction on the main qubits.
Since the final Pauli corrections are only applied to the qubits pertaining to the logical state (and not the auxiliary system), we can classically pre-simulate the action of these corrections if we perform measurements of the main qubits in any Pauli basis. That is, we first ignore the final correction from Eq.~\eqref{eq:FinalCorrection} and directly measure the Pauli-string $O_i$ on the full graph state. By using the Pauli measurement rules (Eq.~\eqref{eq:GraphStatePauliMeasurement}), we can exactly compute the (classical) distribution of the measurement outcomes $o_1, o_2, \dots, o_N$, and also obtain the post-measurement graph state on the auxiliary qubits up to a product of local Cliffords which are independent of the measurement outcomes, and a product of local Paulis which do depend on the classical outcomes. Therefore, we only need to apply the measurement rules once for every $O_i$. We can then simulate a shot by drawing some bitstring from the distribution, execute the measurement-pattern on the post-measurement auxiliary graph state on the quantum computer, compute the final correction based on the auxiliary outcomes $s_1,s_2,\dots, s_M$, and finally apply this correction to the bitstring on the main qubits. As a final note, this procedure can be readily generalized to simultaneously measuring mutually commuting Pauli-strings\footnote{This goes beyond qubit-wise commutation where simultaneous measurements are trivial.}. Such strings can be simultaneously diagonalized using Clifford circuits prior to the measurement \cite{Gokhale2020Measurement}. This Clifford circuit can then be interchanged with the final Pauli corrections (giving rise to new corrections) and absorbed into the resource state.  We next discuss this hybrid simulation scheme in the context of the presented resource state preparation schemes. 

Concerning the LC-MBQC approach, there are two ways to deal with the measurements. First, one can perform the measurements on the initial graph state from Eq.~\eqref{eq:GraphStateResourcePeriodic} prior to the annealing protocol. Then, for Pauli-$Z$ measurements, one can trivially read-off the post-measurement graph state by simply removing the main qubits and incident edges. Computing the Pauli byproducts is also easy given that the neighborhood of the main qubits is known through $\textrm{A}_0$. For Pauli-$Y$ measurements, one has to perform local complementations, which complicates reading off the post-measurement graph state. Still, the graph is guaranteed to be periodic. Only for Pauli-$X$ measurements is the resulting graph no longer periodic on $k=1$ due to edge complementations. Using this approach, the local Paulis arising from byproducts are spread through all auxiliary registers. 
One can instead first perform the simulated annealing, which, if successful, leads to a graph state whose main qubits only connect to the last auxiliary register. As Pauli measurements on a qubit only inflict byproducts on its neighborhood, or on the neighborhood of one of its neighbors, this typically ensures that the byproducts only affect the last time step. Also, simulating the measurements is more efficient this way since the size of the neighborhoods is $\mathcal{O}(N+L)$ instead of $\mathcal{O}(N+KL)$. 

For the AC-MBQC approach, the Pauli measurements can always be performed with $\mathcal{O}(N+L)$ cost, as the main qubits only connect to the first auxiliary register $k=1$. This way, the Pauli byproducts are localized at the beginning of the pattern instead of the end. However, one should keep in mind that these byproducts are applied prior to the forward CNOT ladder. If one were to interchange the Paulis and the ladder, one would again end up with the same Paulis as for LC-MBQC prior to the annealing. 

\begin{figure*}[b!]
    \centering
    \includegraphics[width=\textwidth]{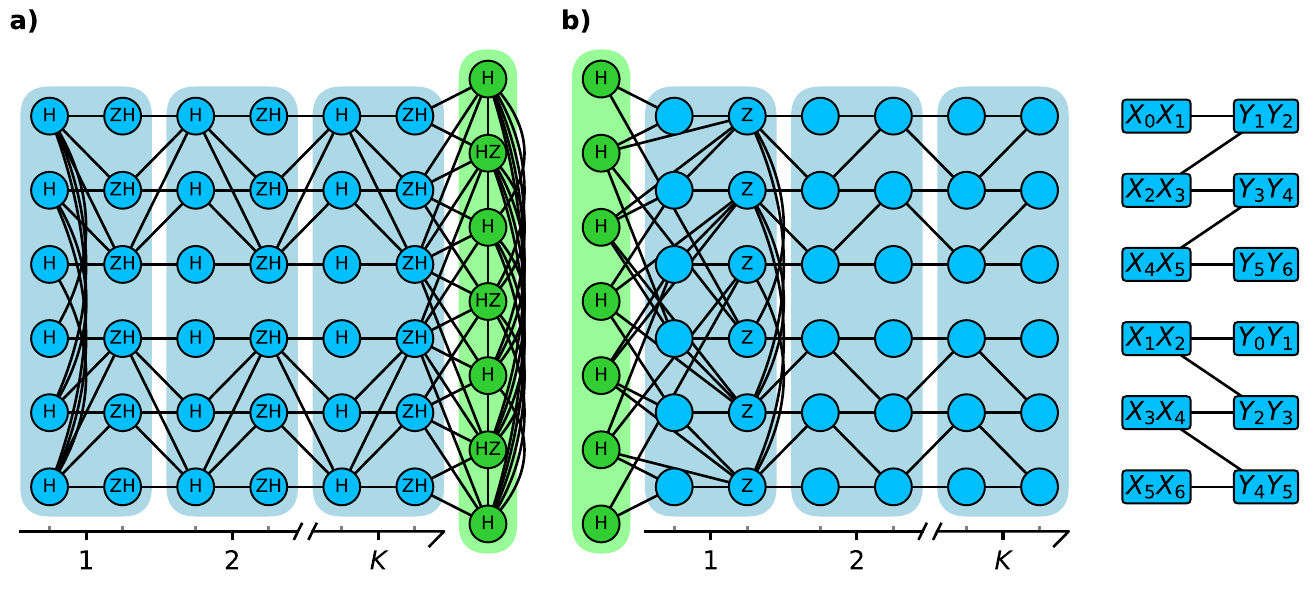}
    \caption{Resource state preparation for Trotterized Hamiltonian simulation of the $XY$ model with $N=7$ in terms of the (optimized) graph state representation and local Clifford operators for a) LC-MBQC, and b) AC-MBQC up to the forward CNOT ladder. 
    The legend (right panel) displays the anticommutation graph of the Hamiltonian’s Pauli-strings and illustrates the correspondence between the auxiliary qubits in the graph states a) and b) and their respective implemented rotations.
    In both cases, owing to the disjoint anticommutation graph (right panel), the measurement patterns separate into two independent parts.}
    \label{fig:xy_odd_gs}
\end{figure*} 

\newpage

\section{Applications} \label{sec:Applications}

\subsection{Trotterized Time Evolution} \label{subsec:TimeEvolution}
In this section we study the patterns that implement the time evolution of different Hamiltonians, namely the $XY$ model and the perturbed toric code Hamiltonian.
\subsubsection{The \texorpdfstring{$XY$}{XY} model}

As our first example, we consider the time evolution of the 1D Heisenberg $XY$-model on a linear chain, whose Hamiltonian is given by
\begin{align}
    H_{XY} = \sum_{n=0}^{N-2} \left[\frac{1+\gamma}{2} X_nX_{n+1} + \frac{1-\gamma}{2}Y_nY_{n+1}\right],
    \label{eq:XY_model}
\end{align}
where $\gamma$ is the anisotropy parameter.
As an initial state, we consider the zero state $\ket{0}^{\otimes N}$, since it is LC-equivalent to all computational basis states representing {\em $n$-particle states} ($n$ spins up, $N-n$ spins down). Therefore, the patterns we derive can be easily adjusted via LC operators and appropriate flips of measurement bases to account for various initial configurations. 

We apply our algorithm to two cases, when $N$ is even and when $N$ is odd.
Our distinction between these two cases is not physically motivated, but rather stems from a structural difference we observe in the LC-MBQC approach. 

\paragraph{Odd sites: }
For the case with odd sites, we consider the $XY$ model with $N=7$ sites and $K=3$ repetitions. As we will see, this provides us with the minimalist instance from which we can straightforwardly extrapolate the solution for arbitrary odd $N$ and integer $K$. 

For the Trotterization, we partition the Hamiltonian into two groups of mutually commuting terms, namely the groups of all $X_nX_{n+1}$, and $Y_nY_{n+1}$ terms. Within these groups, we sort the terms into two brickwall-like structures, which then visually reveals that the anticommutation graph of the $XY$ model separates into two disjoint subgraphs. This has direct implications on the measurement pattern, as intermediate measurement results from auxiliary qubits corresponding to one subgraph cannot inflict corrections onto computations of the other subgraph. 

\begin{figure*}[htb!]
    \centering
    \includegraphics[width=\textwidth]{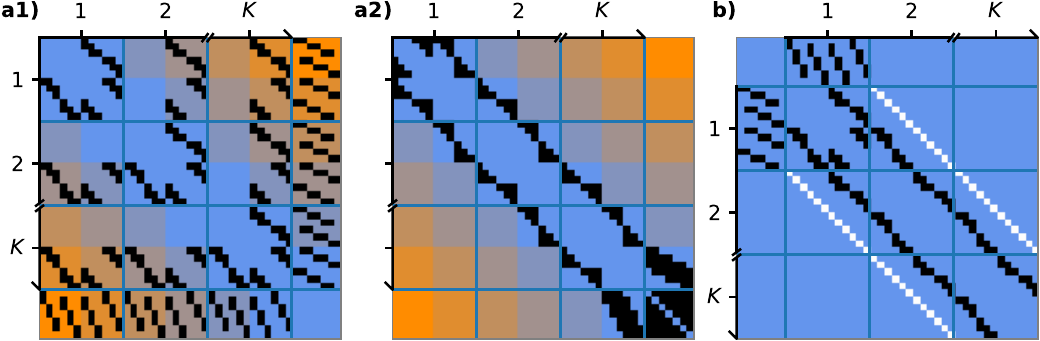}
    \caption{Adjacency matrices of the graph states representations of the resource state for Trotterized Hamiltonian simulation of the $XY$ model with $N=7$ for a) LC-MBQC, and b) AC-MBQC. The background in a) represents the distance matrix on a logarithmic color map. The panels a1) and a2) refer to the pre- and post-optimization graph states, respectively. The off-diagonals (white) in b) represent the forward CNOT ladder which is applied to the underlying graph state (black).}
    \label{fig:xy_odd_adjs}
\end{figure*}

To obtain the graph state and LC operators for the LC-MBQC approach, we employ our annealing scheme with a cooling rate of $\lambda = 0.99995$. Our results for both the AC-MBQC and LC-MBQC are depicted in Fig.~\ref{fig:xy_odd_gs}. It turns out that LC-MBQC has a simple solution, where the LC operators for all auxiliary qubits are $H$ gates. Since we find the periodic graph structure with identical LC operators for the time steps $k=2$ and $k=3$, we can simply repeat the same pattern until some arbitrary $k=K$, without ever having to run the annealing algorithm on a larger graph. 
The output qubits of the LC-MBQC have all-to-all connectivity, which is arguably sub-optimal in terms of edges. This is however not an instance of the optimizer failing, but rather a successful enforcement of the periodic constraints. For the sake of clarity, we additionally depict the adjacency matrices for both protocols in Fig.~\ref{fig:xy_odd_adjs}, especially with the pre- and post-optimization graphs for LC-MBQC. For all subsequent examples, we relegate either the adjacency matrices or graph states to Appendix \ref{app:Adj} to avoid redundancies. 

The $XY$ model highlights pros and cons of both MBQC approaches employed in this work. The time evolution based on the resource state preparation with AC-MBQC requires $4N-3$ active qubits at any intermediate measurement round due to the input qubits being active all the time and the CNOT gates ranging to the next time step. The LC-MBQC comes out with $2N-1$ active qubits at any time, which is due to the absence of the \enquote{long-range} CNOT gates and active input qubits. However, for AC-MBQC the number of edges and CNOTs grows as $\mathcal{O}(KN)$ since the number of edges in the anticommutation graph grows as $\mathcal{O}(N)$, while for LC-MBQC the number of edges grows as $\mathcal{O}(KN^2)$, which becomes a bottleneck when the $\Lambda (Z)$ gates cannot be applied in parallel. The obvious drawback of LC-MBQC lies in the necessity to optimize the graph state itself. 

\begin{figure}[H]
    \centering
    \includegraphics[width=0.85\linewidth]{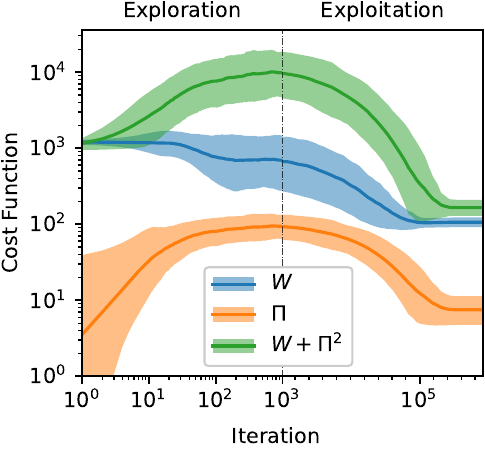}
    \caption{Convergence analysis of the Simulated Annealing algorithm applied to the graph state for the $XY$ model on $N=7$ sites with $K=3$ time steps. Here, $W$ denotes the weight of the graph according to the distance matrix, $\Pi$ is a measure for the periodicity of the graph, and $W+\Pi^2$ is the combined cost function which is minimized by the annealing algorithm. For a better visualization of both the exploration and exploitation phase, we first employ logarithmic and then linear scaling on the x-axis. During the exploration phase, the weak constraint $\Pi$ is strongly violated, showcasing how the optimizer traverses non-periodic graph states during the optimization.}
    \label{fig:xy_annealing}
\end{figure}

In the following, we take a closer look at how the annealing algorithm itself works. 
For that purpose, we consider the same system size as before and perform 20 runs of the annealing algorithm. For each iteration in which a new graph state is accepted, we store the graphs weight $W$, the periodicity score $\Pi$, as well as the total cost function $W+\Pi^2$. Given that all the runs attain new configurations at different iterations, we use linear interpolation to fill the missing values. We then compute the means and the $2\sigma$ confidence intervals across all runs. For visual purpose, we smooth the values using a linear Savitzky–Golay filter \cite{savitzky1964smoothing}. The results are shown in Fig.~\ref{fig:xy_annealing}. 

We can easily distinguish between the exploration- and exploitation-dominant phases of the annealing process. The exploration-dominant phase is comparatively short as it only occurs at high temperatures, which are exponentially suppressed using geometric cooling. During this phase, the optimizer already achieves modest reductions of $W$, at the expense of violating the weak constraints $\Pi$. Once the exploitation-dominant phase kicks in, all quantities are simultaneously reduced, until convergence is achieved. Note that the final variance is low, but not zero, as not every optimization run yields the optimal solution. 

\paragraph{Even sites:} 

\begin{figure*}[htb!]
    \centering
    \includegraphics[width=\textwidth]{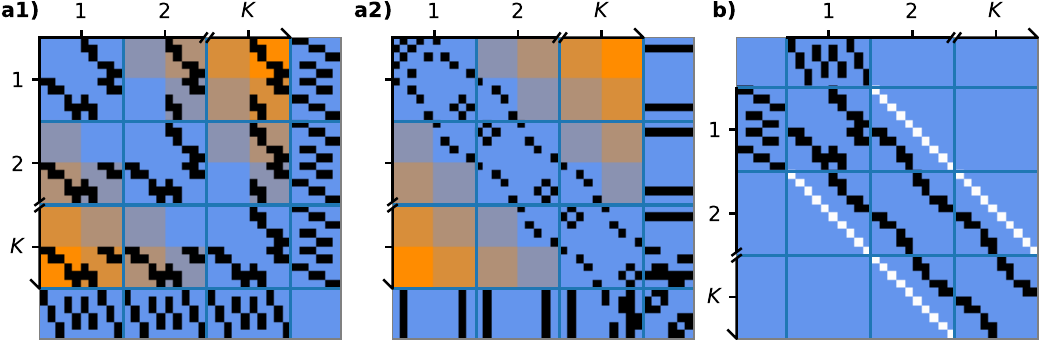}
    \caption{Adjacency matrices of the graph states representations of the resource state for Trotterized Hamiltonian simulation of the $XY$ model with $N=6$ for a) LC-MBQC, and b) AC-MBQC. The background in a) represents the distance matrix on a logarithmic color map. Note that the distance matrix has been modified for this specific example, as discussed in the main text. The panels a1) and a2) refer to the pre- and post-optimization graph states, respectively. The off-diagonals (white) in b) represent the forward CNOT ladder which is applied to the underlying graph state (black).}
    \label{fig:xy_even_adjs}
\end{figure*}

\begin{figure*}[b!]
    \centering
    \includegraphics[width=\textwidth]{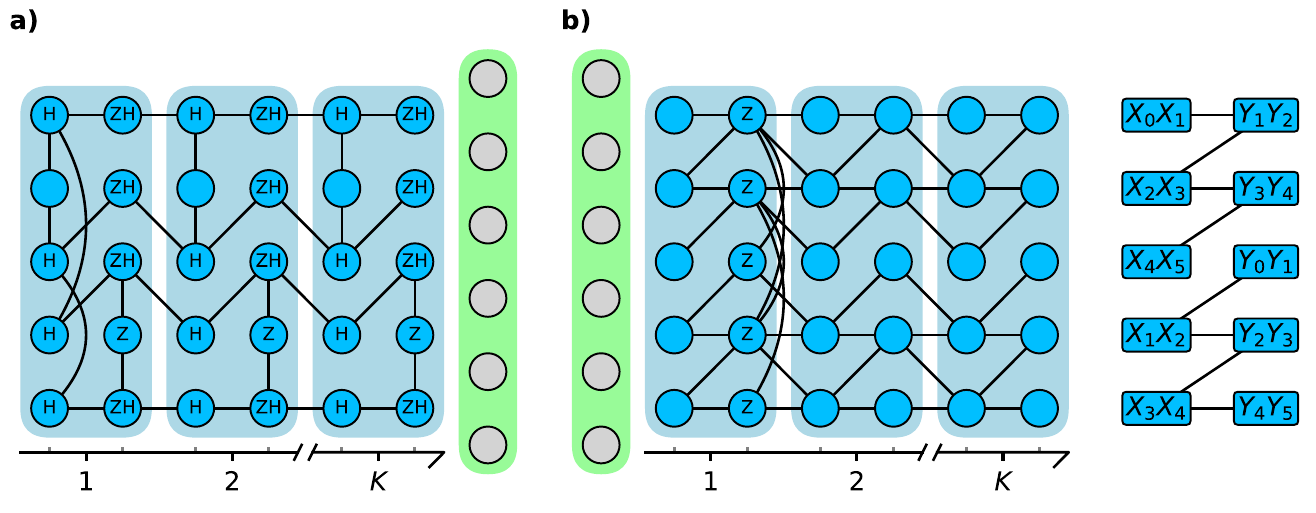}
    \caption{Resource state preparation for the hybrid Trotterized Hamiltonian simulation of $XX$-type spin-correlations in the $XY$ model with $N=6$ in terms of the (optimized) graph state representation and local Clifford operators for a) LC-MBQC, and b) AC-MBQC up to the forward CNOT ladder. All main qubits (gray) have been measured in the Pauli-$X$ basis. This corresponds to the bare Pauli-$Z$ basis after interchanging with the local Clifford operators, hence the graph states are obtained by simply removing the main qubits from the graph states in Fig.~\ref{fig:xy_even_adjs}. The resulting byproducts depend on the measurement outcomes and are not displayed here.
    The legend (right panel) displays the anticommutation graph of the Hamiltonian’s Pauli-strings and illustrates the correspondence between the auxiliary qubits in the graph states a) and b) and their respective implemented rotations.
    In both cases, owing to the disjoint anticommutation graph (right panel), the measurement patterns separate into two independent parts.}
    \label{fig:xy_even_x_gs}
\end{figure*}

When $N$ is even, we find that the annealing algorithm is not able to find a good solution, in a sense that a) the connectivity in-between auxiliary registers should be restricted to adjacent time steps, and b) only the last auxiliary register should connect to the output qubits. By discarding condition b), one can however obtain graph states where a) is satisfied, but every auxiliary register connects to the output, effectively recovering an input-equals-output setting similar to the AC-MBQC approach. We explicitly enforce this solution by modifying the distance matrix $D$ such that all edges between the auxiliary registers and the main register have weight 1. The adjacency matrices for both LC- and AC-MBQC are shown in Fig.~\ref{fig:xy_even_adjs}.

\addtocounter{figure}{1}
\begin{figure*}[b!]
    \centering
    \includegraphics[width=0.75\textwidth]{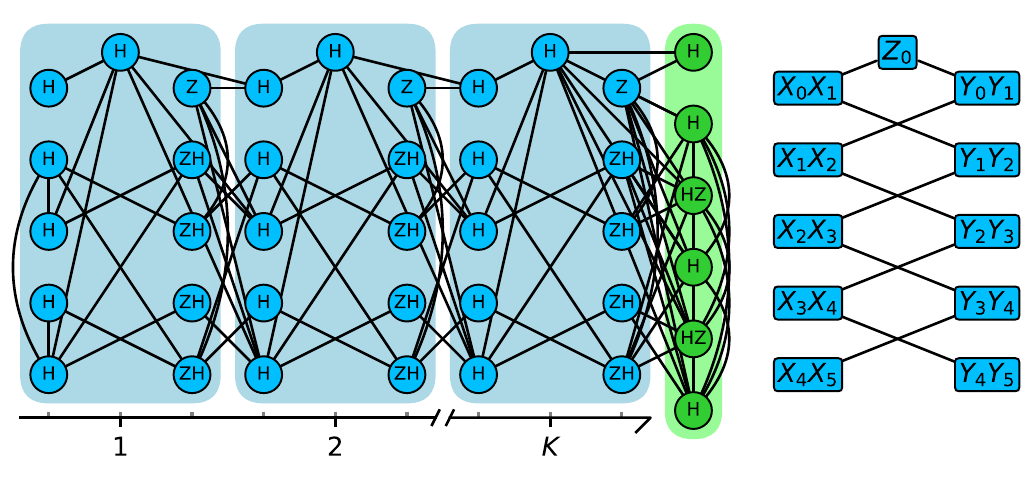}
    \caption{Resource state preparation for the Trotterized Hamiltonian simulation of the perturbed $XY$ model with $N=6$ in terms of the optimized graph state representation and local Clifford operators for LC-MBQC. The legend (right panel) displays the anticommutation graph of the Hamiltonian’s Pauli-strings and illustrates the correspondence between the auxiliary qubits in the graph states a) and b) and their respective implemented rotations.
    Note that the local perturbation $Z_0$ fuses the previously disjoint anticommutation graph.}
    \label{fig:xy_even_imp_gs}
\end{figure*}

In such a scenario, one can recover a good solution by employing the hybrid simulation algorithm from Sec.~\ref{sec:Hybrid}. If one considers the measurement of, e.g., spin-correlations $X_i X_j$, or more generally speaking any measurement in the Pauli-$X$ basis, this turns into a bare Pauli-$Z$ measurement due to the $H$ gates on the main qubits, and therefore simply deletes the edges without rewiring the auxiliary qubits. Such post-measurement graph states for both protocols are depicted in Fig.~\ref{fig:xy_even_x_gs}. One can infer from the VOPs that this solution is indeed structurally different from the odd case. Similar post-measurement graph states are obtained for Pauli-$Y$ measurements. The subtle difference lies in the last slice of auxiliary operations being decoupled from the graph. This is because the $YY$-rotations commute with the $Y$ observables, and therefore have no impact. However, it should be emphasized that the study of spin-dynamics, i.e., the measurement in the Pauli-$Z$ basis, does not result in a convenient post-measurement graph state. That is because it effectively turns into a bare Pauli-$X$ measurement, which can be visualized in terms of an edge complementation prior to a $Z$ measurement. Such edge complementations rewire the auxiliary qubits and induce undesired edges, which our annealing algorithm can not remove. 

\addtocounter{figure}{-2}
\begin{figure}
    \centering
    \includegraphics[width = 0.4\textwidth]{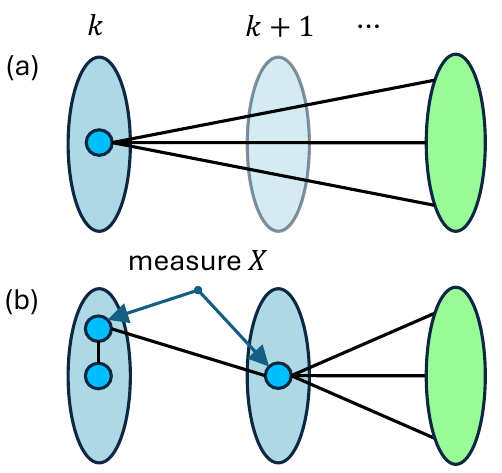}
    \caption{Illustration of a graph state embedding to remove undesired edges. (a) A graph state with undesired edges reaching from some auxiliary register $k$ to the main qubits. (b) By inserting two auxiliary qubits on the registers $k$ and $k+1$, which have to be measured in the Pauli-$X$ basis, one can shift the undesired edges to the auxiliary register $k+1$ and thus reduce the number of active qubits in the pattern.}
    \label{fig:embedding}
\end{figure}
\addtocounter{figure}{1}

Another way to mitigate undesired connections in the graph state lies in vertex-minor embedding theory; see Ref.~\cite{KIM202454} for a survey. 
Let us assume we have an undesired connection from one qubit at time-step $k\neq K$ to the main register.
Then we can introduce two auxiliary qubits, which we have to measure in the $X$ basis at time-step $k$ and $k+1$, as is shown in Fig.~\ref{fig:embedding}. Such embedding to reduce the degree of a given node has been first introduced in Ref.~\cite{Hoyer2006Resources}. Using this embedding, we can transfer the undesired edges to the auxiliary register of time-step $k+1$. This procedure is repeated until only the last auxiliary register $K$ connects to the output.

One can directly achieve such embedding into larger graphs within our framework by artificially growing the Hamiltonian with arbitrary new Pauli-strings of weight 0, thus effectively inserting identity gates into the measurement pattern. In case of the $XY$ model on even sites, we find that it is sufficient to consider some impurity $Z_j$. We give an example in Fig.~\ref{fig:xy_even_imp_gs} where we consider the impurity $Z_0$. Note that the auxiliary qubits for the $R_{Z_0}$ rotations are decorated with $H$ gates, thus they effectively rewire the graph through a Pauli-$X$ measurement, which, again through edge complementations, re-introduces the undesired edges. 

Given that the graph states from Fig.~\ref{fig:xy_even_adjs} contain precisely two qubits with undesired connections per time-step, the iterative application of the embedding algorithm from Ref.~\cite{Hoyer2006Resources} overestimates the required number of Pauli measurements compared to our embedding scheme based on identity terms. How to minimally dress a Hamiltonian such that an efficient graph states comes out remains an open question.

\begin{figure*}[htb!]
    \centering
    \includegraphics[width=\textwidth]{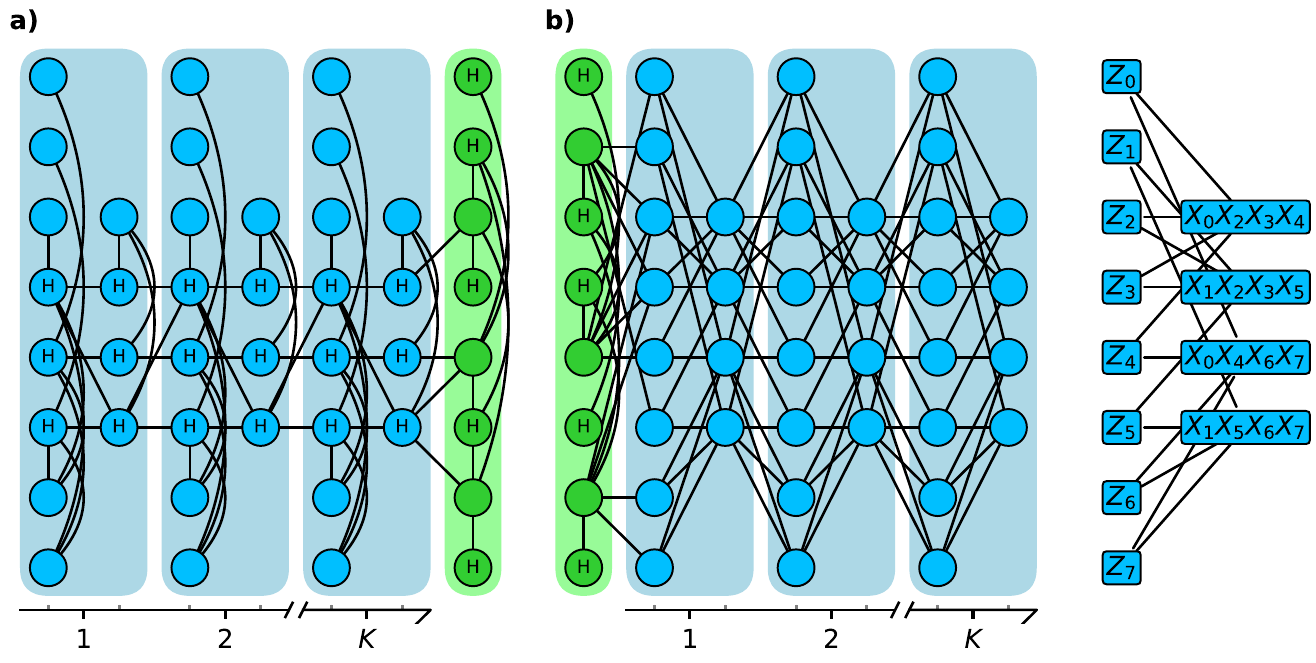}
    \caption{Resource state preparation for Trotterized Hamiltonian simulation of the perturbed toric code Hamiltonian starting from the toric code state in terms of the (optimized) graph state representation and local Clifford operators for a) LC-MBQC, and b) AC-MBQC up to the forward CNOT ladder. 
    The legend (right panel) displays the anticommutation graph of the Hamiltonian’s Pauli-strings and illustrates the correspondence between the auxiliary qubits in the graph states a) and b) and their respective implemented rotations.
    Note that all auxiliary qubits without $H$ gates only act within their respective measurement layers. Consequently, the intermediate state can always be stored on precisely three qubits (the ones with $H$).}
    \label{fig:toric_gs}
\end{figure*}

While the initial motivation of our work lay in the removal of classicality from the measurement patterns, it turns out that a carefully selected fraction of Pauli measurements is sometimes necessary to obtain an efficient pattern based on graph states. Here, one should emphasize that the AC-MBQC approach always works by construction without any additional Pauli measurements.

\subsubsection{Perturbed toric code}
As an example for which the input state is not a trivial product state, we consider perturbations to a system whose undisturbed ground state is given by an entangled stabilizer state. Here, we consider a minimalist instance of the toric code Hamiltonian \cite{Kitaev2003FTQC} on $N=8$ qubits
\begin{align}
    H_{\rm toric}&=Z_0 Z_1 Z_2 Z_6+Z_0 Z_1 Z_3 Z_7\nonumber \\
    &+Z_2 Z_4 Z_5 Z_6+Z_3 Z_4 Z_5 Z_7 \nonumber \\
    &+X_0 X_2 X_3 X_4+X_1 X_2 X_3 X_5\nonumber \\
    &+X_0 X_4 X_6 X_7+X_1 X_5 X_6 X_7.
    \label{eq:toric}
\end{align}
The ground state of this Hamiltonian is a stabilizer state. Its graph state representation can be easily computed given that the 8 mutually commuting Pauli-strings in Eq.~\eqref{eq:toric} can be interpreted as its stabilizers \cite{Huebener2011Tensor}. This type of graph state is also referred to as toric code state \cite{Huebener2011Tensor}. 
Next, we add local perturbations governed by local magnetic fields
\begin{equation}
    H_p=\sum_{i=0}^7 \lambda_i Z_i.      
\end{equation}
The preparation and simulation of a disturbed toric code state has been extensively studied in terms of measurement-based variational quantum eigensolvers \cite{Ferguson2021MBVQE} and tensor networks \cite{Huebener2011Tensor}. 
To prepare the ground state of the perturbed system, one may use adiabatic time evolution, for which one needs the time evolution pattern of the full Hamiltonian $H_{\rm toric}+H_p$. We present patterns for this adiabatic time evolution in Fig.~\ref{fig:toric_gs}. Here, we can discard all the plaquette operators $Z_iZ_jZ_kZ_l$ since they commute with the chosen perturbation and further leave the initial state invariant. 

Using LC-MBQC, we find that during any intermediate stage of the measurement-pattern, only three qubits are required to store the entire quantum information. Only in the end is it unfolded onto the 8 qubits. Such qubit reductions can not be detected using the AC-MBQC graph state. Given that the goal of this computation would be to measure the expectation value of the perturbed toric code Hamiltonian, one could employ the hybrid simulation algorithm and get rid of these 8 qubits by performing measurements in the Pauli-$X$ and $Z$ bases.  Interestingly, this is an example where even the auxiliary register of $k=1$ matches the periodic constraints perfectly. 

\subsection{Minimal Generating Sets} \label{subsec:MinimalGeneratingSets}

\begin{figure*}[htb!]
    \centering
    \includegraphics[width=\textwidth]{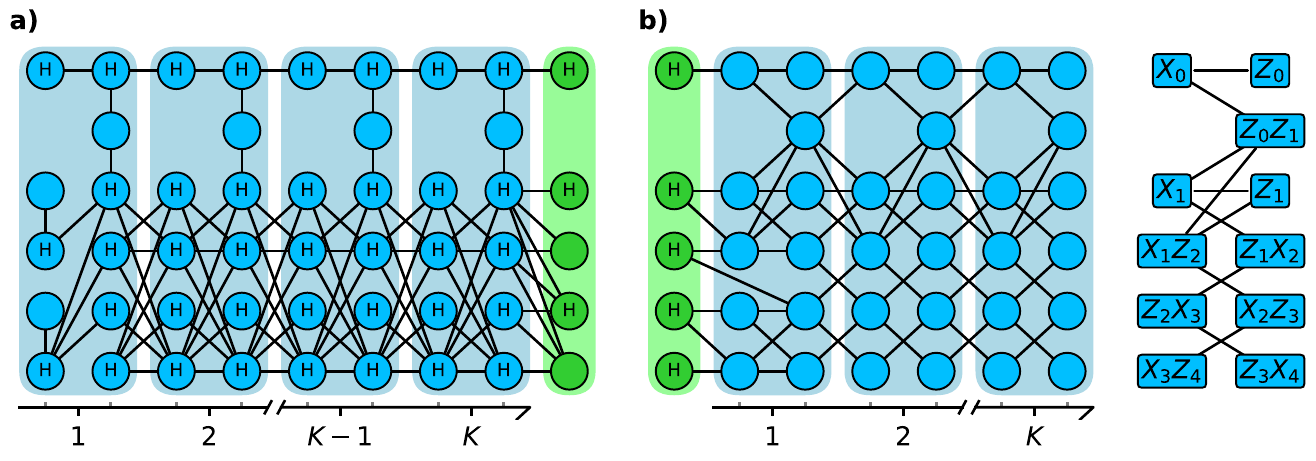}
    \caption{Resource state preparation for universal quantum computation generated by the minimal 2-local set with $N=5$ in terms of the (optimized) graph state representation and local Clifford operators for a) LC-MBQC, and b) AC-MBQC up to the forward CNOT ladder. 
    The legend (right panel) displays the anticommutation graph of the Hamiltonian’s Pauli-strings and illustrates the correspondence between the auxiliary qubits in the graph states a) and b) and their respective implemented rotations.}
    \label{fig:2-local_gs}
\end{figure*}

Above, we considered unitaries $U(\bm\theta)$ corresponding to the Trotterization of the time-evolution according to a Hamiltonian $H$. The resultant unitaries consisted of a product of rotations of Pauli-strings from a given set. It is natural to wonder: which sets of Pauli-strings can produce {\em any} unitary in this way? In Ref.~\cite{smith2025optimally}, it was demonstrated that any such universal set of Pauli-strings must contain at least $2N+1$ elements and, moreover, provided various examples of universal sets that obtain this bound and analysed their compilation efficiency. Here, we consider several examples of universal sets, both minimal and not, in the context of the deterministic graph state algorithm presented above.

Given such a universal set of at least $2N+1$ Pauli-strings $\mathcal{A} = \{\mathcal{P}^{(1)}, \dots, \mathcal{P}^{(2N+1)}, \dots\}$, one can then implement any Pauli rotation $R_{\mathcal{P}}(\theta)$ from Eq.~\eqref{eq:PauliRotation} using a sequence $G_1, \dots, G_l \in \mathcal{A}$ as follows:
\begin{align}
    R_\mathcal{P}(\theta) &= R_{G_1}\left(-\frac{\pi}{2}\right) \cdots R_{G_{l-1}}\left(-\frac{\pi}{2}\right) R_{G_{l}}\left(2^{l-1}\theta \right) \nonumber \\ 
    &\hphantom{=~} \times 
    R_{G_{l-1}}\left(\frac{\pi}{2}\right) \cdots R_{G_1}\left(\frac{\pi}{2}\right)
    \label{eq:PauliRotationDecomposition}
\end{align}
The value $l$ relates to the number of gates (i.e., the unitaries $e^{i\alpha G_l}$ in the above expression) required to compile $R_\mathcal{P}(\theta)$, and hence also relates to the circuit depth for implementing $R_\mathcal{P}(\theta)$ in this context (the exact details of the connection to the circuit depth depend on the nature of the $G_{l}$). For different choices of $\mathcal{A}$, the maximum value of $l$ for compiling to any Pauli-string $\mathcal{P} \in \mathbb{P}_{N}$ differs. In Ref.~\cite{smith2025optimally}, an algorithm was provided which, for generating sets $\mathcal{A}$ exhibiting a certain structure, produces the optimal length sequence of $G_{l}$ for implementing $R_\mathcal{P}(\theta)$ as above, with the guarantee that $l=\mathcal{O}(N)$. Furthermore, Ref.~\cite{smith2025optimally} also provided an analysis of how much of $\mathbb{P}_{N}$ can be generated as a function of sequence length $l$ for different choices of $\mathcal{A}$, a quantity that was referred to as {\em compilation rate} (see Ref.~\cite{smith2025optimally} for further details).

Similarly to the case of Trotterized dynamics considered earlier, let us consider the following unitary defined as a parametrized, repeating product of rotations of the elements of $\mathcal{A}$, i.e., 
\begin{align}
U_{\mathcal{A}}(\bm \theta, K) := \prod_{k =1}^{K} \prod_{j=1}^{|\mathcal{A}|} R_{\mathcal{P}^{(i)}}(\theta_{kj}).
\label{eq:UnitaryDecomposition}
\end{align}
The fact that $\mathcal{A}$ is universal in particular means that, for any desired $N$-qubit unitary $U$ there exists some $K \in \mathbb{N}$ and some $\bm{\theta} \in [-\pi, \pi)^{K|\mathcal{A}|}$ such that $U_{\mathcal{A}}(\bm \theta, K) \approx U$ to any desired level of approximation. Given that Eq.~\eqref{eq:UnitaryDecomposition} has the same structure as the unitary for Trotterized Hamiltonian simulation, namely as a product of generalized Pauli rotations, we can now use our algorithms to compute resource states for implementing $U_{\mathcal{A}}(\bm \theta, K)$ via MBQC.

\subsubsection{2-Local Set}

One such minimal universal generating set, presented as Example $2$ in Ref.~\cite{smith2025optimally}, is the following:
\begin{align}
\{X_{1}, Z_{1}, X_{2}, Z_{2}, Z_{1} \otimes Z_{2} \} \cup \{X_{i} \otimes Z_{i+1}, Z_{i} \otimes X_{i+1} \}_{i = 2}^{N-1}.
\end{align}
If the $N$ qubits are considered to be arranged linearly, then this generating set satisfies the criteria of being universal, minimal and consisting of at most $2$-local, nearest-neighbour interactions. However, in terms of compilation rate, this set is known to be sub-optimal (cf.~Fig.~1 of Ref.~\cite{smith2025optimally}). We give an example on the Trotterization of such set on $N=5$ qubits in Fig.~\ref{fig:2-local_gs}. The optimized LC-MBQC graph state has some structural similarities the $XY$ model, and in particular, also suffers from an $\mathcal{O}(N^2)$ edge scaling. Given that the $R_{ZZ}$ rotation is realized by an auxiliary qubit which only connects to qubits of the same measurement layer, we can infer that the entire quantum information can be stored by precisely $N$ qubits, as one would expect for a universal set on $N$ qubits. The AC-MBQC approach achieves an $\mathcal{O}(N)$ scaling in the number of entangling gates per time step, at the expense of more active qubits. 

\subsubsection{Minimal Sets Containing All Weight 1 Pauli-\texorpdfstring{$X$}{X} and Pauli-\texorpdfstring{$Z$}{Z} strings}

\begin{figure*}[htb!]
    \centering
    \includegraphics[width=\textwidth]{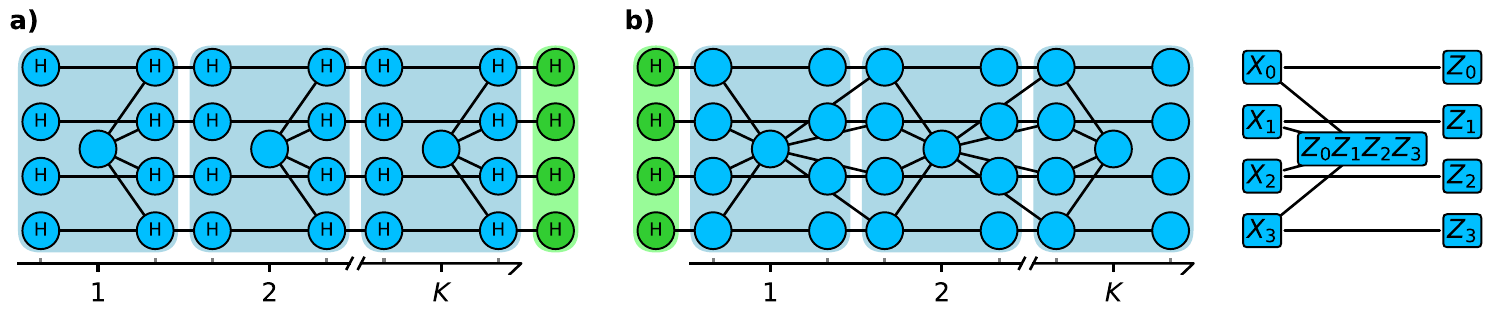}
    \par\vspace{1em}
    \includegraphics[trim=0 0 0 20, clip, width=\textwidth]{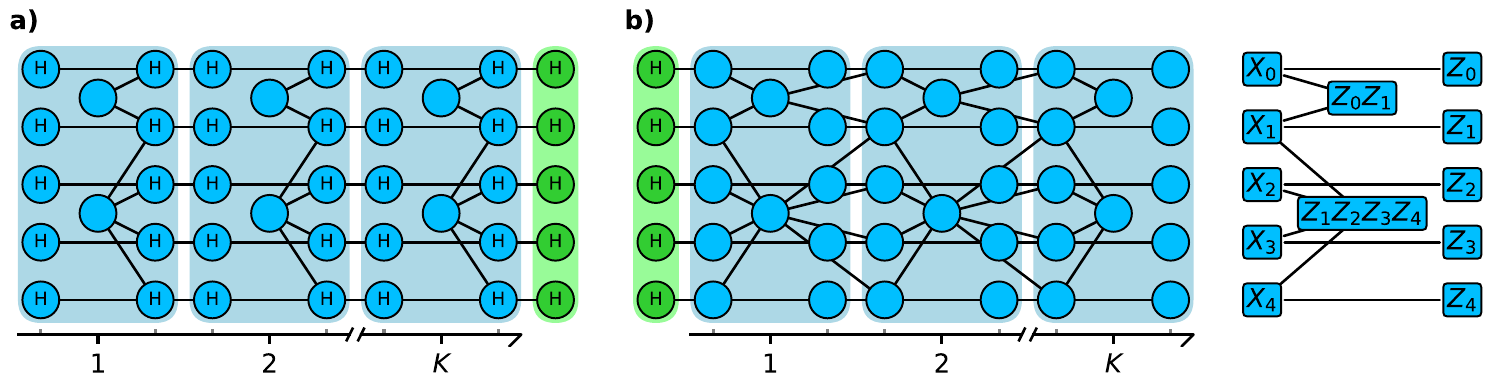}
    \caption{Resource state preparation for universal quantum computation generated by the minimal set containing all weight 1 Pauli-strings with (top panel) the even case $N=4$, and (bottom panel) the odd case $N=5$ and $s=2$, in terms of the (optimized) graph state representation and local Clifford operators for a) LC-MBQC, and b) AC-MBQC up to the forward CNOT ladder. 
    The legend (right panel) displays the anticommutation graph of the Hamiltonian’s Pauli-strings and illustrates the correspondence between the auxiliary qubits in the graph states a) and b) and their respective implemented rotations.}
    \label{fig:1-local_gs}
\end{figure*}

In the context of generating sets of Pauli-strings, the ability to perform any single-qubit unitary on any of the $N$ qubits corresponds to a generating set containing the set of strings $\{X_{1},Z_{1},X_{2},Z_{2},\dots, X_{N},Z_{N} \}$. Clearly, this set alone contains $2N$ elements, so one could rightfully ask: is it always possible to append just a single Pauli-string to obtain a universal set that obtains the minimal bound $2N+1$? As demonstrated in Ref.~\cite{smith2025minimally}, the answer to this question in fact depends on $N$: if $N$ is even, then yes, while if $N$ is odd, at least $2$ additional Pauli-strings are required for universality. In the latter case, there are, in fact, several distinct possible choices of additional Pauli-strings that ensure universality. 
Explicitly, throughout this subsection, we will consider the universal generating sets given, in the case where $N$ is even, by
\begin{align}
    \{X_{1},Z_{1},\dots,X_{N},Z_{N},Z_{1}\otimes \cdots \otimes Z_{N} \},
\label{eq:1-local_even}
\end{align}
and, in the case where $N$ is odd, by
\begin{align}
\begin{adjustbox}{width=\linewidth}$
    \{X_{1},Z_{1},\dots,X_{N},Z_{N},Z_{1}\otimes \cdots \otimes Z_s, Z_s \otimes \cdots \otimes Z_{N} \},
$\end{adjustbox}
\label{eq:1-local_odd}
\end{align}
where $1 < s < N$ is even.
The graph states to implement the sets from Eqs.~\eqref{eq:1-local_even} and \eqref{eq:1-local_odd} are presented in Fig.~\ref{fig:1-local_gs}. It quickly becomes apparent here that the solutions for LC-MBQC could have been straightforwardly obtained using basic building blocks of MBQC for local $X$ and $Z$ rotations and non-local $Z$ rotations. We further note that our solutions here are equivalent to the ones provided in Ref.~\cite{smith2025minimally}. This serves as a convenient confirmation of our results, but also highlights that more straightforward approaches than our annealing algorithm exist for certain problems.      

\subsubsection{Structured Set with Close-to-Optimal Generating Rate}

\begin{figure*}[htb!]
    \centering
    \includegraphics[width=\textwidth]{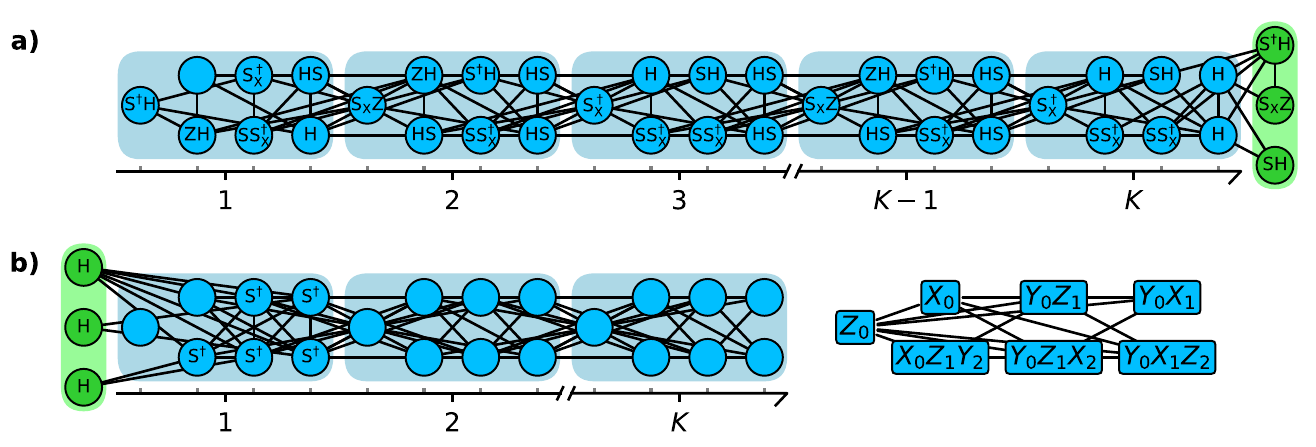}
    \caption{Resource state preparation for universal quantum computation generated by the CQCA with $N=3$ in terms of the (optimized) graph state representation and local Clifford operators for a) LC-MBQC, and b) AC-MBQC up to the forward CNOT ladder. 
    The legend (bottom right panel) displays the anticommutation graph of the Hamiltonian’s Pauli-strings and illustrates the correspondence between the auxiliary qubits in the graph states a) and b) and their respective implemented rotations. Note that the graph state a) has a period of $L$, whereas the local Clifford operators have a period of $2L$.}
    \label{fig:cqca_gs}
\end{figure*}

As discussed in Ref.~\cite{smith2025optimally}, there is a close connection between the optimal compilation rate and the number of pairs of strings within the generating set that anticommute. In particular, there is a ``sweetspot'' where the fraction of such anticommuting pairs compared to all pairs of strings is neither too high nor too low. While it remains an open question to obtain a family of Pauli-string generating sets, one for each value of $N$, that obtain this optimal rate, Example $3$ of Ref.~\cite{smith2025optimally} presents a family that gets close. It is to this family of generating sets that we turn next.

The close-to-optimal generating set of Ref.~\cite{smith2025optimally} are defined in terms of Clifford Quantum Cellular Automata (CQCAs). For our present purposes, a CQCA is a unitary operation $T$ that (i) consists entirely of nearest-neighbour Clifford gates, (ii) has finite circuit depth, (iii) is translation invariant and (iv) locality preserving. Informally, the latter two criteria can be understood as $T$ acting locally the same on all qubits and that the ``spread'' of information by $T$ is bounded (see, e.g., \cite{Stephen2019subsystem,schlingemann2008structure} for a more formal treatment). There is a strong connection between CQCAs and MBQC (see, e.g., \cite{HPN_24,Raussendorf_05,Raussendrof_19}) which has been leveraged in, for example, quantum machine learning \cite{Arun_24}.

The specific CQCA relevant to this subsection was first introduced by Ref.~\cite{Stephen_24}. For $N=3$ qubits, the generating set corresponding to this CQCA is given by
\begin{align}
    \{Z_0, X_0, Y_0Z_1, Y_0X_1, X_0Y_1Y_2, Y_0Z_1X_2, Y_0 X_1 Z_2\}.
\end{align}
We show the corresponding graph states in Fig.~\ref{fig:cqca_gs}. From the examples considered so far in this work, this is the only instance where the LC-MBQC approach spans edges from one auxiliary qubit to its next repetition. Therefore, ignoring the main qubits, which one can easily justify as any observable can be realized via sufficiently large repetitions $K$ and Pauli-$Z$ measurements due to universality, this is the first instance where both approaches require the same number of active qubits. It is also worth highlighting that unlike the previous examples, the LC-MBQC graph state requires local Clifford gates which repeat with a period of $2L$, despite the graph having period $L$. Interestingly, for larger $N \geq 6$, we have only been able to find LC-MBQC patterns where the number of active qubits is larger than $2N+2$. This is an instance where the AC-MBQC approach is favorable. 

\section{Discussion} \label{sec:Discussion}

In this work, we considered two schemes for preparing resource states for periodic measurement-based computations without Pauli measurements. At first we considered LC-MBQC, for which we derived an exact description of the resource state in terms of a graph state with local Cliffords, which, in essence, is the solution to the algorithm from \cite{Nest2004Graphical}. Based on this graph state, we provide a simulated annealing approach, which unlike Ref.~\cite{Kaldenbach2025Mapping}, not only minimizes the number of edges to reduce the entangling gate count, but also the number of active qubits while ensuring periodicity of the graph for temporal extrapolation of the pattern. 
We find that our optimizer works reliably across numerous examples, and, in instances like the $XY$ model or perturbed toric code Hamiltonian reduces the number of active qubits. 
However, there exist examples in which an efficient graph state with respect to the number of active qubits can not be obtained from our optimizer. It is worth highlighting that we can find such examples both in Hamiltonian simulation and universal quantum computation, thus likely ruling out universality of the generating set as a criterion for efficient graph state representations in terms of active qubits. Using hybrid simulation and embedding techniques, we provide some workarounds for situations in which efficient graphs do not straightforwardly arise. 

Also, given that for the CQCA we find that the local Cliffords follow a larger period than the graph state itself, it might be insightful to optimize for larger periods when the regular scheme fails. Such solutions would be inferior to AC-MBQC in terms of active qubits, but insightful from a theoretical point of view. 

Another degree of freedom which we have not considered in this work arises from the {\em local unitary equivalence} (LU-equivalence) of certain graph states. It is proven that LC- and LU-equivalence coincide for graph states up to 19 qubits \cite{Claudet2025Deciding}. The smallest known example where LC $\neq$ LU is 27 qubits \cite{Ji2008LULC}. Given that any practically relevant application of our techniques will likely entail more than 27 qubits, LU equivalence might prove to be a viable option to include in the search space.

Our second resource state preparation scheme entails a graph state with local Cliffords and a CNOT ladder on top. Given that the underlying graph state structurally mostly boils down to the anticommutation graph of the generating set, we name this scheme anticommutation (AC-)MBQC. Therefore, using this scheme, the circuit complexity is determined by the number of edges in the anticommutation matrix. We like to point out that the anticommutation fraction simultaneously characterizes the efficiency of a minimal universal generating set. This naturally leads to a future research question: which anticommutation fraction is truly optimal for MBQC? A smaller anticommutation fraction requires more repetitions (time steps) of the pattern to generate an arbitrary Pauli-string, but at the same time, each pattern needs fewer gates.
AC-MBQC typically leads to more active qubits than LC-MBQC, since the CNOT-ladder and anticommutation graphs span $\mathcal{O}(L)$ qubits. Meanwhile, in cases where LC-MBQC works, it achieves $\mathcal{O}(N)$ active qubits, which for most applications with $N \leq L$ is preferable. To combine the guarantees from AC-MBQC with the qubit reductions from LC-MBQC, one may need to consider more sophisticated decompositions of Clifford circuits \cite{de2025graph}.

Given that the removal of Pauli measurements from MBQC ideally boils down the computation to the bare minimum number of qubits required, our work contributes to  making MBQC more accessible on near-term hardware where the qubit number requirements of MBQC are still considered a limiting factor. Depending on the type of hardware, the graph state preparation is further hindered by the hardware topology, which is why in this work we have illustrated a protocol to perform AC-MBQC on a linear topology. More sophisticated resource state preparation specifically tailored towards limited topologies such as linear chains, heavy-hex, or square-grid, will in the end be necessary to efficiently deploy our approaches on real hardware. Given that there already exists convincing evidence for the advantage of measurement-based approaches, i.e. dynamic circuits outperforming their unitary counterparts on real hardware in terms of fidelity \cite{Baeumer2024Efficient}, we believe this is a research direction worth pursuing.  

\section*{Acknowledgments}
\addcontentsline{toc}{section}{Acknowledgments}
This project was made possible by the DLR Quantum Computing Initiative and the Federal Ministry for Economic Affairs and Climate Action; \url{https://qci.dlr.de/quanticom} (T.N.K.). 
This research was funded in part by the Austrian Science Fund (FWF) [SFB BeyondC F7102, DOI: \href{https://doi.org/10.55776/F71}{10.55776/F71}; WIT9503323, DOI: \href{https://doi.org/10.55776/WIT9503323}{10.55776/WIT9503323} (I.D.S, H.P.N., and H.J.B.), and in part by the research project Zentrum für Angewandtes Quantencomputing (ZAQC), which is funded by the Hessian Ministry for Digital Strategy and Innovation and the Hessian Ministry of Higher Education, Research and the Arts (M.H.). For open access purposes, the authors have applied a CC BY public copyright license to any author-accepted manuscript version arising from this submission. We gratefully acknowledge support from the European Union (ERC Advanced Grant, QuantAI, No. 101055129) (I.D.S, H.P.N., and H.J.B.). The views and opinions expressed in this article are however those of the authors only and do not necessarily reflect those of the European Union or the European Research Council. Neither the European Union nor the granting authority can be held responsible for them.
T.N.K. extends his gratitude to Robert Raußendorf for mediating the guest research stay that resulted in this research project.

\vfill\null
\newpage

\printbibliography

\clearpage
\appendix

\onecolumn

\section{Block-Row Reduction} \label{app:BlockRow}

\paragraph{The Auxiliary Block Row:} \mbox{}\\
To perform the block-row reduction required to obtain the graph state representation of the resource state up to local Cliffords, we introduce an auxiliary block row as an intermediate step. We explicitly keep track of the phases which arise due to multiplication of (stabilizer) strings to ensure that every row in the auxiliary block row is a valid stabilizer string itself. This is necessary to not only compute a valid adjacency matrix, but also the correct local Clifford operators. We represent our auxiliary strings using the tableau
\begin{align}
    &\hphantom{~}\textrm{X} 
    \left[\begin{array}{cc|cc|c}
    \textrm{I} & 
    0 & 
    \Gamma_0 & 
    \textrm{A}_0 & 
    0
    \end{array}\right] \nonumber
    \\
    &= [\begin{array}{cc|cc|c}
    \textrm{X} & 
    0 & 
    \textrm{X} \Gamma_0 & 
    \textrm{X} \textrm{A}_0 & 
    \textrm{R}_\textrm{aux.}
    \end{array}] \nonumber \\
    &= [\begin{array}{cc|cc|c}
    \textrm{X} & 
    0 & 
    \textrm{X} \Gamma_0 & 
    \textrm{X} \textrm{Z}^T \oplusr \textrm{X} \Gamma_0 \textrm{X}^T & 
    \textrm{R}_\textrm{aux.}
    \end{array}] \nonumber \\
    &= [\begin{array}{cc|cc|c}
    \textrm{X} & 
    0 & 
    \textrm{X} \Gamma_0 & 
    \textrm{X} \textrm{Z}^T \oplusr \Gamma_{\textrm{X}} & 
    \textrm{R}_\textrm{aux.}
    \end{array}].
    \label{eq:AuxBlockRow}
\end{align}
Here, we used the definition $\textrm{A}_0 = \textrm{Z}^T \oplusr \Gamma_0 \textrm{X}^T$, and then abbreviated $\Gamma_{\textrm{X}} \coloneqq \textrm{X}\Gamma_0 \textrm{X}^T$. Note that the entries of the matrices encoding the strings directly arise from binary matrix multiplication (cf.~Eq.~\eqref{eq:matmul_bin}) with $\textrm{X}$, while the phase exponents $\textrm{R}$ follow the more complicated rule from Eq.~\eqref{eq:StringExponentFunction}. More specifically, given that we compute products of stabilizer strings which always yields a real phase, we can use Eq.~\eqref{eq:ExponentFunctionCommuting}. 
The tricky part in the computation of the auxiliary block-row remains in determining the new phases $(-1)^{\textrm{R}_\textrm{aux.}}$. 

\paragraph{Phases of the Auxiliary Block Row:} \mbox{}\\
Note that each row in $[\textrm{X}| \textrm{X} \Gamma_0]$ is simply a product of (multiple) stabilizer strings of the initial graph state $\ket{G_0}$. If we consider such a product of graph state stabilizer strings for some arbitrary graph state $\ket{G}$ with $G=(E, V)$, we find
\begin{align}
    (-1)^r \mathcal{P}
    &= \prod_{v \in V} \Biggl(\bigotimes_{\substack{w<v \\(v, w)\in E}} Z_w \Biggr) X_v \Biggl(\bigotimes_{\substack{w>v \\(v, w)\in E}} Z_w \Biggr) \nonumber \\
    &= \bigotimes_{v\in V} \Biggl(\prod_{\substack{w<v \\(v, w)\in E}} Z_v\Biggr) X_v \Biggl(\prod_{\substack{w>v \\ (v, w)\in E}} Z_v\Biggr) \nonumber \\
    &= \prod_{v\in V}(-1)^{|N_G^\text{R}(v)|} \bigotimes_{v\in V} Z_v^{\text{deg}(v)} X_v \nonumber \\
    &= \underbrace{\prod_{v\in V}(-1)^{|N_G^\text{R}(v)|} \prod_{v \in V} i^{\text{deg}(v) \bmod{2}}}_{(-1)^r} \underbrace{\bigotimes_{v\in V} Y_v^{\text{deg}(v)} X_v^{\text{deg}(v)+1}}_{\mathcal{P}},
\end{align}
where $N_G^\text{R}(v)$ is the right-sided neighborhood of $v$ (one can define any order on the vertices $v$ here), and $\text{deg}(v)\coloneqq |N_G(v)|$ is the degree of $v$, i.e.,~the size of its entire neighborhood. 
Note that for the first step, we used $\{X_v, Z_v\} = 0$, and for the second one we employed $Z_v X_v = iY_v$. These manipulations permit us to explicitly separate the phase $(-1)^r$ from the Pauli-string $\mathcal{P}$. Next, we explicitly evaluate the phase
\begin{align}
    (-1)^r &= \prod_{v\in V}(-1)^{|N_G^\text{R}(v)|} \prod_{v \in V} i^{\text{deg}(v) \bmod{2}} \nonumber \\
    &= (-1)^{\sum_{v\in V} |N_G^\text{R}(v)|} i^{\sum_{v \in V} \text{deg}(v) \bmod{2}} \nonumber \\
    &= (-1)^{\frac{1}{2}\sum_{v\in V} \text{deg}(v)} (-1)^{\frac{1}{2}\sum_{v \in V} \text{deg}(v) \bmod{2}} \nonumber \\
    &= (-1)^{\sum_{v\in V} \frac{\text{deg}(v) - \text{deg}(v) \bmod{2}}{2}} \nonumber \\
    &= (-1)^{\sum_{v\in V} \left\lfloor \frac{\text{deg}(v)}{2} \right\rfloor},
\end{align}
where $\lfloor \cdot \rfloor$ denotes the floor operation.
Here, we used that the sum of all right-sided neighborhoods $N_G^\text{R}(v)$ amounts to the total number of edges $|E|$. The other sum $\sum_v \text{deg}(v) \bmod{2}$ counts the number of vertices with an odd degree. Given that the sum of all degrees amounts to an even number $2|E|$, we know that the number of vertices with odd degree has to be even. Therefore, we can safely factor out a $2$ while keeping an integer exponent, which permits us to get rid of the imaginary unit. In the last step, we employed the identity $x \bmod{y} = x - y \lfloor x/y \rfloor$ for integers $x, y$. Finally, we have 
\begin{align}
    r = \bigoplus_{v\in V} \left\lfloor \frac{\text{deg}(v)}{2}\right\rfloor 
    \label{eq:PhaseExpGraphStateStabProd}
\end{align}
A simple example can be inferred from a 3-qubit linear cluster state, whose stabilizers are given by $X_1 Z_2 I_3, Z_1 X_2 Z_3$, and $I_1 Z_2 X_3$. Clearly, the product of all strings is $-Y_1 X_2 Y_3$ and thus has a phase of $-1$. Using Eq.~\eqref{eq:PhaseExpGraphStateStabProd}, we find $r=\lfloor 1/2\rfloor + \lfloor 2/2\rfloor + \lfloor 1/2\rfloor = 1$, which yields the correct phase $(-1)^r=-1$. 

Based on this result, we can now derive the phase exponents for the auxiliary block row from Eq.~\eqref{eq:AuxBlockRow}. Each of the rows $m=1, \dots, M$ contains the product of the stabilizer strings of vertices $n$ with $\textrm{X}_{m, n} = 1$. Thus, we have $\textrm{R}_\textrm{aux.}=[r_1, r_2, \dots, r_M]^T$ with
\begin{align}
    r_m &= \bigoplus_{n=1}^N   \left\lfloor \frac{\sum_{n'=1}^N \textrm{X}_{mn} \textrm{X}_{mn'} (\Gamma_0)_{nn'}}{2}\right\rfloor \nonumber \\
    &= \bigoplus_{n=1}^N \textrm{X}_{mn} \left\lfloor \frac{\sum_{n'=1}^N   (\Gamma_0)_{nn'} \textrm{X}^T_{n' m}}{2}\right\rfloor.
    \label{eq:PhaseExpAuxRow}
\end{align}
We will come back to this equation later to compute the phases of the reduced block row.

\paragraph{The Block-Row Reduction:}\mbox{}\\
We now perform the block-row reduction by adding the two block rows, which cancels out the $X$ on the main qubits,
\begin{align}
    &\hphantom{+~}\left[\begin{array}{cc|cc|c}
    \textrm{X} & 
    0 & 
    \textrm{X} \Gamma_0 & 
    \textrm{X} \textrm{Z}^T \oplusr \Gamma_{\textrm{X}} &
    \textrm{R}_\textrm{aux.}
    \end{array}\right] 
    \nonumber \\
    &\oplusr
    \left[\begin{array}{cc|cc|c}
    \textrm{X} & 
    \textrm{I} & 
    \textrm{Z} & 
    \text{UT}(\textrm{A}) & 
    \textrm{R}
    \end{array}\right] \nonumber \\
    &= 
    \left[\begin{array}{cc|cc|c}
    0 & 
    \textrm{I} & 
    \textrm{Z} \oplusr \textrm{X} \Gamma_0 & 
    \text{UT}(\textrm{A}) \oplusr \textrm{X}\textrm{A}_0 &
    \textrm{R} \oplusr \textrm{R}_\textrm{aux.} \oplusr \textrm{R}'
    \end{array}\right] \nonumber \\
    &=
    \left[\begin{array}{cc|cc|c}
    0 & 
    \textrm{I} & 
    \textrm{A}_0^T & 
    \text{UT}(\textrm{Z}\textrm{X}^T) \oplusr \text{LT}(\textrm{X}\textrm{Z}^T) \oplusr \text{D}(\textrm{X}\textrm{Z}^T) \oplusr \Gamma_{\textrm{X}} &
    \textrm{R} \oplusr \textrm{R}_\textrm{aux.} + \textrm{R}'
    \end{array}\right].
    \label{eq:BlockRowReduction}
\end{align}
\paragraph{Phases of the Reduced Block Row:} \mbox{}\\
The additional phase $\textrm{R}'$ is obtained by applying Eq.~\eqref{eq:StringExponentFunction} for each row-wise addition performed in Eq.~\eqref{eq:BlockRowReduction}. As an intermediate step, we define the $M\times (N+M)$ exponent matrix $\mathcal{E}$, where each row stores the exponents to which $i$ is raised (either $0, 1$, or $-1$), which we compute according to Eq.~\eqref{eq:ExponentFunction}:
\begin{align}
    \mathcal{E}_{ij} \coloneqq g(x_1=[\textrm{X}~~~\textrm{I}]_{ij}, z_1=[\textrm{Z}~~~\text{UT}(\textrm{A})]_{ij}, x_2 = [\textrm{X}~~~0]_{ij}, z_2 = [\textrm{X}\Gamma_0~~~\textrm{X}\textrm{Z}^T + \Gamma_{\textrm{X}}]_{ij}).
\end{align}
We now evaluate this expression explicitly, which can be compactly achieved by using the element-wise matrix product $(A\cdot B)_{ij} \coloneqq A_{ij} B_{ij}$, and the logical NOT operation $(\bar A)_{ij} \coloneqq A_{ij} \oplus 1$. We can use these two operations to construct the logical binary operations which account for the different cases in Eq.~\eqref{eq:ExponentFunction}: 
\begin{align}
    \mathcal{E} &=  [(\textrm{X} \cdot \textrm{Z})\cdot (\textrm{X}\Gamma_0-\textrm{X})
    + (\textrm{X} \cdot \bar{\textrm{Z}})\cdot (\textrm{X}\Gamma_0) \cdot (2\textrm{X}-1^{M\times N}) + (\textrm{Z}\cdot \bar{\textrm{X}})\cdot \textrm{X}\cdot (1^{M\times N}-2\textrm{X}\Gamma_0), \nonumber\\
    &\hphantom{=~} \textrm{I}\cdot (\textrm{X}\textrm{Z}^T + \Gamma_{\textrm{X}}) \cdot (0^{M\times M} - 1^{M\times M}) + \text{UT}(\textrm{A}) \cdot 0^{M\times M} \cdot (1^{M\times M}-2(\textrm{X}\textrm{Z}^T + \Gamma_{\textrm{X}}))] \nonumber 
    \\
    &= [-(\textrm{X} \cdot \textrm{Z}) + (\textrm{X} \cdot \textrm{Z})\cdot (\textrm{X}\Gamma_0) + (\textrm{X} \cdot \bar{\textrm{Z}})\cdot (\textrm{X}\Gamma_0), -\textrm{D}(\textrm{X}\textrm{Z}^T)] \nonumber \\
    &= [-(\textrm{X} \cdot \textrm{Z}) + (\textrm{X} \cdot \textrm{Z} + \textrm{X} \cdot \bar{\textrm{Z}})\cdot (\textrm{X}\Gamma_0), -\textrm{D}(\textrm{X}\textrm{Z}^T)] \nonumber \\
    &= [-(\textrm{X} \cdot \textrm{Z}) + \textrm{X} \cdot (\textrm{X}\Gamma_0), -\textrm{D}(\textrm{X}\textrm{Z}^T)].
    \label{eq:ExponentMatrix}
\end{align}
Following Eq.~\eqref{eq:StringExponentFunction}, we obtain the new additional phases $\textrm{R}'$ by computing the row sum $\text{RS}(\mathcal{E}) \bmod{2} $. We note that Eq.~\eqref{eq:ExponentMatrix} consists of two separate structure, where one only depends on the generators encoded by $\textrm{X}, \textrm{Z}$, and the other instead takes into account $\textrm{X}$ and the adjacency matrix $\Gamma_0$ of the initial graph $G_0$. Using Eq.~\eqref{eq:ExponentFunctionCommuting} and the linearity of the row sum operation, we can write the additional phase exponent $\textrm{R}'$ as:
\begin{align}
    \textrm{R}' = \text{RS}(\mathcal{E}) \bmod{2}= \left(- \frac{\text{RS}(\textrm{X}\cdot \textrm{Z}) + \text{RS}(\text{D}(\textrm{X}\textrm{Z}^T))}{2} + \frac{\text{RS}(\textrm{X}\cdot (\textrm{X}\Gamma_0))}{2}\right) \bmod{2}.
\end{align}
The term $\text{RS}(\textrm{X}\cdot \textrm{Z})=\textrm{N}_Y$ simply counts the numbers of Pauli-$Y$s encoded by each the new generators $\mathcal{P}'$. In addition, we have $\text{RS}(\text{D}(\textrm{X}\textrm{Z}^T))$, which turns out to be 0 if the number of $Y$s is even, and $1$ if the number is odd. Consequently, $\text{RS}(\textrm{X} \cdot  \textrm{Z}) + \text{RS}(\text{D}(\textrm{X}\textrm{Z}^T))$ will always be an even number. Thus, we may simplify $\textrm{R}'$ as follows:
\begin{equation}
    \textrm{R}' = \left\lceil \frac{\textrm{N}_Y}{2} \right\rceil \oplus \frac{\text{RS}(\textrm{X}\cdot (\textrm{X}\Gamma_0))}{2},
\end{equation}
where $\lceil \cdot \rceil$ denotes the ceiling operation, and $\textrm{N}_Y$ is the $m$-component vector, with $(\textrm{N}_{Y})_{m}$ counting the number of Pauli-$Y$s in the $m$-th Pauli generator $\mathcal{P}^{(m)}$. 
The second term $\text{RS}(\textrm{X}\cdot (\textrm{X}\Gamma_0))$ consequently yields even numbers as well. We rewrite it as follows:
\begin{align}
    \text{RS}(\textrm{X} \cdot (\textrm{X}\Gamma_0))_m &= \sum_{n=1}^N [\textrm{X}\cdot (\textrm{X}\Gamma_0)]_{mn} = \sum_{n=1}^N \textrm{X}_{mn} (\textrm{X}\Gamma_0)_{mn} \nonumber \\
    &= \sum_{n=1}^N \textrm{X}_{mn} \left(\bigoplus_{n'=1}^N \textrm{X}_{mn'} (\Gamma_0)_{nn'}\right) \nonumber \\
    &= \sum_{n=1}^N \textrm{X}_{mn} \left(\bigoplus_{n'=1}^N (\Gamma_0)_{nn'} \textrm{X}^T_{n'm} \right).
\end{align}
As a final step, we need to incorporate the phases $\textrm{R}_{\textrm{aux.}}$ (Eq.~\eqref{eq:PhaseExpAuxRow}) from the auxiliary block-row into $\textrm{R}\oplusr \textrm{R}'$. The for sake of simplicity, we only consider the contributions from the initial graph $\Gamma_0$ here:
\begin{align}
    (\textrm{R}_\textrm{aux.})_m \oplusr \frac{\text{RS}(\textrm{X} \cdot (\textrm{X}\Gamma_0))_m}{2} &=
    \left\{\bigoplus_{n=1}^N \textrm{X}_{mn} \left\lfloor \frac{\sum_{n'=1}^N   (\Gamma_0)_{nn'} \textrm{X}^T_{n' m}}{2}\right\rfloor\right\} \oplusr \left\{\sum_{n=1}^N \textrm{X}_{mn} \frac{\bigoplus_{n'=1}^N (\Gamma_0)_{nn'} \textrm{X}^T_{n'm}}{2} \right\}
    \nonumber \\
    &=\left\{\sum_{n=1}^N \textrm{X}_{mn} \left[\left\lfloor \frac{\sum_{n'=1}^N   (\Gamma_0)_{nn'} \textrm{X}^T_{n'm}}{2}\right\rfloor + \frac{\left(\sum_{n'=1}^N (\Gamma_0)_{nn'} \textrm{X}^T_{n'm} \right)\bmod{2} }{2} \right]\right\} \bmod{2} \nonumber \\
    &= \left\{\sum_{n=1}^N \textrm{X}_{mn} \frac{\sum_{n'=1}^N (\Gamma_0)_{nn'} \textrm{X}^T_{n'm} }{2} \right\} \bmod{2} \nonumber \\
    &= \frac{(\textrm{X} \circ \Gamma_0 \circ \textrm{X}^T)_{mm}}{2} \bmod{2}\nonumber \\
    &\eqqcolon \frac{(\Gamma_{\textrm{X}}^\circ)_m}{2} \bmod{2},
\end{align}
where we abbreviated the final result using $(\Gamma_{\textrm{X}}^\circ)_{m} \coloneqq (\textrm{X} \circ \Gamma_0 \circ \textrm{X}^T)_{mm}$. Note that the difference between $\Gamma_{\textrm{X}}$ and $\Gamma_{\textrm{X}}^\circ$ lies in using regular matrix multiplication instead of using the symplectic inner product (and only taking the diagonal entries of course). We can finally write down the phases of the graph states stabilizers as
\begin{align}
    \textrm{R} \oplusr \textrm{R}_\textrm{aux.} \oplusr \textrm{R}' = \textrm{R} \oplusr \left\lceil \frac{\textrm{N}_Y}{2} \right\rceil \oplusr \frac{\Gamma_{\textrm{X}}^\circ}{2}.
    \tag{Eq.~\eqref{eq:GraphStateResourcePhases} revisited}
\end{align}

\section{Local Clifford Operators} \label{app:VOPs}

Following the block-row reduction from Appendix \ref{app:BlockRow}, we have the tableau $[{\bf I}|{\bf Z}|{\bf R} ]$, with 
\begin{equation}
    {\bf Z} =
    \begin{bmatrix}
    \Gamma_0 & \textrm{A}_0 \\
    \textrm{A}_0^T & \Gamma_{\textrm{X}} \oplusr \text{D}(\textrm{X}\textrm{Z}^T) \oplusr \text{UT}(\textrm{Z}\textrm{X}^T) \oplusr \text{LT}(\textrm{X}\textrm{Z}^T)
    \end{bmatrix},
    \tag{Eq.~\eqref{eq:GraphStateResourceLoops} revisited}
\end{equation}
Note that ${\bf Z}$ is now symmetric, with the only diagonal entries given by $\text{D}(\textrm{X}\textrm{Z}^T)$. Further note that $(\text{D}(\textrm{X}\textrm{Z}^T))_{mm}=1$ if and only if the generator $\mathcal{P}^{(m)}$ contains an odd number of Pauli $Y$'s. 
To obtain the proper adjacency matrix of the graph (without self-loops), we have to shift these entries into the local Clifford layer. 
This is achieved by applying an $S$ gate to every auxiliary qubit $m$ whose generator $\mathcal{P}'^{(m)}$ contains an odd number of $Y$'s. 
Besides the removal of self-loops, we also need to remove the phases $\textrm{R}$ from Eq.~\eqref{eq:GraphStateResourcePhases}.
This is achieved by applying Pauli-$Z$ to every auxiliary qubits $m$ with $\textrm{R}_m=1$.  
Conveniently, we can simplify the VOPs arising due to $\textrm{N}_Y$ as follows:
\begin{align}
    S_m^{(\textrm{N}_Y)_m \bmod{2}} Z_m^{\left\lceil \frac{(\textrm{N}_Y)_m}{2} \right\rceil} = S_m^{-(\textrm{N}_Y)_m}
\end{align}
For the VOPs arising due to $\Gamma_{\textrm{X}}^\circ$, we can use that all entries of $\Gamma_{\textrm{X}}^\circ$ are even, and thus write
\begin{align}
    Z_m^{\frac{(\Gamma_{\textrm{X}}^\circ)_m}{2}} = S_m^{(\Gamma_{\textrm{X}}^\circ)_m}.
\end{align}
Finally, we can state that the local Clifford operators compensating for the phases and self-loops are given by 
\begin{align}
    \mathcal{C}_\textrm{aux.} = \prod_{m=1}^M S_m^{(2\textrm{R} + \Gamma_{\textrm{X}}^\circ - \textrm{N}_Y)_m}.
    \tag{Eq.~\eqref{eq:GraphStateResourceVOPs} revisited}
\end{align}

\section{The Distance Matrix} \label{app:DistanceMatrix}

In our work, we compute the distance matrices $\textrm{D}$ as follows:

\begin{enumerate}
    \item Define a target memory size $N_\textrm{targ.}$. This number describes the desired number of qubits to store the quantum state during any intermediate stage of the measurement pattern (apart from the final output). A good choice is $N_\textrm{targ.} = N$, but in case of possible qubit reductions, a smaller target can be chosen.
    \item Partition the generators $\mathcal{A} = \{\mathcal{P}^{(m)} | 1 \leq m \leq M\}$ into $J \leq M$ groups of mutually commuting strings $\mathcal{A}_j \subset \mathcal{A}$ such that these are carried out in order $\mathcal{A}_1, \dots, \mathcal{A}_J$.
    \item Initialize the $(J+1) \times (J+1)$ auxiliary distance matrix $D=\bm 0$. For every $j = 1, \dots, J-1$, initialize some qubit counter $N_j = 0$. Then iterate through $l = j+1, \dots, J$, and compute $N_j \leftarrow N_j + |\mathcal{A}_{l}|$, until $N_j \geq N_\textrm{targ.}$. Then set $\textrm{D}_{j,l+1} = 1$. 
    \item $\textrm{D}$ now is a directed distance matrix which stores all paths of length 1. From this, we can compute the longer paths using $\textrm{D} \leftarrow \sum_{j=1}^{J-1} j \textrm{D}^j$, since each power $\textrm{D}^j$ gives us the paths of length $j$. 
    \item We can recover the symmetric distance matrix via $\textrm{D} \leftarrow \textrm{D} + \textrm{D}^T$. Doing this as a last step is crucial, as otherwise the matrix powers would count loops.
    \item We replace $\textrm{D}_{ij} \leftarrow 2^{\textrm{D}_{ij}}$. This way, all edges have a weight of at least 1, and large distances are exponentially weighted. 
    \item Finally, we expand $\textrm{D}$ into the larger $(N+M)\times (N+M)$ distance matrix, which for any pair of qubits simply stores the distance of the groups they belong to. This corresponds to $\textrm{D}_{ij} \leftarrow 1^{|\mathcal{A}_i| \times |\mathcal{A}_j|} \cdot \textrm{D}_{ij}$, where $|\mathcal{A}_{J+1}| \coloneqq N$.  
\end{enumerate}
Note that with $M=KL$, this algorithm readily generalizes to the periodic case. 
\paragraph{Example: The $XY$ Model} \mbox{} \\
We consider the $XY$ model Hamiltonian on a linear chain with $N=3$ sites and $K=3$ time steps.
\begin{enumerate}
    \item We set our desired number of qubits to $N_\textrm{targ.}=N-1$. This reduction below $N$ is motivated by a possible qubit reduction using qubit tapering.
    \item We can trivially partition the Pauli-strings of the Hamiltonian into two mutually commuting group $\mathcal{A}_X = \{X_1 X_2, \dots, X_{N-1}X_N\}$, and $\mathcal{A}_Y = \{Y_1 Y_2, \dots, Y_{N-1} Y_N\}$. We further consider $K=3$ time steps, which amounts to the sequence of groups $\mathcal{A}_X, \mathcal{A}_Y, \mathcal{A}_X, \mathcal{A}_Y, \mathcal{A}_X, \mathcal{A}_Y$, which has length $J=6$.
    \item We initialize the $7\times 7$ matrix $\textrm{D}=\bm 0$. For every $j=1,\dots, 5$, we find that $l=j+1$ already satisfies the stopping criterion $N_j = |\mathcal{A}_{X/Y}|=N-1$, and thus $\textrm{D}_{j, j+2}=1$. At this point, we have
    \begin{align}
        \textrm{D} = \begin{bmatrix}
        0 & 0 & 1 & 0 & 0 & 0 & 0 \\
        0 & 0 & 0 & 1 & 0 & 0 & 0 \\
        0 & 0 & 0 & 0 & 1 & 0 & 0 \\
        0 & 0 & 0 & 0 & 0 & 1 & 0 \\
        0 & 0 & 0 & 0 & 0 & 0 & 1 \\
        0 & 0 & 0 & 0 & 0 & 0 & 0 \\
        0 & 0 & 0 & 0 & 0 & 0 & 0 \\
        \end{bmatrix}.
    \end{align}
    \item We now fill the upper triangular with the longer paths:
    \begin{align}
        \textrm{D} = \begin{bmatrix}
        0 & 0 & 1 & 2 & 3 & 4 & 5 \\
        0 & 0 & 0 & 1 & 2 & 3 & 4 \\
        0 & 0 & 0 & 0 & 1 & 2 & 3 \\
        0 & 0 & 0 & 0 & 0 & 1 & 2 \\
        0 & 0 & 0 & 0 & 0 & 0 & 1 \\
        0 & 0 & 0 & 0 & 0 & 0 & 0 \\
        0 & 0 & 0 & 0 & 0 & 0 & 0 \\
        \end{bmatrix}.
    \end{align}
    \item Next, we perform the symmetrization:
        \begin{align}
        \textrm{D} = \begin{bmatrix}
        0 & 0 & 1 & 2 & 3 & 4 & 5 \\
        0 & 0 & 0 & 1 & 2 & 3 & 4 \\
        1 & 0 & 0 & 0 & 1 & 2 & 3 \\
        2 & 1 & 0 & 0 & 0 & 1 & 2 \\
        3 & 2 & 1 & 0 & 0 & 0 & 1 \\
        4 & 3 & 2 & 1 & 0 & 0 & 0 \\
        5 & 4 & 3 & 2 & 1 & 0 & 0 \\
        \end{bmatrix}.
    \end{align}
    \item Now, we perform the exponentiation:
    \begin{align}
        \textrm{D} = \begin{bmatrix}
        2^0 & 2^0 & 2^1 & 2^2 & 2^3 & 2^4 & 2^5 \\
        2^0 & 2^0 & 2^0 & 2^1 & 2^2 & 2^3 & 2^4 \\
        2^1 & 2^0 & 2^0 & 2^0 & 2^1 & 2^2 & 2^3 \\
        2^2 & 2^1 & 2^0 & 2^0 & 2^0 & 2^1 & 2^2 \\
        2^3 & 2^2 & 2^1 & 2^0 & 2^0 & 2^0 & 2^1 \\
        2^4 & 2^3 & 2^2 & 2^1 & 2^0 & 2^0 & 2^0 \\
        2^5 & 2^4 & 2^3 & 2^2 & 2^1 & 2^0 & 2^0 \\
        \end{bmatrix} =  \begin{bmatrix}
        1 & 1 & 2 & 4 & 8 & 16 & 32 \\
        1 & 1 & 1 & 2 & 4 & 8 & 16 \\
        2 & 1 & 1 & 1 & 2 & 4 & 8 \\
        4 & 2 & 1 & 1 & 1 & 2 & 4 \\
        8 & 4 & 2 & 1 & 1 & 1 & 2 \\
        16 & 8 & 4 & 2 & 1 & 1 & 1 \\
        32 & 16 & 8 & 4 & 2 & 1 & 1 \\
        \end{bmatrix}
    \end{align}
    \item Last, we expand the matrix to
    \begin{align}
    \textrm{D} = 
        \begin{bmatrix}
        1^{2\times 2} \cdot 1 & 1^{2\times 2} \cdot 1 & 1^{2\times 2} \cdot 2 & 1^{2\times 2} \cdot 4 & 1^{2\times 2} \cdot 8 & 1^{2\times 2} \cdot 16 & 1^{2\times 3} \cdot 32 \\
        1^{2\times 2} \cdot 1 & 1^{2\times 2} \cdot 1 & 1^{2\times 2} \cdot 1 & 1^{2\times 2} \cdot 2 & 1^{2\times 2} \cdot 4 & 1^{2\times 2} \cdot 8 & 1^{2\times 3} \cdot 16 \\
        1^{2\times 2} \cdot 2 & 1^{2\times 2} \cdot 1 & 1^{2\times 2} \cdot 1 & 1^{2\times 2} \cdot 1 & 1^{2\times 2} \cdot 2 & 1^{2\times 2} \cdot 4 & 1^{2\times 3} \cdot 8 \\
        1^{2\times 2} \cdot 4 & 1^{2\times 2} \cdot 2 & 1^{2\times 2} \cdot 1 & 1^{2\times 2} \cdot 1 & 1^{2\times 2} \cdot 1 & 1^{2\times 2} \cdot 2 & 1^{2\times 3} \cdot 4 \\
        1^{2\times 2} \cdot 8 & 1^{2\times 2} \cdot 4 & 1^{2\times 2} \cdot 2 & 1^{2\times 2} \cdot 1 & 1^{2\times 2} \cdot 1 & 1^{2\times 2} \cdot 1 & 1^{2\times 3} \cdot 2 \\
        1^{2\times 2} \cdot 16 & 1^{2\times 2} \cdot 8 & 1^{2\times 2} \cdot 4 & 1^{2\times 2} \cdot 2 & 1^{2\times 2} \cdot 1 & 1^{2\times 2} \cdot 1 & 1^{2\times 3} \cdot 1 \\
        1^{3\times 2} \cdot 32 & 1^{3\times 2} \cdot 16 & 1^{3\times 2} \cdot 8 & 1^{3\times 2} \cdot 4 & 1^{3\times 2} \cdot 2 & 1^{3\times 2} \cdot 1 & 1^{3\times 3} \cdot 1 
        \end{bmatrix},
    \end{align}
where $1^{n\times m}$ denotes the matrix filled with ones everywhere of size $n\times m$.
\end{enumerate}

\section{CNOT Ladders} \label{app:Ladders}

In this Appendix, we derive the graph states after absorbing the backward or forward CNOT ladders. Given that the absorption of the backward CNOT ladder is the protocol we actually employ in our work, we start with an extensive derivation of this formalism in Sec.~\ref{subapp:Backward}. Thereafter, we also provide analogous  results for the absorption of the forward CNOT ladder in Sec.~\ref{subapp:Forward}. 

\subsection{First Backwards - Then Forwards} \label{subapp:Backward}

We start from the stabilizer tableau of the resource state after expressing the initial state by a graph state $\ket{G_0}$, i.e.~Eq.~\eqref{eq:StabilizerTableauGraph_STEP0}.
By generalizing Eq.~\eqref{eq:StabilizerTableauGraph_STEP0} to a periodic sequence of $M=KL$ generators, we have:
\begin{align}
    \begin{array}{c|c!{\vrule width 2pt}c|c|c|c|}
    \cline{2-6}
    \multirow{6}{*}{${\bf X}=$} & \textrm{I} & 0 & 0 & \cdots & 0 \\ \Xcline{2-6}{2pt} \rule{0pt}{10pt}
    & \textrm{X} & \textrm{I} & 0 & \cdots & 0 \\ \cline{2-6}
    & \textrm{X} & 0 & \textrm{I} & \cdots & 0 \\ \cline{2-6}
    & \vdots & \vdots & \vdots & \ddots & \vdots \\ \cline{2-6} 
    & \textrm{X} & 0 & 0 & \cdots & \textrm{I} \\ \cline{2-6} 
    \multicolumn{6}{c}{} \\ \cline{2-6} 
    \multirow{6}{*}{${\bf Z}=$} & \Gamma_0 & \textrm{A}_0 & \textrm{A}_0 & \cdots & \textrm{A}_0 \\ \Xcline{2-6}{2pt} \rule{0pt}{10pt}
    & \textrm{Z} & \text{UT}(\textrm{A}) & \textrm{A} & \cdots & \textrm{A} \\ \cline{2-6}
    & \textrm{Z} & 0 & \text{UT}(\textrm{A}) & \cdots & \textrm{A} \\ \cline{2-6}
    & \vdots & \vdots & \vdots & \ddots & \vdots \\ \cline{2-6}
    &  \textrm{Z} & 0 & 0 & \cdots & \text{UT}(\textrm{A}) \\ \cline{2-6}
    \end{array}
    \label{eq:StabilizersResource}
\end{align}
We now show that almost all of the stabilizers $[{\bf X}| {\bf Z}| \bm 0]$ can be localized to act on at most $L+1$ qubits.  
For that purpose, we perform the row sum for each auxiliary block row $k=1,\dots, K-1$ with its subsequent block-row $k+1$, that is, the second row of the block form of the matrix from Eq.~\eqref{eq:StabilizersResource} is updated to be the block-wise sum of the second and third rows, the third row becomes the sum of the third and fourth rows, and so on until the final row. These additions will result in cancellations of identical entries between $k$ and $k+1$, and thereby localize the stabilizer strings:
\begin{align}
    \begin{array}{c|c!{\vrule width 2pt}c|c|c|c|c|}
    \cline{2-7}
    \multirow{6}{*}{${\bf X}=$} & \textrm{I} & 0 & 0 & 0 & \cdots & 0 \\ \Xcline{2-7}{2pt} \rule{0pt}{10pt}
    & 0 & \textrm{I} & \textrm{I} & 0 & \cdots & 0 \\ \cline{2-7}
    & 0 & 0 & \textrm{I} & \textrm{I} & \cdots & 0 \\ \cline{2-7}
    & \vdots & \vdots & \vdots & \vdots & \ddots & \vdots \\ \cline{2-7} 
    & \textrm{X} & 0 & 0 & 0 & \cdots & \textrm{I} \\ \cline{2-7} \multicolumn{6}{c}{} \\ \cline{2-7} 
    \multirow{6}{*}{${\bf Z}=$} & \Gamma_0 & \textrm{A}_0 & \textrm{A}_0 & \textrm{A}_0 & \cdots & \textrm{A}_0 \\ \Xcline{2-7}{2pt} \rule{0pt}{10pt}
    & 0 & \text{UT}(\textrm{A}) & \text{LT}(\textrm{A}) & 0 & \cdots & 0 \\ \cline{2-7}
    & 0 & 0 & \text{UT}(\textrm{A}) & \text{LT}(\textrm{A}) & \cdots & 0 \\ \cline{2-7}
    & \vdots & \vdots & \vdots & \vdots & \ddots & \vdots \\ \cline{2-7}
    &  \textrm{Z} & 0 & 0 & 0 & \cdots & \text{UT}(\textrm{A}) \\
    \cline{2-7}
    \end{array}
    \label{eq:LocalStabilizersResource}
\end{align}
This construction gives rise to $K-1$ block-rows with at most $(L+1)$-local strings, then one block row with at most $(N+L)$-local strings, and only $N$ up-to-global (at most $(N+M)$-local) stabilizer strings. Note that this manipulation of the tableau does not change the phases ${\bf R}= 0$.
We now apply the reverse CNOT ladder to the auxiliary qubits to localize the remaining stabilizer strings from Eq.~\eqref{eq:LocalStabilizersResource}: 
\begin{align}
    \begin{array}{c|c!{\vrule width 2pt}c|c|c|c|c|c|}
    \cline{2-8}
    \multirow{8}{*}{${\bf X}=$} & \textrm{I} & 0 & 0 & 0 & \cdots & 0 & 0 \\ \Xcline{2-8}{2pt} \rule{0pt}{10pt}
    & 0 & 0 & \textrm{I} & 0 & \cdots & 0 & 0\\ \cline{2-8}
    & 0 & 0 & 0 & \textrm{I} & \cdots & 0 & 0\\ \cline{2-8}
    & \vdots & \vdots & \vdots & \vdots & \ddots & \vdots & \vdots \\ \cline{2-8} 
    & 0 & 0 & 0 & 0 & \cdots & \textrm{I} & 0 \\ \cline{2-8}
    & 0 & 0 & 0 & 0 & \cdots & 0 & \textrm{I} \\ \cline{2-8}
    & \textrm{X} & \textrm{I} & \textrm{I} & \textrm{I} & \cdots & \textrm{I} & \textrm{I} \\ \cline{2-8}\multicolumn{6}{c}{} \\ \cline{2-8} 
    \multirow{8}{*}{${\bf Z}=$} & \Gamma_0 & \textrm{A}_0 & 0 & 0 & \cdots & 0 & 0\\ \Xcline{2-8}{2pt} \rule{0pt}{10pt}
    & 0 & \text{UT}(\textrm{A}) & \textrm{A} & \text{LT}(\textrm{A}) & \cdots & 0 & 0\\ \cline{2-8}
    & 0 & 0 & \text{UT}(\textrm{A}) & \textrm{A} & \cdots & 0 & 0\\ \cline{2-8}
    & \vdots & \vdots & \vdots & \vdots & \ddots & \vdots & \vdots \\ \cline{2-8}
    & 0 & 0 & 0 & 0 & \cdots & \textrm{A} & \text{LT}(\textrm{A})\\ \cline{2-8}
    & 0 & 0 & 0 & 0 & \cdots & \text{UT}(\textrm{A}) & \textrm{A}\\ \cline{2-8}
    & \textrm{Z} & 0 & 0 & 0 & \cdots & 0 & \text{UT}(\textrm{A}) \\
    \cline{2-8}
    \end{array}
    \label{eq:StabilizersBackwardLadder}
\end{align}
Note that this step does not change the phases. 
Replacing the $K$-th auxiliary block row by the sum of all $K$ auxiliary block rows, and then swapping the $1$-st and the $K$-th block row, yields: 
\begin{align}
    \begin{array}{c|c!{\vrule width 2pt}c|c|c|c|c|c|}
    \cline{2-8}
    \multirow{8}{*}{${\bf X}=$} & \textrm{I} & 0 & 0 & 0 & \cdots & 0 & 0 \\ \Xcline{2-8}{2pt} \rule{0pt}{10pt}
    & \textrm{X} & \textrm{I} & 0 & 0 & \cdots & 0 & 0 \\ \cline{2-8}
    & 0 & 0 & \textrm{I} & 0 & \cdots & 0 & 0\\ \cline{2-8}
    & 0 & 0 & 0 & \textrm{I} & \cdots & 0 & 0\\ \cline{2-8}
    & \vdots & \vdots & \vdots & \vdots & \ddots & \vdots & \vdots \\ \cline{2-8} 
    & 0 & 0 & 0 & 0 & \cdots & \textrm{I} & 0 \\ \cline{2-8}
    & 0 & 0 & 0 & 0 & \cdots & 0 & \textrm{I} \\ \cline{2-8} \multicolumn{6}{c}{} \\ \cline{2-8} 
    \multirow{8}{*}{${\bf Z}=$} & \Gamma_0 & \textrm{A}_0 & 0 & 0 & \cdots & 0 & 0\\ \Xcline{2-8}{2pt} \rule{0pt}{10pt}
    & \textrm{Z} & \text{UT}(\textrm{A}) & \text{LT}(\textrm{A}) & 0 & \cdots & 0 & 0 \\ \cline{2-8}
    & 0 & \text{UT}(\textrm{A}) & \textrm{A} & \text{LT}(\textrm{A}) & \cdots & 0 & 0\\ \cline{2-8}
    & 0 & 0 & \text{UT}(\textrm{A}) & \textrm{A} & \cdots & 0 & 0\\ \cline{2-8}
    & \vdots & \vdots & \vdots & \vdots & \ddots & \vdots & \vdots \\ \cline{2-8}
    & 0 & 0 & 0 & 0 & \cdots & \textrm{A} & \text{LT}(\textrm{A})\\ \cline{2-8}
    & 0 & 0 & 0 & 0 & \cdots & \text{UT}(\textrm{A}) & \textrm{A}\\ \cline{2-8}
    \end{array}
    \label{eq:LocalStabilizersBackwardLadder}
\end{align}
What we find in the top-left corner of both matrices ${\bf X}$ and ${\bf Z}$ is exactly the same tableau as in Eq.~\eqref{eq:StabilizerTableauGraph_STEP0} for $K=1$. Conveniently, we have already derived the result of the block-row reduction which turns this tableau into a graph state in Appendices \ref{app:BlockRow} and \ref{app:VOPs}. Thus, the resulting graph state is given by 
\begin{align}
    \Gamma = 
    \begin{array}{|c!{\vrule width 2pt}c|c|c|c|c|}
        \hline 
        \Gamma_0 & \textrm{A}_0 & 0 & 0 & \cdots & 0 \\ \Xcline{1-6}{2pt} \rule{0pt}{10pt}
        & \Gamma_{\textrm{X}} & & & & \\ 
        \textrm{A}_0^T & \oplusr \text{UT}(\textrm{Z}\textrm{X}^T) & \text{LT}(\textrm{A}) & 0 & \cdots & 0\\ 
        & \oplusr\text{LT}(\textrm{X}\textrm{Z}^T) & & & & \\ \hline 
        0 & \text{UT}(\textrm{A}) & \textrm{A} & \text{LT}(\textrm{A}) & \cdots & 0  \\ \hline 
        0 & 0 & \text{UT}(\textrm{A}) & \textrm{A} & \cdots & 0  \\ \hline 
        \vdots & \vdots & \vdots & \vdots & \ddots & \vdots \\ \hline 
        0 & 0 & 0 & 0 & \cdots & \textrm{A}  \\
        \hline
    \end{array}.
    \tag{Eq.~\eqref{eq:GraphState_BW} revisited}
\end{align}
The local Clifford operators compensating for the phases and self-loops are given by 
\begin{align}
    \mathcal{C}_\textrm{aux.} = \prod_{l=1}^L S_{1,l}^{(2\textrm{R}+\Gamma_{\textrm{X}}^\circ - \textrm{N}_Y)_l},
    \tag{Eq.~\eqref{eq:GraphStateLadderVOPs} revisited}
\end{align}
where all operators act solely on the first time step $k=1$. 

\subsection{First Forwards - Then Backwards} \label{subapp:Forward}
We now instead apply a forward CNOT ladder to the localized stabilizers encoded by the tableau in Eq.~\eqref{eq:LocalStabilizersResource} in the previous section. 
Note that we interchange the roles of the controls and targets, meaning that we effectively compute
\begin{align}
    (I^{\otimes N}\otimes H^{\otimes M}) \Lambda(X)_{\textrm{FW}} (I^{\otimes N}\otimes H^{\otimes M}) \ket{\mathcal{S}} = \mathcal{C} \ket{G}.
\end{align}
From here, we skip the intermediate steps as they are analogous to Sec.~\ref{subapp:Backward}, and present the adjacency matrix $\Gamma$ of the resultant graph state $\ket{G}$
\begin{equation}
    \Gamma = 
    \begin{array}{|c!{\vrule width 2pt}c|c|c|c|c|c|}
        \hline 
        \Gamma_0 & 0 & 0 & 0 & \cdots & 0 & \textrm{A}_0 \\ \Xcline{1-7}{2pt} \rule{0pt}{10pt}
        0 & \textrm{A} & \text{LT}(\textrm{A}) & 0 & \cdots & 0 & 0\\ \hline 
        0 & \text{UT}(\textrm{A}) & \textrm{A} & \text{LT}(\textrm{A}) & \cdots & 0 & 0 \\ \hline 
        0 & 0 & \text{UT}(\textrm{A}) & \textrm{A} & \cdots & 0 & 0 \\ \hline 
        \vdots & \vdots & \vdots & \vdots & \ddots & \vdots & \vdots \\ \hline 
        0 & 0 & 0 & 0 & \cdots & \textrm{A} & \text{LT}(\textrm{A}) \\ \hline 
        & & & & & & \Gamma_{\textrm{X}}  \\
        \textrm{A}_0^T & 0 & 0 & 0 & \cdots & \text{UT}(\textrm{A}) & \oplusr \text{UT}(\textrm{Z}\textrm{X}^T) \\
        & & & & &  & \oplusr \text{LT}(\textrm{X}\textrm{Z}^T) \\
        \hline
    \end{array}.
\label{eq:GraphState_FW}
\end{equation}
This graph state is the same as in Eq.~\eqref{eq:GraphState_BW}, with the first and last auxiliary registers being swapped. In the same manner, we obtain the local Clifford operators compensating for the phases and self-loops as
\begin{align}
    \mathcal{C}_\textrm{aux.} = \prod_{l=1}^L S_{K,l}^{(2\textrm{R}+\Gamma_{\textrm{X}}^\circ - \textrm{N}_Y)_l},
\end{align}
where all operators act solely on the last time step $k=K$. However, $\mathcal{C} \ket{G}$ has to be followed by the backward CNOT ladder $H^{\otimes M}\Lambda^\dagger(X)_{\text{FW}}H^{\otimes M}$, which does not comply with the temporal order of the measurement pattern, as discussed in Sec.~\ref{subsec:AC-MBQC}. While in this formalism, unlike in the previous section, we can avoid keeping the main qubits active until the last time step of the computation, this comes at the expense of all $KL$ auxiliary qubits being active at once. This is the reason why we opt for the \enquote{first backwards- then forwards} approach in our work. 

\section{Adjacency Matrices} \label{app:Adj}

In this Appendix, we collect all the adjacency matrices corresponding to graph states which were discussed in Sec.~\ref{sec:Applications}. This is mostly intended for the sake of completeness, and to avoid ambiguities when the edges appear too crowded in the graphs.

\begin{figure*}
    \centering
    \includegraphics[width=\textwidth]{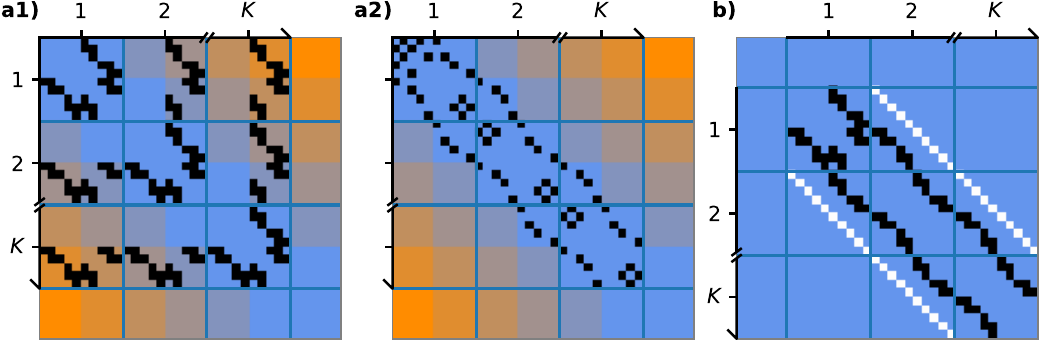}
    \caption{Adjacency matrices of the graph states representations of the resource state for the hybrid Trotterized Hamiltonian simulation of $XX$-type spin-correlations in the $XY$ model with $N=6$ for a) LC-MBQC, and b) AC-MBQC. The background in a) represents the distance matrix on a logarithmic color map. The panels a1) and a2) refer to the pre- and post-optimization graph states, respectively. The off-diagonals (white) in b) represent the forward CNOT ladder which is applied to the underlying graph state (black).}
    \label{fig:xy_even_x_adjs}
\end{figure*}

\begin{figure*}
    \centering
    \includegraphics[width=\textwidth]{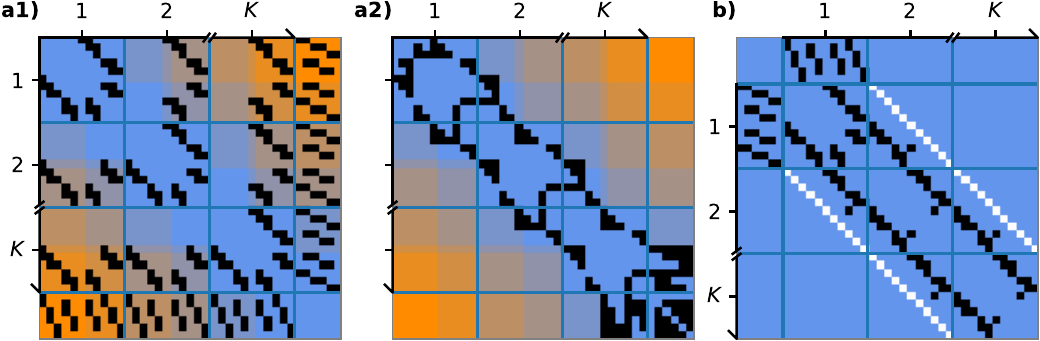}
    \caption{Adjacency matrices of the graph states representations of the resource state for Trotterized Hamiltonian simulation of the perturbed $XY$ model with $N=6$ for a) LC-MBQC, and b) AC-MBQC. The background in a) represents the distance matrix on a logarithmic color map. The panels a1) and a2) refer to the pre- and post-optimization graph states, respectively. The off-diagonals (white) in b) represent the forward CNOT ladder which is applied to the underlying graph state (black).}
    \label{fig:xy_even_imp_adjs}
\end{figure*}

\begin{figure*}
    \centering
    \includegraphics[width=\textwidth]{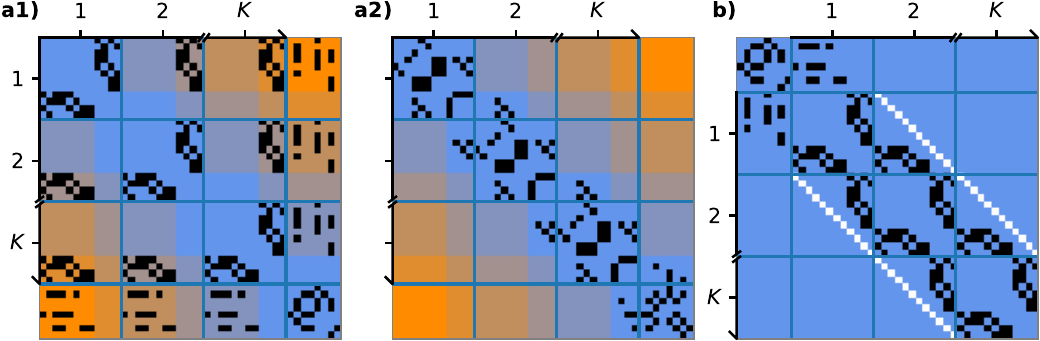}
    \caption{Adjacency matrices of the graph states representations of the resource state for Trotterized Hamiltonian simulation of the perturbed toric code Hamiltonian starting from the toric code state for a) LC-MBQC, and b) AC-MBQC. The background in a) represents the distance matrix on a logarithmic color map. The panels a1) and a2) refer to the pre- and post-optimization graph states, respectively. The off-diagonals (white) in b) represent the forward CNOT ladder which is applied to the underlying graph state (black).}
    \label{fig:toric_adjs}
\end{figure*}

\begin{figure*}
    \centering
    \includegraphics[width=\textwidth]{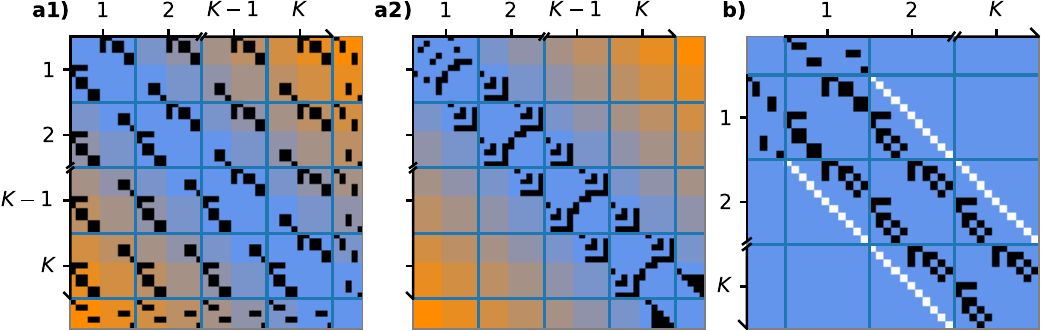}
    \caption{Adjacency matrices of the graph states representations of the resource state for universal quantum computation generated by the minimal 2-local set with $N=5$ for a) LC-MBQC, and b) AC-MBQC. The background in a) represents the distance matrix on a logarithmic color map. The panels a1) and a2) refer to the pre- and post-optimization graph states, respectively. The off-diagonals (white) in b) represent the forward CNOT ladder which is applied to the underlying graph state (black).}
    \label{fig:2-local_adjs}
\end{figure*}

\begin{figure*}
    \centering
    \includegraphics[width=\textwidth]{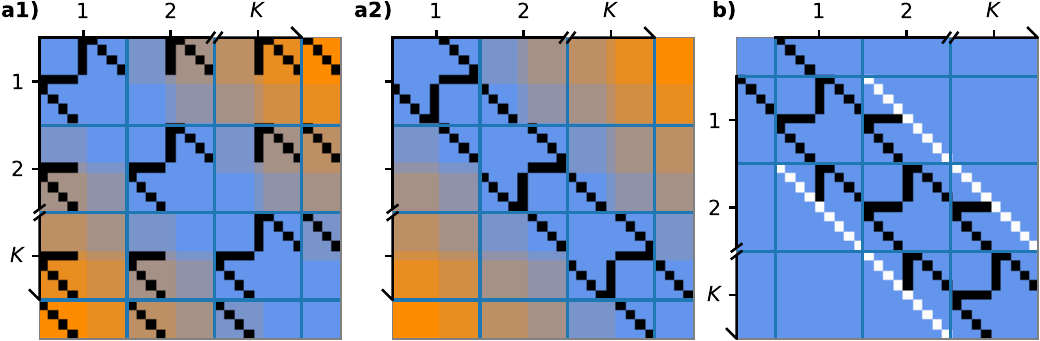}
    \caption{Adjacency matrices of the graph states representations of the resource state for universal quantum computation generated by the minimal set containing all weight 1 Pauli-strings with $N=4$ (even) for a) LC-MBQC, and b) AC-MBQC. The background in a) represents the distance matrix on a logarithmic color map. The panels a1) and a2) refer to the pre- and post-optimization graph states, respectively. The off-diagonals (white) in b) represent the forward CNOT ladder which is applied to the underlying graph state (black).}
    \label{fig:1-local_even_adjs}
\end{figure*}

\begin{figure*}
    \centering
    \includegraphics[width=\textwidth]{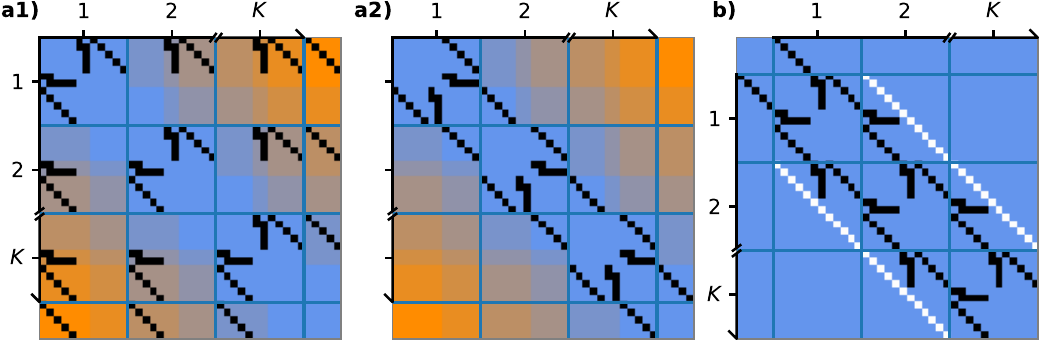}
    \caption{Adjacency matrices of the graph states representations of the resource state for universal quantum computation generated by the minimal set containing all weight 1 Pauli-strings with $N=5$ (odd) and $s=2$ for a) LC-MBQC, and b) AC-MBQC. The background in a) represents the distance matrix on a logarithmic color map. The panels a1) and a2) refer to the pre- and post-optimization graph states, respectively. The off-diagonals (white) in b) represent the forward CNOT ladder which is applied to the underlying graph state (black).}
    \label{fig:1-local_odd_adjs}
\end{figure*}

\begin{figure*}[htb!]
    \centering
    \includegraphics[width=\textwidth]{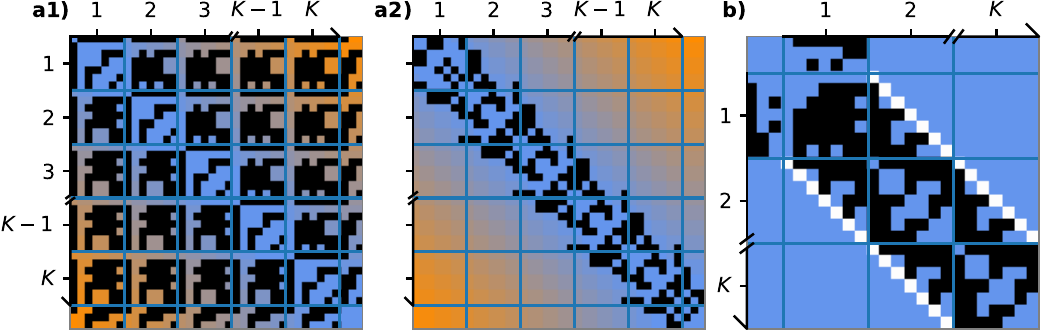}
    \caption{Adjacency matrices of the graph states representations of the resource state for universal quantum computation generated by the CQCA with $N=3$ for a) LC-MBQC, and b) AC-MBQC. The background in a) represents the distance matrix on a logarithmic color map. The panels a1) and a2) refer to the pre- and post-optimization graph states, respectively. The off-diagonals (white) in b) represent the forward CNOT ladder which is applied to the underlying graph state (black).}
    \label{fig:cqca_adjs}
\end{figure*}

\end{document}